\documentclass[aps,prf,reprint,superscriptaddress,onecolumn,notitlepage]{revtex4-1}

\usepackage[english]{babel}
\usepackage[T1]{fontenc}
\usepackage[title]{appendix}
\usepackage[a4paper,top=3cm,bottom=2cm,left=3cm,right=3cm,marginparwidth=1.75cm]{geometry}
\usepackage{etoolbox}
\usepackage{amsmath,amssymb,amsthm}
\usepackage{subcaption}
\usepackage{tikz}
\usepackage{graphicx}
\usepackage{xcolor}
\usepackage[colorlinks=true, allcolors=blue]{hyperref}
\usepackage{mathtools}
\usepackage{MnSymbol}
\usepackage{cleveref}
\usepackage{ragged2e}

\DeclareCaptionJustification{justified}{\justifying}

\crefname{appsec}{appendix}{appendices}
\Crefname{appsec}{Appendix}{Appendices}

\makeatletter
\newcommand*{\declarecommand}{%
  \@star@or@long\declare@command
}
\newcommand*{\declare@command}[1]{%
  \provide@command{#1}{}%
  \renew@command{#1}%
}
\makeatother

\newtheorem{remark}{Remark}

\let\oldsigma\sigma
\declarecommand{\sigma}{\boldsymbol{\oldsigma}}

\declarecommand{\u}{{\mathbf{u}}}
\declarecommand{\U}{{\mathbf{U}}}
\declarecommand{\f}{{\mathbf{f}}}
\declarecommand{\x}{{\mathbf{x}}}
\declarecommand{\n}{{\mathbf{n}}}
\declarecommand{\N}{{\mathbf{N}}}
\declarecommand{\X}{{\mathbf{X}}}
\declarecommand{\V}{{\mathbf{V}}}
\declarecommand{\v}{{\mathbf{v}}}
\declarecommand{\b}{{\mathbf{b}}}
\declarecommand{\a}{{\mathbf{a}}}
\declarecommand{\d}{{\mathbf{n}}}
\declarecommand{\c}{{\mathbf{c}}}
\declarecommand{\l}{{\mathbf{l}}}
\declarecommand{\s}{{\alpha}} 
\declarecommand{\bmu}{{\boldsymbol{\mu}}}
\declarecommand{\balpha}{{\boldsymbol{\alpha}}}
\declarecommand{\bbeta}{{\boldsymbol{\beta}}}
\declarecommand{\grad}{\nabla}
\declarecommand{\Id}{\mathbb{I}}
\declarecommand{\F}{{\boldsymbol{\mathcal{F}}}}
\declarecommand{\A}{{\mathcal{A}}}
\declarecommand{\au}{{\langle\u\rangle}}
\declarecommand{\ap}{{\langle p\rangle}}
\declarecommand{\af}{{\langle\f\rangle}}
\declarecommand{\cu}{{\overline\u}}
\declarecommand{\cp}{{\overline p}}
\declarecommand{\auj}{{\langle\cu^{j}\rangle}}
\declarecommand{\apj}{{\langle\cp^{j}\rangle}}
\declarecommand{\IEL}{\mathcal{I}_{E\to L}}
\declarecommand{\ILE}{\mathcal{I}_{L\to E}}
\declarecommand{\AJ}[1]{\mathcal{C}^{#1}}
\declarecommand{\NDU}{\textnormal{U}}
\DeclarePairedDelimiter{\discreteaverage}{\llangle}{\rrangle}

\definecolor{mypurple}{HTML}{BF77F6}

\makeatletter
\patchcmd{\@makecaption}{\@ifdim{\wd\@tempboxa >\hsize}}{\@firstoftwo}{}{}
\makeatother

\begin{document}

\title{Coarse-graining the dynamics of immersed and driven fiber assemblies}
\author{David B. Stein}
\email{dstein@flatironinstitute.org}
\affiliation{Center for Computational Biology, Flatiron Institute, New York, NY 10010, USA}
\author{Michael J. Shelley}
\affiliation{Center for Computational Biology, Flatiron Institute, New York, NY 10010, USA}
\affiliation{Courant Institute, New York University,
New York, NY 10012, USA}
  
\begin{abstract}
    An important class of fluid-structure problems involve the dynamics of ordered arrays of immersed, flexible fibers. While specialized numerical methods have been developed to study fluid-fiber systems, they become infeasible when there are many, rather than a few, fibers present, nor do these methods lend themselves to analytical calculation. Here, we introduce a coarse-grained continuum model, based on local-slender body theory, for elastic fibers immersed in a viscous Newtonian fluid. It takes the form of an anisotropic Brinkman equation whose skeletal drag is coupled to elastic forces. This model has two significant benefits: (1) the density effects of the fibers in a suspension become analytically manifest, and (2) it allows for the rapid simulation of dense suspensions of fibers in regimes inaccessible to standard methods. As a first validation, without fitting parameters, we achieve very reasonable agreement with 3D Immersed Boundary simulations of a bed of anchored fibers bent by a shear flow. Secondly, we characterize the effect of density on the relaxation time of fiber beds under oscillatory shear, and find close agreement to results from full numerical simulations. We then study buckling instabilities in beds of fibers, using our model both numerically and analytically to understand the role of fiber density and the structure of buckling transitions. We next apply our model to study the flow-induced bending of inclined fibers in a channel, as has been recently studied as a flow rectifier, examining the nature of the internal flows within the bed, and the emergence of inhomogeneous permeability. Finally, we extend the method to study a simple model of metachronal waves on beds of actuated fibers, as a model for ciliary beds. Our simulations reproduce qualitatively the pumping action of coordinated waves of compression through the bed.
\end{abstract}

\maketitle

\section{Introduction}

Many fundamental hydrodynamic phenomena, particularly in biology, involve the interaction of structured arrays of immersed fibers, often anchored to a substrate \cite{dRLNS2019}. For example, in eukaryotic cells, arrays of aligned microtubules in the spindle orchestrate the segregation of chromosomes, while those around centrosomes help position the spindle prior to cell division \cite{howard2001mechanics}. During mid-oogenesis of \emph{Drosophila}, kinesin motors interacting with the microtubule cytoskeleton drive large-scale coherent flows known as cytoplasmic streaming \cite{GWPG2012}.  Another example is the beds of driven cilia that pump fluid, move mucus, or propel microorganisms \cite{brennen1977fluid}. In microfluidic engineering, fabricated arrays of flexible fibers are the elements of proposed soft flow rectifiers \cite{Alvarado2017}. A central aspect of all of these examples is that the relevant dynamics is collective and not well-described by the dynamics of a single fiber. 

Given the importance of fluid-fiber systems, specialized computational methods have been developed to treat them, most especially in the zero Reynolds limit where flows are governed by the Stokes equation. These approaches include the use of nonlocal slender body theory \cite{TS2004,NRZS2017}, the immersed boundary method \cite{Peskin2002,Stockie1998,Lim2004,Nguyen2014}, bead-spring or -rod models \cite{HLHN2011,LU2009,SK2004,Delmotte2015}, the regularized Stokeslet method \cite{Cortez2001,Smith2009,BLY2011,OLC2013}, and overlapping grid methods \cite{Mitran2007}. See \cite{LS2015} for a recent review. Because of computational constraints these methods are typically applied to simulating one or a few fibers. A few exceptions exist, including the works of \cite{NRZS2017,rostami2016kernel}, which use slender-body hydrodynamics and regularized Stokeslets, respectively, and to which fast solution methodologies for the Stokes equations (such as the Fast Multipole Method \cite{NRZS2017,NRNS2017}) can be applied. Even so, these simulations are restricted to $O(1000)$ fibers, and are lengthy in execution even in massively parallel computing environments. Further, discrete fiber formulations do not provide a good basis for analytical investigations of many-fiber systems. In \cite{hussong2011continuum}, the authors develop a continuum model, based on volume-averaging the finite Reynolds number Navier-Stokes equations, to study metachronal waves in fiber beds attached to a wall. They consider the specific case of rigid fibers beating in a prescribed pattern that is stationary in the frame co-moving with the wave. For many systems, the emergent deformation of the fiber bed is of primary interest, and studying these systems requires a fully coupled theory.

Here, we develop a coarse-grained continuum model for the motion of aligned assemblies of elastic fibers moving through a Newtonian fluid. The model takes the form of an anisotropic Brinkman equation for coarse-grained fluid velocity and pressure fields driven by the drag forces of the immersed fibers. Solution of the Brinkman equation gives the background fluid velocity and velocity gradients that move and deform the fiber assembly, where fiber velocities are approximated via slender-body theory and internal fiber forces are those for Euler elasticae. This model has two significant benefits: {\bf (1)} the density effects of the fibers in a suspension become analytically manifest and exploitable, leading to the generalization of approaches to study basic instabilities of single fibers in flows; {\bf (2)} The model allows for the rapid simulation of dense suspensions of fibers in regimes inaccessible to standard methods. For this we have developed numerical methods that solve the coupled Brinkman-Elastica (BE) equations, as well as couple them to an outer fluid.

In Sect. \ref{section:derivation} we derive of our BE model using formal asymptotics and recent theoretical work justifying slender-body theory \cite{mori2018theoretical}, and in Sect.~\ref{section:numerics} we discuss numerical methods for its solution. As a first validation and without fitting parameters, we achieve very reasonable agreement of the BE model with 3D Immersed Boundary simulations of a bed of anchored fibers bent by a shear flow (Sect. \ref{section:deflection}). Secondly, we characterize the effect of density within the BE model on the relaxation time of fiber beds under oscillatory shear, and find close agreement to previous results from the full numerical simulations of Nazockdast et al \cite{NRZS2017} (Sect.~\ref{section:rheology}). Aside from their particulars, these first two studies show that hydrodynamic interactions within the bed generally stiffen fiber response as fiber density increases. We then study buckling instabilities in beds of fibers, using the BE model both numerically and analytically to understand the role of fiber density and the structure of buckling transitions (Sect.~\ref{section:buckling}). We next apply our model to study the flow-induced bending of inclined fibers in a channel, as has been recently studied as a flow rectifier, examining the nature of the internal flows within the bed, and the emergence of inhomogeneous permeability (Sect.~\ref{section:rectification}). Finally, we extend the method to study more complex flows and fiber deformations, in this case through a simple model of metachronal waves on beds of actuated fibers (Sect.~\ref{section:waves}). This is motivated by internally driven waves in ciliary beds, and our simulations reproduce qualitatively the observed pumping action of coordinated waves of bed compression \cite{NawrothEtAl2017}.

\section{Model formulation and derivation}
\label{section:derivation}

We assume that the immersing fluid is Newtonian and that inertial forces are negligible, so that the fluid flow between the fibers is described by the Stokes equations:
  \begin{align}
    -\mu\Delta\u + \grad p  = 0,\qquad \grad\cdot\u= 0,
  \end{align}
where $\u$ is the fluid velocity, $p$ is the pressure, and $\mu$ is the viscosity. The immersed fibers are taken to be inextensible Euler-Bernoulli elasticae, although the ensuing derivation does not depend on this and a different constitutive model could be used. Thus, for a fiber with configuration $\X(\s)$ where $\s$ is an arclength parameterization, the fiber generates a force-per-unit-length of:
\begin{equation}
    \F(\X, T)  = -E\X_{\s\s\s\s} + (T\X_\s)_\s.
    \label{equation:Euler_Bernoulli}
\end{equation}
Here $E$ is the bending rigidity of the fiber, and $T$ is the tension, a Lagrange multiplier that enforces the inextensibility of the fiber ($\s$ is thus material to the fiber flow $\X_t$). Variables with subscripted $\s$'s denote differentiation with respect to the arclength $\s$. The arclength derivative of the position $\X$ gives the unit-tangent to the curve, which we will denote by $\d=\X_\s$.

Consider a collection of $K$ fibers, with $i$ indexing the individual fibers, for $1\leq i\leq K$. Let $\Sigma^i$ denote the volume occupied by that fiber, with $\Gamma^i$ its boundary. The outward-oriented normal and Jacobian are denoted by $\N^i$ and $\mathcal{J}^i$ respectively. Furthermore, define $\Gamma^i_\s$ to be the one-dimensional ring defined by the intersection of $\Gamma^i$ and the plane orthogonal to $\n^i(\s)$ that passes through $\X^i(\s)$. We assume that the velocity $\u$ is given everywhere in the fluid domain $\Omega=\mathbb{R}^3\setminus\cup_{i}\Sigma^i$ by the solution to:
\begin{subequations}
    \label{equation:u_not_missing}
    \begin{align}
        -\mu\Delta\u + \grad p &= 0,\qquad\grad\cdot\u = 0\qquad\textnormal{in }\Omega = \mathbb{R}^3\setminus\cup_{i}\Sigma^i,\label{equation:u_not_missing:a}\\
        \int_{\Gamma^i_\s}\sigma \N^i\mathcal{J}^i\, d\theta &= \F^i(\s),\qquad\textnormal{ for }1\leq i\leq K,\label{equation:u_not_missing:b}\\
        \u|_{\Gamma^i_\s} &= \mathmakebox[0pt][l]{\u^i(\s),}\hphantom{\F^i(\s),\,}\qquad\textnormal{ for }1\leq i\leq K.\label{equation:u_not_missing:constraint}
    \end{align}
\end{subequations}
along with appropriate far-field boundary conditions for $\u$ (or no-slip conditions on walls). In these equations, $\sigma=\mu(\grad\u + \grad\u^\intercal) - p\boldsymbol{\mathbb{I}}$ denotes the viscous stress tensor, and the fiber centerlines move at an unknown velocity $\partial_t\X^i(\s)=\u^i(\s)$ determined by balancing the hydrodynamic stress integrated over $\Gamma^i_\alpha$ against $\F^i(\alpha)$. \Cref{equation:u_not_missing:constraint} is a constraint that the fibers move semi-rigidly: all points in $\Gamma^i_\s$ must move with the fiber centerline velocity $\u^i(\s)$; this constraint is sufficient to fix $\sigma$ on $\Gamma^i$. The fact that this system is well-posed and that the difference between its solution and the classical slender-body theory representation in the $L^2(\Omega)$ norm is small (proportional to $\epsilon\log\epsilon$) is proven in \cite{mori2018theoretical}.

\Cref{equation:u_not_missing}, when coupled with appropriate boundary conditions for the fibers, provides a full framework for the evolution of the fully coupled fluid-fiber system. In the examples presented in this paper, the fiber will be clamped at $\s=0$ and free at $\s=L$. Many other boundary conditions are physically relevant; they require only minimal modifications to the theory and numerics as presented. As will be shown in \Cref{section:derivation:coarse-graining}, \Cref{equation:u_not_missing:a,equation:u_not_missing:b} are relatively simple to coarse-grain; \Cref{equation:u_not_missing:constraint}, however, is not. The fiber moves with the local velocity $\u$ of the fluid, but will move \emph{relative} to the local coarse-grained fluid velocity. To capture this we turn to classical slender-body theory.

The motion of a slender fiber moving in a background velocity field $\U(\x)$ may be described using slender-body theory \cite{KR1976,Johnson1980,Gotz2000,TS2004,mori2018theoretical,koens2018boundary}. For a fiber of length $L$, we assume that the fibers have a circular cross-section in the plane normal to $\d(\s)$ with radius $r_\textnormal{fiber}(\s)=2\epsilon\sqrt{\s(L-\s)}$, where $\epsilon=r_\textnormal{fiber}(L/2)/L\ll 1$ describes the slenderness of the fiber. To leading order in $\epsilon$, the motion of the fiber is given by local slender-body theory:
\begin{equation}
	8\pi\mu(\partial_t\X(\s) - \U(\X(\s))) = c(\Id + \d(\s)\d(\s))\F(\s).
\end{equation}
  Here $c=-\log(\epsilon^2e)>0$. For brevity we define $\eta=8\pi\mu/c$, $\mathcal{A}=\Id+\d\d$, and suppress all functional dependence unless required for clarity, writing this relationship as:
\begin{equation}
	\eta(\partial_t\X - \U) = \A\F.
\end{equation}
Higher-order expansions lead to non-local slender-body theory; for more details see \cite{KR1976,Johnson1980,Gotz2000,TS2004,mori2018theoretical,koens2018boundary}.

We neglect the nonlocal contributions from SBT to fiber self-induction. Hence, when there are multiple fibers present, each fiber sees only the fluid velocity due to the forces generated by the other fibers. Thus the evolution for each fiber is given by 
\begin{align}
\label{ImLocal}
    \eta(\partial_t\X^j - \cu^j) = \A^j\F^j
\end{align}
where $\cu^j$ is the \emph{complementary velocity}, i.e. the velocity field induced by all fibers \emph{except} the $j^\textnormal{th}$ fiber. This velocity is defined by a modified version of \Cref{equation:u_not_missing}, with the contribution of the $j^\textnormal{th}$ fiber removed:
\begin{subequations}
    \label{equation:u_missing}
    \begin{align}
        -\mu\Delta\cu^j + \grad\cp^j &= 0,\qquad\grad\cdot\cu^j = 0\qquad\textnormal{in }\overline{\Omega}^j = \mathbb{R}^3\setminus\cup_{i\neq j}\Sigma^j,\\
        \int_0^{2\pi}\overline\sigma^j \N^i\mathcal{J}^i\, d\theta &= \F^i(\s),\qquad\textnormal{ for }1\leq i\leq K,\ i\neq j,\\
        \cu^j|_{\Gamma^i} &= \mathmakebox[0pt][l]{\cu^i(\s),}\hphantom{\F^i(\s),\,}\qquad\textnormal{ for }1\leq i\leq K,\ i\neq j,
    \end{align}
\end{subequations}
along with appropriate far-field boundary conditions for $\cu^j$. Equation~(\ref{ImLocal}) is solved together with the inextensibility constraint, clamped boundary conditions at the base of the fibers, ($\n^j(0)=\n_0^j$, $\X^j(0)=\X_0^j$), and free-end boundary conditions at the tips ($\X^j_{\s\s}(L)=\X^j_{\s\s\s}(L)=T^j(L)=0$), and an initial condition $(\X^j(t=0)= \X_{\textnormal{init}}^j)$. Equation~(\ref{equation:u_missing}) is solved together with whatever boundary conditions on $\cu^j$ are relevant (i.e., no-slip on a wall). Note that when a global constraint exists in the fiber constitutive model (e.g. inextensibility), the individual problems for $\cu^j$ are not decoupled, as the tensions $T^i$ in each fiber are not known \emph{a priori}.

\begin{remark}
A version of this system, Eqs.~(\ref{ImLocal}) \& (\ref{equation:u_missing}), based on slender-body theory, has been used, in conjunction with the Fast Multipole Method, to simulate many-fiber systems, including the transport and positioning of the mitotic spindle \cite{NRZS2017,NRNS2017} by centrosomal microtubules. That formulation used line distributions of Stokeslets to represent the induced velocity fields from the fibers and boundary integral representations to capture the backflows from other immersed bodies and bounding surfaces.
\end{remark}
\begin{remark} 
The inclusion of velocity contributions from surrounding fibers but exclusion of nonlocal self-induction is not asymptotically consistent as both contributions arise at the same order in slender body theory. However, Nazockdast {\it et al.} \cite{NRZS2017} studied the contribution of nonlocal self-induction to fiber motion in many-fiber systems and found that its contribution was negligible in comparison to the flows induced by surrounding fibers. 
\end{remark}
Many-fiber simulations are very time-consuming, and the many-fiber model is challenging to draw analytic conclusions from. When $K$ is large and the initial fiber positions and parameters are coherent across length scales significantly larger than the inter-fiber spacing, it is feasible to coarse-grain many fiber systems.

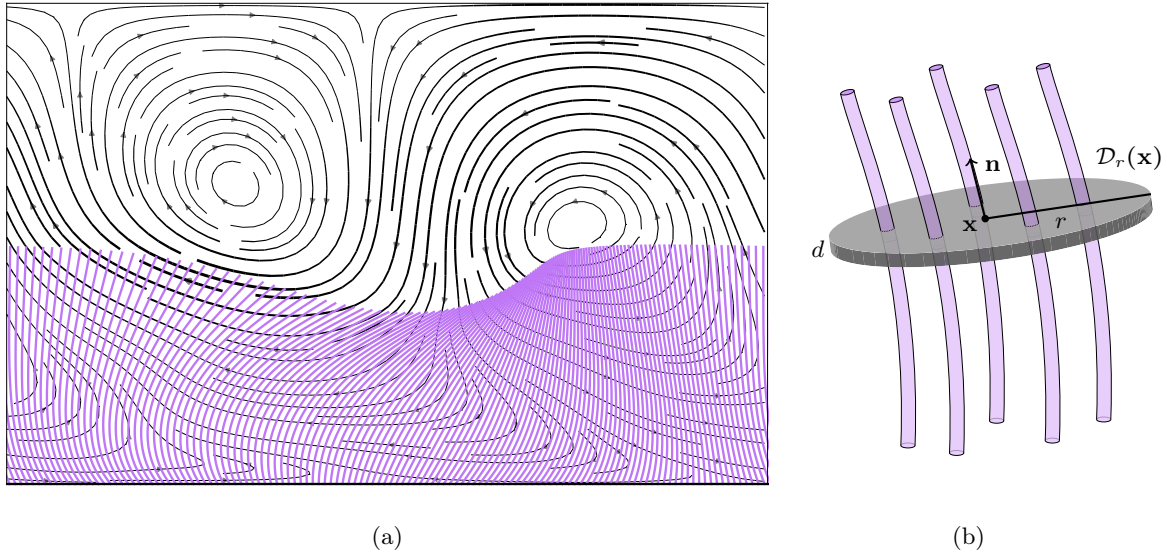
\begin{figure}[h!]
    \centering
    \subcaptionbox{\label{figure:diagram:1}}[0.69\textwidth]{
        \includegraphics[width=0.69\textwidth]{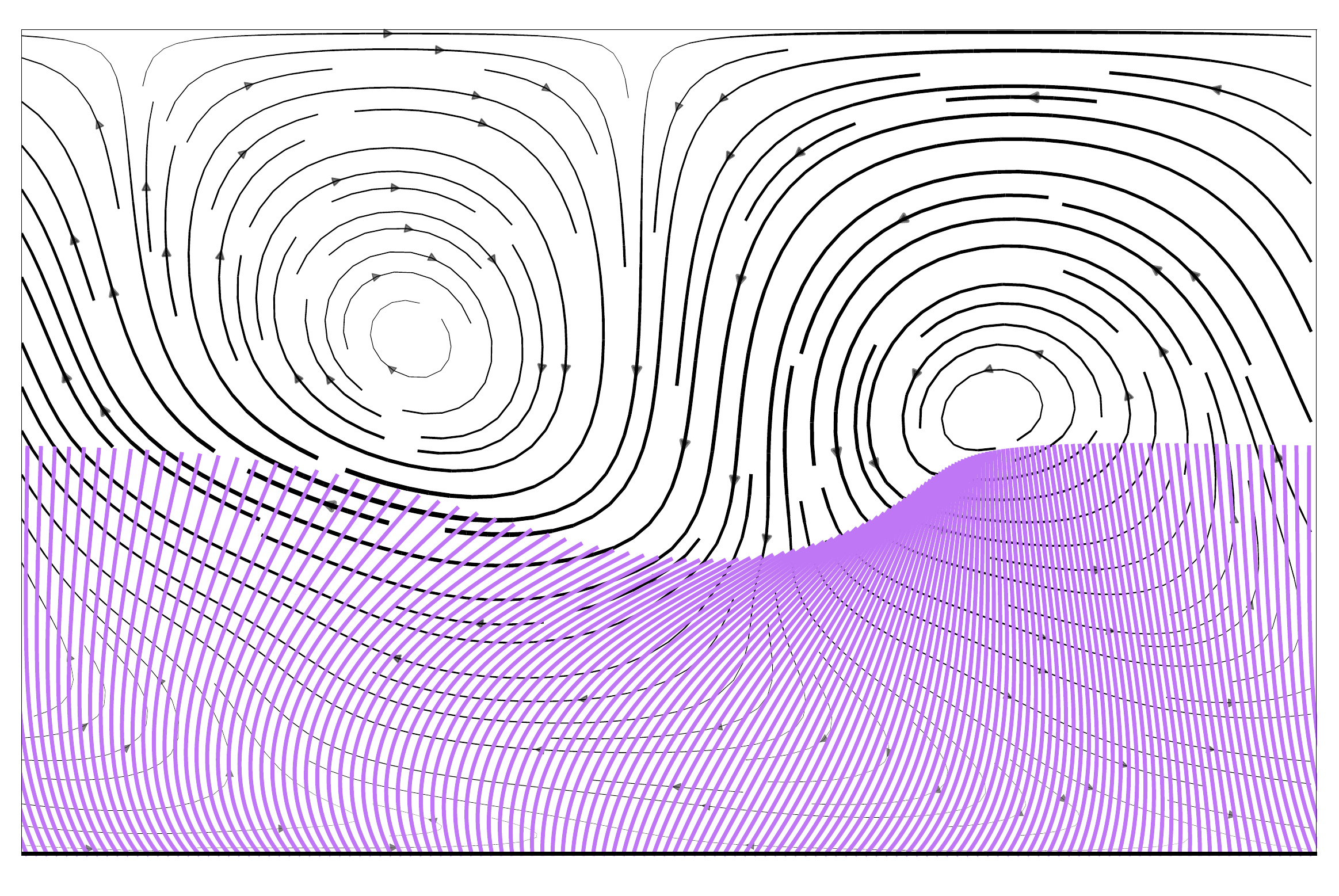}
    }
    \hfill
    \subcaptionbox{\label{figure:diagram:2}}[0.29\textwidth]{
        \begin{tikzpicture}
    	    \node[anchor=south west,inner sep=0] at (0,0){\includegraphics[width=0.29\textwidth]{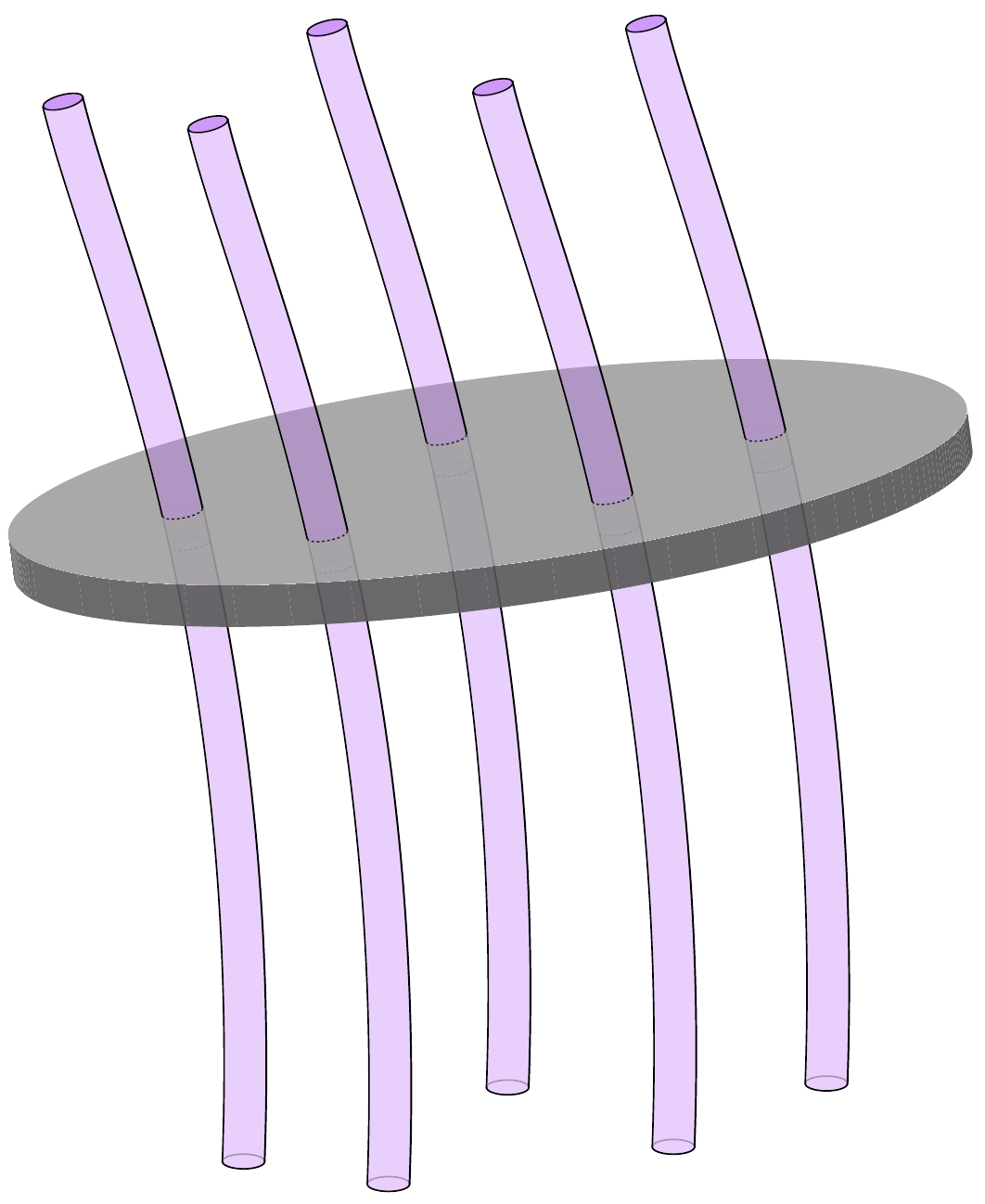}};
    	    \node at (4,4) {$\mathcal{D}_r(\x)$};
    	    \node at (2.1,3.2)[circle,fill,inner sep=1pt,fill=black]{};
    	    \node at (1.9,3.1) {$\x$};
    	    \node at (-0.1,2.8) {$d$};
    	    \node at (2.2,3.9) {$\n$};
    	    \node at (3.1,3.15) {$r$};
    	    \draw [-,thick] (2.1,3.2) -- (4.28,3.525);
    	    \draw [->,thick] (2.1,3.2) -- (1.9,4.0);
    	\end{tikzpicture}%
    	\vspace{2em}
    }
    \caption{Many biological and industrial problems involve a regular bed of fibers that are clamped to a boundary, as shown in Panel (a) (which depicts the boundary to which the fibers are fixed as flat, though it need not be). When fiber orientation and motion are coherent, it is viable to coarse-grain the fluid-fiber system. The effect of the fibers on the fluid is coarse-grained by averaging the forces exerted by the fibers on the fluid in a disk of radius $r$ and thickness $d$ that lies in the plane normal to the unit-tangent $\n$ at $\x$, denoted by $\mathcal{D}_r(\x)$, shown as the gray disk in Panel (b).}
    \label{figure:diagram}
\end{figure}


\subsection{Coarse-graining of the fluid-fiber system}
\label{section:derivation:coarse-graining}

For simplicity, we fix a physical situation and consider a regular bed of $K$ fibers with uniform inter-fiber base spacing $\delta$, length $L$, bending rigidity $E$, and slenderness ratio $\epsilon$. The fibers are clamped at their bases on the plane $z=0$, and at an orientation that may vary as a function of time. They may be subjected to pressure or shear driven flows, or internally driven through fiber boundary conditions. This basic configuration is shown in \Cref{figure:diagram}. We can denote any position within the fiber assembly by the fiber index $j$ and arclength $\s$ along the $j^\textnormal{th}$ fiber. To feasibly coarse grain the system, we assume that there exists a length scale $l$ over which average fluid velocities vary in the direction transverse to the fibers, and that the fiber radius, fiber separation, and length scale of variation satisfy $r_\textnormal{fiber}\ll\delta\ll l$.

Assuming that the fibers do not intersect with one another, we may define a coordinate system on the fiber field using the fiber base coordinates. In particular, we introduce a base coordinate $\bbeta$, and fiber field coordinate $\balpha=(\alpha,\bbeta)$, and a smooth(ed) fiber field $\tilde\X(\balpha,t)$ such that $\tilde\X\left(\alpha,\bbeta=\X_j(0)\right) = \X_j(\alpha)$ for all $\alpha$ and $1\leq j\leq K$. The field $\tilde\X(\balpha)$ is the Lagrangian flow map \emph{for the fiber field}. We assume that $\tilde\X(\balpha)$ is smooth, invertible, and that the Jacobian $J=|\partial\X/\partial\balpha|$ is bounded from above and below. Note that the Jacobian $J$ is distinct from $\mathcal{J}$ used in \Cref{equation:u_not_missing}. Since $\tilde\X$ is assumed to be smooth, we can define the orientation field $\tilde\n=\partial_\alpha\tilde\X$. Although we consider only fiber collections with constant length $L$ and bending rigidity $E$, collections of fibers in which these quantities vary slowly with respect to the inter-fiber spacing $\delta$ may be handled naturally with minimal modification by defining appropriate fields, as functions of $\bbeta$.

We now define a moving average operator, $\langle\cdot\rangle$, by its action on a function $g(\x)$ relative to the local fiber orientation field $\tilde{\n}$:
\begin{equation}
\langle g\rangle(\x) = \frac{1}{\pi r^2d}\int_{\mathcal{D}^r(\x;\tilde\n)}dV_{\x'} g(\x'),
\end{equation}
where $\mathcal{D}^r(\x;\tilde\n)$ is the cylindrical disk of radius $r\gg\delta$ and thickness $d\ll r$, centered at $\x$ with normal $\tilde\n(\x)$, excised of the volume occupied by the fibers (see \Cref{figure:diagram:2} for a depiction of $\mathcal{D}^r(\x;\tilde\n)$). Note that the volume of $\mathcal{D}^r(\x;\tilde\n)$ is $\pi r^2 d$, up to $\mathcal{O}((r_\textnormal{fiber}/\delta)^2)$. This averaging operator is appropriate for Eulerian functions defined in the fluid region. For functions defined on the centerline of fibers, we define a discrete averaging operator $\discreteaverage{\cdot}$ as:
\begin{equation}
    \discreteaverage{G}(\balpha) = \frac{1}{N(\tilde\X(\balpha))d}\sum_{j\in\mathfrak{J}}\int_{\AJ{j}}G^j(\s')\,d\s',
\end{equation}
where $\mathfrak{J}=\left\{j\,|\,\X(\s^j)\in\mathcal{D}^r(\tilde\X(\balpha);\tilde\n)\right\}$ is the set of all fiber indices $j$ such that the fiber intersects $\mathcal{D}^r(\tilde\X(\balpha);\tilde\n)$, $\AJ{j}$ denotes the set of values $\s$ such that $\X^j(\s)\in\mathcal{D}^r(\tilde\X(\balpha);\tilde\n)$, and $N(\tilde\X(\balpha))$ is the number of elements in $\mathfrak{J}$. The fiber density-per-unit-area $\rho(\tilde\X(\balpha))$ in the plane orthogonal to $\tilde\n(\balpha)$ is defined to be:
\begin{equation}
  \rho(\tilde\X(\balpha)) = \frac{N(\tilde\X(\balpha))}{\pi r^2}.
  \label{equation:rho_definition}
\end{equation}
Since $\rho(\tilde\X(\balpha), t)J(\balpha)=\rho(\balpha, 0)$, the initial density $\rho_0=\rho(\balpha, 0)$ may be thought of as a \emph{parameter} of the system. In generic settings, $\rho_0$ is field-valued, but when the initial geometry is simple, $\rho_0$ will often be a constant, or a simple function depending on the geometry of the fiber bed. As a closure condition, we assume that the fiber field $\tilde \X$ satisfies $\tilde\X(\balpha)=\discreteaverage{\X}(\balpha)$.

We now derive a few useful properties of these operators. As a consistency check, we may compute that:
\begin{equation}
    \partial_\s\tilde\X(\balpha) = \frac{1}{N(\tilde\X(\balpha))d}\sum_{j\in\mathfrak{J}}\partial_\s\int_{\AJ{j}}\X^j(\s')\,d\s' + \mathcal{O}(r),
\end{equation}
where we note that although $N$ changes as a function of $\s$, this is a normalization constant to the changing index set $\mathfrak{J}$, and the $\mathcal{O}(r)$ error is due to the fact that new fibers added to the index set $\mathfrak{J}$ may differ from those already in the index set by $\mathcal{O}(r)$. For an individual fiber $j$, we have that:
\begin{equation}
    \partial_\s\int_{\AJ{j}}\X^j(\s')\,d\s' = \lim_{\s\to0}\s^{-1}\left[\int_{\s_1+\s+\mathcal{O}(r)}^{\s_2+\s+\mathcal{O}(r)}\X^j(\s')\,d\s' - \int_{\s_1}^{\s_2}\X^j(\s')\,d\s'\right] = \int_{\AJ{j}}\n^j(\s')\,d\s'+\mathcal{O}(r),
\end{equation}
and thus $\tilde\n(\balpha)=\partial_\s\tilde\X(\balpha)=\discreteaverage{\n}(\balpha)$, to leading order in $r$. In general, differentiation with respect to $\s$ commutes with the discrete averaging operator $\discreteaverage{\cdot}$.

Additionally, we will need to take the averages of certain products. For smooth $F$ and $G$ with bounded derivatives as a function of $\alpha$ and that further satisfy $|F^j(\alpha)-F^k(\alpha)|\leq C|\X^j(\alpha)-\X^k(\alpha)|$, then we have:
\begin{equation}
    \discreteaverage{F}\discreteaverage{G} = \frac{1}{N^2d^2}\sum_{j\in\mathfrak{J}}\sum_{k\in\mathfrak{J}}\int_{\AJ{j}} F^j(\alpha')\,d\alpha'\int_{\AJ{k}} G^k(\alpha')\,d\alpha' = F^j(\tilde\alpha)G^j(\tilde\alpha) + \mathcal{O}(r),
\end{equation}
for any $j\in\mathfrak{J}$ and $\tilde\alpha\in\AJ{j}$. Thus $\discreteaverage{F}\discreteaverage{G}=\discreteaverage{FG}+\mathcal{O}(r)$. Finally, we note that for a smooth function $g$ defined everywhere, the discrete and continuous averaging operators may be related by $\langle g\rangle(\x)=\discreteaverage{g}(\tilde\X(\balpha)) + \|\grad g\|_{L^\infty}\mathcal{O}(\delta)$.
\begin{remark}
The two-dimensional, fiber-aligned nature of the averaging operator is necessary because averaged fluid velocities and forces may vary rapidly along the fiber length. For example, in a dense bed of immersed fibers, the penetration of the outer velocity field into the fiber bed decays exponentially. This choice of averaging operator does not smooth over that direction of possibly rapid variation, and commutes with differentiation with respect to $\s$.
\end{remark}
In applying the averaging operator $\langle\cdot\rangle$ to \Cref{equation:u_not_missing}, the main difficulty is in commutation of the averaging operation with differentiation. Consider taking a derivative in the $\mathbf{\hat x}$ direction. Let $\Omega(h)=\mathcal{D}^r(\x+h\mathbf{\hat x};\,\tilde\n(\x+h\mathbf{\hat x}))$. Then:
\begin{equation}
    \partial_x\langle g\rangle(\x) = \frac{1}{\pi r^2d}\frac{d}{dh}\left[\int_{\Omega(h)}dV_{\x'} g(\x')\right]_{h=0} = \frac{1}{\pi r^2d}\int_{\partial^e\mathcal{D}(\x;\tilde\n)}dS(\v\cdot\N) g,
\end{equation}
where $\v$ is the velocity (with respect to $h$) of points on the boundary, $\N$ is the normal vector to the boundary, and $\partial^e\mathcal{D}^r(\x;\tilde\n)$ denotes the \emph{exterior} boundary, that is, the boundary of $\mathcal{D}^r(\x;\tilde\n)$ excluding any fiber boundaries. Because the orientation angle of the disc, $\tilde\n$, depends on $\x$, $\v$ does not exactly equal $\mathbf{\hat x}$, but is instead $\v=\mathbf{\hat x} + \mathcal{O}(r)|\grad\tilde\n|$. Integrating by parts and keeping leading order terms in $r$, we find that:
\begin{equation}
    \partial_x\langle g\rangle(\x) = \langle\partial_x g\rangle(\x) - \frac{1}{\pi r^2d}\int_{\partial^f\mathcal{D}^r(\x;\tilde\n)}dS (\N\cdot\mathbf{\hat x}) g,
\end{equation}
where $\partial^f\mathcal{D}^r(\x;\tilde\n)$ denotes the collection of all fiber boundaries intersecting $\mathcal{D}^r(\x;\tilde\n)$. Applying this recipe to \Cref{equation:u_not_missing}, and again dropping terms of size $\mathcal{O}(r)$ gives:
\begin{equation}
    -\mu\grad\cdot\langle\grad\u + \grad\u^\intercal\rangle + \grad\langle p\rangle - \frac{1}{\pi r^2d}\sum_{j\in\mathfrak{J}}\int_{\Gamma^j\cap\mathcal{D}^r(\x;\tilde\n)}\sigma\N\,dS=\mathbf{0}.
    \label{equation:averaging:first}
\end{equation}
Applying the averaging operation once more to derivatives of $\u$ gives:
\begin{equation}
    \langle\grad\u\rangle=\grad\langle\u\rangle + \frac{1}{\pi r^2d}\sum_{j\in\mathfrak{J}}\int_{\Gamma^j\cap\mathcal{D}^r(\x;\tilde\n)}\u\N\,dS.
    \label{equation:commute_around_grad_u}
\end{equation}
If the fiber field were straight (that is, $\tilde\n$ is constant), then each of the integrals in \Cref{equation:commute_around_grad_u} may be rewritten as:
\begin{equation}
    \int_{\Gamma^j\cap\mathcal{D}^r(\x;\tilde\n)}\u\N\,dS = \int_{\AJ{j}}\int_ 0^{2\pi}\u(\s)\N(\s,\theta)\mathcal{J}^j\,d\theta\,d\s.
    \label{equation:commute_around_grad_u:rewritten}
\end{equation}
and are thus exactly $0$ due to the constraint that the fiber moves semi-rigidly (see \Cref{equation:u_not_missing:constraint}). When the fiber field is not straight, the surface $\Gamma^j\cap\mathcal{D}^r(\x;\tilde\n)$ need not coincide with the simple cylindrical decomposition used in \Cref{equation:commute_around_grad_u:rewritten}. Instead, for the $j^\textnormal{th}$ fiber, define $\balpha^j_+$, $\balpha^j_-$, and $\balpha^j$ so that $\tilde\X(\balpha^j_+)$, $\tilde\X(\balpha^j_-)$, and $\tilde\X(\balpha^j)$ are the locations where $\X^j$ intersect the two caps of the cylindrical disk and the midplane of the disk, respectively. Because $J$ is bounded, $|\x-\tilde\X(\balpha^j_\pm)|=\mathcal{O}(r)$ implies that $|\balpha-\balpha^j_\pm|=\mathcal{O}(r)$ and $|\tilde\n(\balpha)-\tilde\n(\balpha^j_\pm)|=|\grad\tilde\n|\mathcal{O}(r)$. Thus error terms in the integrals in \Cref{equation:commute_around_grad_u:rewritten} are confined to wedges with area $\mathcal{O}(r)$. Because $\u$ is bounded and there are $\mathcal{O}(r^2)$ fibers in the sum in \Cref{equation:commute_around_grad_u}, we find that:
\begin{equation}
    \langle\grad\u\rangle=\grad\langle\u\rangle + \mathcal{O}(r).
    \label{equation:commute_around_grad_u:finished}
\end{equation}
We finally must deal with the integrals involving the stress $\sigma$ in \Cref{equation:averaging:first}. As in the commutation of $\langle\cdot\rangle$ around the $\grad$ operator in the term $\grad\u$ described in \Cref{equation:commute_around_grad_u:rewritten,equation:commute_around_grad_u:finished}, errors exist due to the fact that the fibers need not be exactly orthogonal to the averaging disks, but these errors are of size $\mathcal{O}(r)$, again, for the same reasons, and thus we may rewrite each of these integrals as:
\begin{equation}
    \int_{\Gamma^j\cap\mathcal{D}^r(\x;\tilde\n)}\sigma\N\,dS = \int_{\AJ{j}}\int_ 0^{2\pi}\sigma(\s,\theta)\N(\s,\theta)\mathcal{J}^j\,d\theta\,d\s + \mathcal{O}(r) = \int_{\AJ{j}}\mathcal{\F}^j(\alpha)\,d\s + \mathcal{O}(r),
\end{equation}
where the final equality holds due to \Cref{equation:u_not_missing:b}. Plugging this back into the final term of \Cref{equation:averaging:first}, we find that:
\begin{equation}
    \frac{1}{\pi r^2d}\sum_{j\in\mathfrak{J}}\int_{\Gamma^j\cap\mathcal{D}^r(\x;\tilde\n)}\sigma\N\,dS = \rho(\tilde\X(\balpha))\discreteaverage{\F}(\balpha) + \mathcal{O}(r).
    \label{equation:Fs_averaged}
\end{equation}

Combining \Cref{equation:averaging:first} with \Cref{equation:commute_around_grad_u:finished,equation:Fs_averaged,equation:rho_definition}, and dropping any terms of $\mathcal{O}(r)$, gives an equation for the averaged velocity field:
\begin{equation}
	-\mu\Delta\au(\x) + \grad\ap(\x) = (\rho_0 J^{-1} \tilde\F)(\tilde\X^{-1}(\x));\qquad
	\grad\cdot\au(\x) = 0,
	\label{equation:averaged_velocity}
\end{equation}
where $\tilde\F(\balpha)=-E\tilde\X_{\s\s\s\s} + (\tilde T\tilde \X_\s)_\s$ and we have replaced $\balpha$ by $\tilde\X^{-1}(\x)$, to make apparent the dependence on the Eulerian coordinate $\x$. The averaged incompressibility condition follows from the arguments used to derive \Cref{equation:commute_around_grad_u:finished}. We remark that the average incompressibility holds only because we have assumed $r_\textnormal{fiber}\ll\delta$, and thus the fibers take up a negligible proportion of the volume in the averaging disk $\mathcal{D}^r(\x,\tilde\n)$. This assumption restricts the formal applicability of the theory to fiber beds with a low volume fraction occupied by fibers. If this condition did not hold, the divergence of the averaged velocity field $\au$ would be offset by changes in the fiber density field $\rho$.

To determine how the Lagrangian flow map $\tilde\X$ evolves, we examine the evolution equation for the $j^\textnormal{th}$ fiber:
\begin{equation}
	\eta(\partial_t\X^j(\s) - \cu^{j}(\X(\s)) = \A^j(\s)\F^j(\s).
	\label{equation:single_evolution}
\end{equation}
We assume that the fiber $\X^j$ sees the velocity field obtained by coarse-graining the complementary velocity $\cu^j$. Applying the continuous averaging operator to the velocity field $\cu^{j}$, defined by \Cref{equation:u_missing}, gives the same result as applying the averaging operator to the velocity $\u$, except that the the sum excludes a single fiber. Thus the result is the same as $\u$ but with a modified density:
\begin{subequations}
  \begin{align}
    -\mu\Delta\auj(\x) + \grad\apj(\x) &= \rho_-(\x)\tilde\F(\tilde\X^{-1}(\x)) + \mathcal{O}(r),\qquad\grad\cdot\auj  = 0,
  \end{align}
\end{subequations}
where $\rho_-(\x) = \frac{N(\x)-1}{\pi r^2}$. Since $\delta\ll r$, then $N\gg1$, and so comparing to \Cref{equation:rho_definition}, we have that $\rho_-\approx\rho$, and hence $\auj\approx\au$. Applying the discrete averaging operator to \Cref{equation:single_evolution} and applying derived identities now gives:
\begin{equation}
    \eta(\partial_t\tilde\X(\balpha)-\auj(\tilde\X(\balpha)) = \tilde\A(\balpha)\tilde\F(\balpha),
    \label{equation:discrete_averaged_evolution}
\end{equation}
where $\tilde\A(\balpha)=\mathbb{I}+\tilde\n(\balpha)\tilde\n(\balpha)$. Combining \Cref{equation:averaged_velocity,equation:discrete_averaged_evolution} gives the full coarse-grained system that the averaged quantities $\au$ and $\ap$, along with the fiber field $\tilde\X$, satisfy to leading order (dropping explicit reference to the averaged and field quantities):
\begin{subequations}
	\label{equation:simplified_model}
	\begin{align}
    	-\mu\Delta\u(\x) + \grad p(\x) &= \rho(\x)\F(\X^{-1}(\x)),\qquad
        \grad\cdot\u(\x)= 0, \label{equation:simplified_model:1} \\
        \eta(\V(\balpha)-\u(\X(\balpha))            &= \A(\balpha)\F(\balpha),         \label{equation:simplified_model:3}
    \end{align}
\end{subequations}
where $\V=\partial_t\X$. We will refer to the coarse-grained model given by the system in \Cref{equation:simplified_model} as the Brinkman-Elasticae (BE) model. In some of the forthcoming material, we will drop explicit reference to coordinates for convenience; one must be careful to keep in mind that the $\F$ term in \Cref{equation:simplified_model:1} and the $\u$ term in \Cref{equation:simplified_model:3} require the coordinate transformations $\X^{-1}$ and $\X$, respectively. The Lagrangian transformation in \Cref{equation:simplified_model:1} brings forces generated by the deformed fiber field from the Lagrangian frame to the Eulerian frame and defines the instantaneous velocity field; \Cref{equation:simplified_model:3} evolves the fiber flow map according to local-slender body theory, relative to the background flow that is generated both by the fiber field itself as well as externally generated flows and backflows due to boundary conditions.

\begin{remark}
	It may not be immediately obvious that the Brinkman-Elasticae model defined in \Cref{equation:simplified_model} is a coarse-grained equation in a spirit similar to the Brinkman equation for flow through an isotropic porous media. To see this, we rewrite this set of equations as:
\begin{subequations}
	\begin{align}
    	-\mu\Delta\u + \rho\eta\mathcal{A}^{-1}(\u - \V) + \grad p &= 0,	\qquad\grad\cdot\u                                                 = 0,	\\
        \eta(\V-\u)                                                &= \A\F,
    \end{align}
\end{subequations}
from which the analogy to Brinkman is apparent: the equations describe the (relative) motion of a fluid through a porous media with an anisotropic permittivity $\rho\eta\mathcal{A}^{-1}=\rho\eta(\Id-\d\d/2)$.
\end{remark}

\begin{remark}
For the case of Euler-Bernoulli elasticae, the force density $\F=-E\X_{\s\s\s\s} + (T\X)_\s$ depends on the fourth-derivative of $\X$, as well as a global constraint due to the inextensibility of the fibers. The Brinkman Elasticae model given by \Cref{equation:simplified_model} is thus fourth-order, nonlinear, and non-local. From a numerical point of view, this system of equations is best evolved implicitly, and inversion of the discrete operators is challenging due to the hydrodynamic interactions between the fibers. An efficient numerical scheme for solving this system of equations is presented in \Cref{section:numerics}.
\end{remark}

\begin{remark}
It will often be convenient to work with the unit-tangent $\n$, rather than $\X$, as our fundamental field. Differentiating \Cref{equation:simplified_model:3} with respect to $\s\textbf{}$, we find:
\begin{equation}
	\eta(\n_t - \partial_\s\u) = \partial_\s\mathcal{A}\F = \partial_\s\mathcal{A}(-E\n_{\s\s\s} + (T\n)_\s).
\end{equation}
The analogous clamped and free boundary conditions are $\n_\s=\n_{\s\s}=0$ at free ends, and $\n=\n_0$, $\F=0$ at clamped ends. The boundary condition that $\F=0$ at a clamped end comes from time differentiating the equation $\X=\X_0$ to give $\V=\V_\textnormal{clamp}$, where $\V_\textnormal{clamp}$ is the velocity of the surface to which the fiber is clamped. The fiber clamp location is typically coincident with a no-slip boundary for the fluid $\u$, and thus evaluating the evolution equation $\eta(\V-\u)=\mathcal{A}\F$ at this location gives $\F=0$. For an Euler-Bernoulli elasticae, this boundary condition takes the form $-E\n_{\s\s\s} + (T\n)_\s = 0$. The fiber field $\X$ may be recovered from $\n$ and its base coordinate $\X_0$ by computing:
\begin{equation}
	\X(\alpha, \bbeta) = \X_0(\bbeta) + \int_0^\alpha \n(\alpha')\,d\alpha'.
\end{equation}
\end{remark}

\section{Numerical Methods}
\label{section:numerics}

Despite its significant simplification to Eqs.~(\ref{ImLocal}) \& (\ref{equation:u_missing}), the Brinkman-Elasticae model is still fourth-order, nonlinear, and non-local, with a free boundary at the interface between fiber and non-fiber regions and complex boundary conditions that must be imposed on this free boundary. A purely Eulerian formulation is possible, but requires the imposition of the equivalent boundary condition to $\X_{\s\s\s}=0$ on the free interface between fluid-only and fluid-fiber regions. This is a significant challenge, and it is simpler to work in a mixed Eulerian-Lagrangian framework, with the fiber configuration tracked in the natural Lagrangian coordinate $\balpha$ and the fluid velocity solved for in Eulerian coordinates.

\subsubsection{Spatial Discretization}

To fix the details of the spatial discretization, we consider the domain shown in \Cref{figure:diagram}, with periodic boundary conditions in the $x$-direction, and no-slip boundary conditions for the fluid velocity $\u$ at the top and bottom of the domain, corresponding to a bed of fibers in a channel. We discretize the problem in only two-dimensions, but need to be careful in the interpretation: the BE model is derived from local slender-body theory, which is a fundamentally 3D result, and thus this should be considered as discretizing a uniform infinite bed of fibers in the neglected $y$-direction.

\begin{figure}[h!]
	\centering
    \begin{subfigure}[c]{0.48\textwidth}
    	\centering
        \scalebox{1.35}{%
        \begin{tikzpicture}
            \draw [fill=white] (0,0) rectangle (5,2.5);
            \begin{scope}
                \clip (0,0) rectangle (5,2.5);
                \foreach \i in {-12,...,120}
                {
                    \draw [mypurple, thin, domain=0.1:{0.45+0.2*sin(2*pi*\i/2)}] plot ({\x+\i/24}, {ln(\x)-ln(0.1)});
                }
            \end{scope}
            \draw (0,0) -- (0, 2.5);
            \draw (5,0) -- (5, 2.5);
            \draw [line width=0.5mm] (0,0) -- (5, 0);
            \draw [line width=0.5mm] (0,2.5) -- (5, 2.5);
        \end{tikzpicture}
        }
    	\caption{Actual fiber field}
        \label{figure:sparse_discretization:real}
    \end{subfigure}
    \begin{subfigure}[c]{0.48\textwidth}
    	\centering
    	\scalebox{1.35}{
        \begin{tikzpicture}
            \draw [fill=white] (0,0) rectangle (5,2.5);
            \begin{scope}
                \clip (0,0) rectangle (5,2.5);
                \foreach \i in {-3,...,30}
                {
                    \draw [thick, mypurple, domain=0.1:{0.45+0.2*sin(2*pi*\i*2)}] plot ({\x+\i/6}, {ln(\x)-ln(0.1)});
                }
                \end{scope}
            \draw (0,0) -- (0, 2.5);
            \draw (5,0) -- (5, 2.5);
            \draw [line width=0.5mm] (0,0) -- (5, 0);
            \draw [line width=0.5mm] (0,2.5) -- (5, 2.5);
        \end{tikzpicture}
        }
    	\caption{Sparser discretization}
        \label{figure:sparse_discretization:sparse}
    \end{subfigure}
    \caption{Panel (a) shows a dense fiber bed, with all physical fibers represented. Because the bed is extremely dense, a direct discretization (e.g. via Slender-Body Theory or the Immersed Boundary Method) would be computationally expensive. However, the orientation field of the fibers is relatively smooth, and can be accurately approximated using the Brinkman-Elasticae model given in \Cref{equation:simplified_model} and a sparser discretization, such as the one shown in Panel (b).}
    \label{figure:sparse_discretization}
\end{figure}
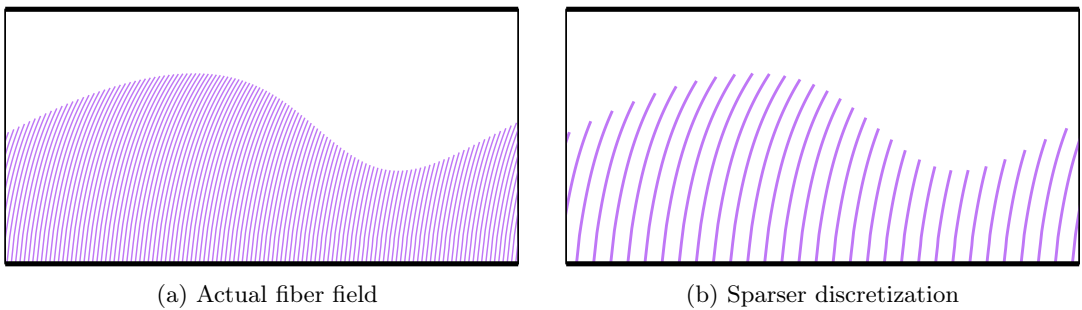

The fiber tangent field $\n$ is discretized in Lagrangian coordinates. For this geometry, both $\alpha$ and $\bbeta$ may be discretized using equispaced grids, with $\beta_j=x_\textnormal{min}+j\,\delta\beta$, for $0\leq j<N_f$, and $\alpha_{i+1/2}=(i+1/2)\delta\alpha$ for $0\leq i<N$, where $N_f$ is the number of discrete `fibers' used to discretize the system, $N$ is the number of points used to discretize individual discrete fibers, $\delta\beta=(x_\textnormal{max}-x_\textnormal{min})/N_f$, and $\delta\alpha=L/N$. When $\n$ is known at these discrete nodes, the fiber field $\X$ can be reconstructed as:
\begin{equation}
	\X(\alpha_i, \beta_j) = \X_0(\beta_j) + \sum_{k=0}^{i-1} \n(\alpha_{k+1/2})\delta\alpha + \mathcal{O}(\delta\alpha^2).
\end{equation}
We emphasize that discretization of $\bbeta$ does not have to be coincident with the actual fibers of the physical system. Rather, it should be chosen so that variations in fiber-field geometry are well described. This distinction is shown in \Cref{figure:sparse_discretization}.

In the mixed Eulerian-Lagrangian formulation, the primary spatial discretization is simple. For a single fiber, the discretization of the differential operators in \Cref{equation:simplified_model:3} is done using second-order finite differences, with appropriate ghost cells used at the endpoints defined by the boundary conditions. The unit-tangent $\n$ and $T$ are discretized at half-integer nodes, and the position $\X$ and force density $\F$ are computed at whole-integer nodes. The Stokes equations may be solved in a variety of ways depending on the domain; for general domains we use the Immersed Boundary Smooth Extension method with $C^2$ extensions, described in \cite{Stein2017}, which provides third-order accuracy of the velocity field in $L^\infty$. The primary challenges in this formulation arise from \emph{communication between frames} (described in \Cref{section:numerics:communication}), and the \emph{efficient time-stepping of the semi-discretized system} (described in \Cref{section:numerics:timestepping}).

\begin{remark}
The discretization defined above for the fiber tangent field $\n$ is reminiscent of discretizing $N_f$ discrete fibers. This may make this method seem similar to a direct simulation. However, because we are discretizing the coarse-grained BE equations rather than the full fluid-fiber system, the discretization of the fiber field needs to be chosen only so as to resolve variations in the fiber field, and does not need to accurately represent the true density of fibers in the system (which is instead controlled through the density field $\rho$). An example making clear this distinction is shown in \Cref{figure:sparse_discretization}. As an extreme example of this, consider the case of a uniform bi-infinite array of fibers, as will be discussed in \Cref{section:deflection,section:rheology,section:buckling,section:rectification}. In this case, we need only take $N_f=1$ to fully describe the BE system, regardless of the fiber density!
\end{remark}

\subsubsection{Communication between Eulerian and Lagrangian frames}
\label{section:numerics:communication}

Because the fiber tangent field $\n$ is discretized using Lagrangian coordinates, the forces $\F$ due to fiber deformation are easily computable in this frame. However, these forces act on the fluid through the Stokes equation that appears as part of the BE model \Cref{equation:simplified_model}. It is most natural to solve for this velocity field in Eulerian coordinates, requiring communication between the two frames. This communication goes in both directions: fluid velocities, computed on the regular nodes of a grid, will need to be evaluated at the discrete nodes $\X(\alpha_i, \beta_j)$, and the force density $\F$, known at $\X(\alpha_i, \beta_j)$, will have to be evaluated at the regular nodes of the fluid grid that lie within the fiber region.

The first of these problems is simple: since $\u$ is known everywhere on a regular grid, and is smooth (except across the fluid-fiber interface, where it is $C^1$), standard methods, such as bi-linear or bi-cubic interpolation, or interpolation using the non-uniform Fast-Fourier transform, provide an efficient and accurate ($\mathcal{O}(\Delta x^2)$) means to evaluate $\u$ at the locations of the fiber grid. The action of this interpolation operator will be denoted by $\IEL$, that is:
\begin{equation}
	(\IEL\u)(\balpha) = \u(\X(\balpha)).
\end{equation}
The time at which $\X$ is evaluated in the operator $\IEL$ will be denoted with a superscript when necessary for clarity, e.g. $\IEL^{t}$ means that the interpolation operator interpolates the function $\u$ from the regular nodes of the grid to the discrete nodes of $\X(t)$. Note that $\u$ need not be evaluated at the same time as $\X$.

\begin{figure}[h!]
  \centering
  \begin{subfigure}[c]{0.48\textwidth}
  	\includegraphics[width=\textwidth]{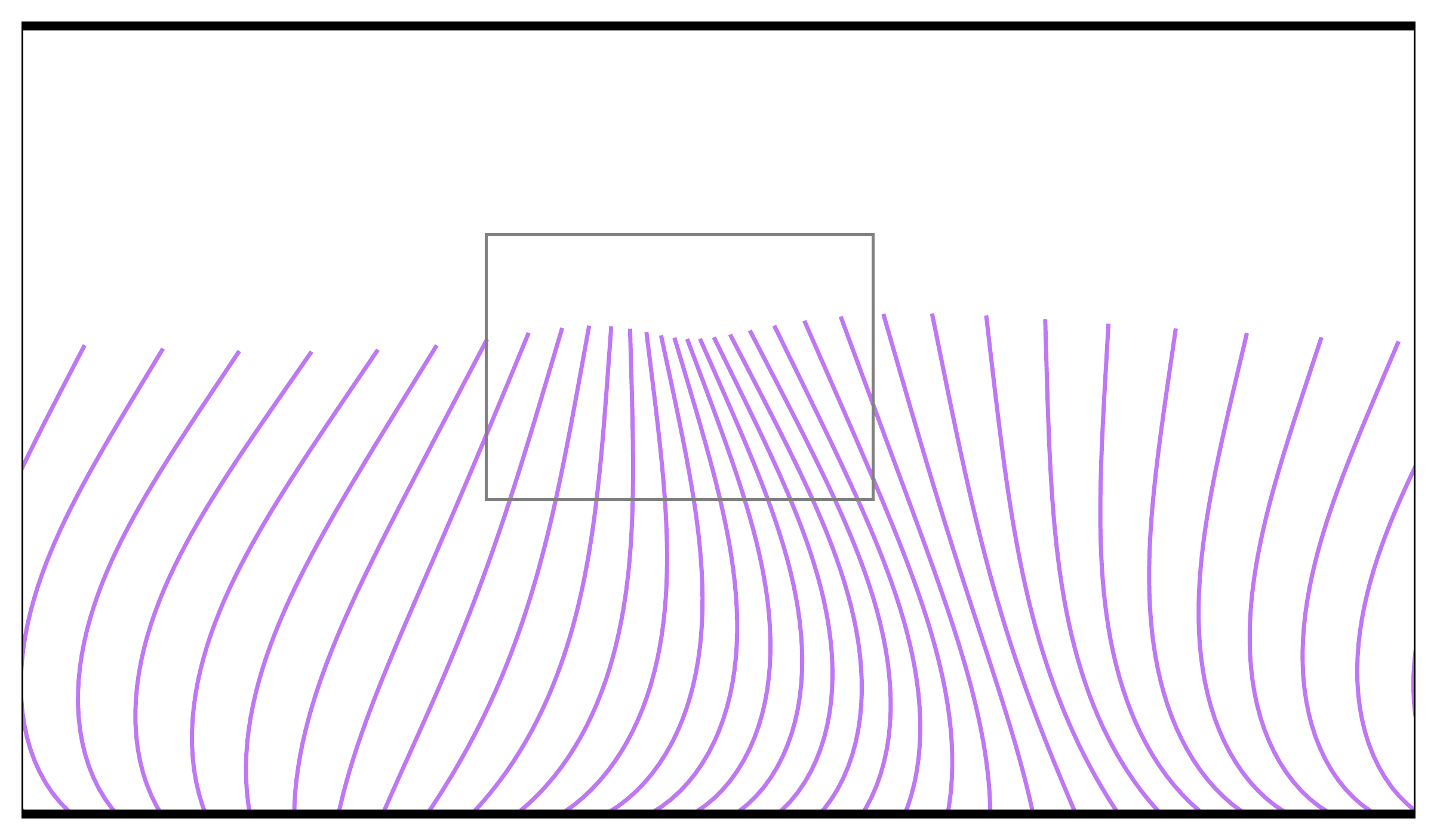}
    \caption{ }
    \label{figure:interpolation:zoom_out}
  \end{subfigure}
  \begin{subfigure}[c]{0.48\textwidth}
  	\includegraphics[width=\textwidth]{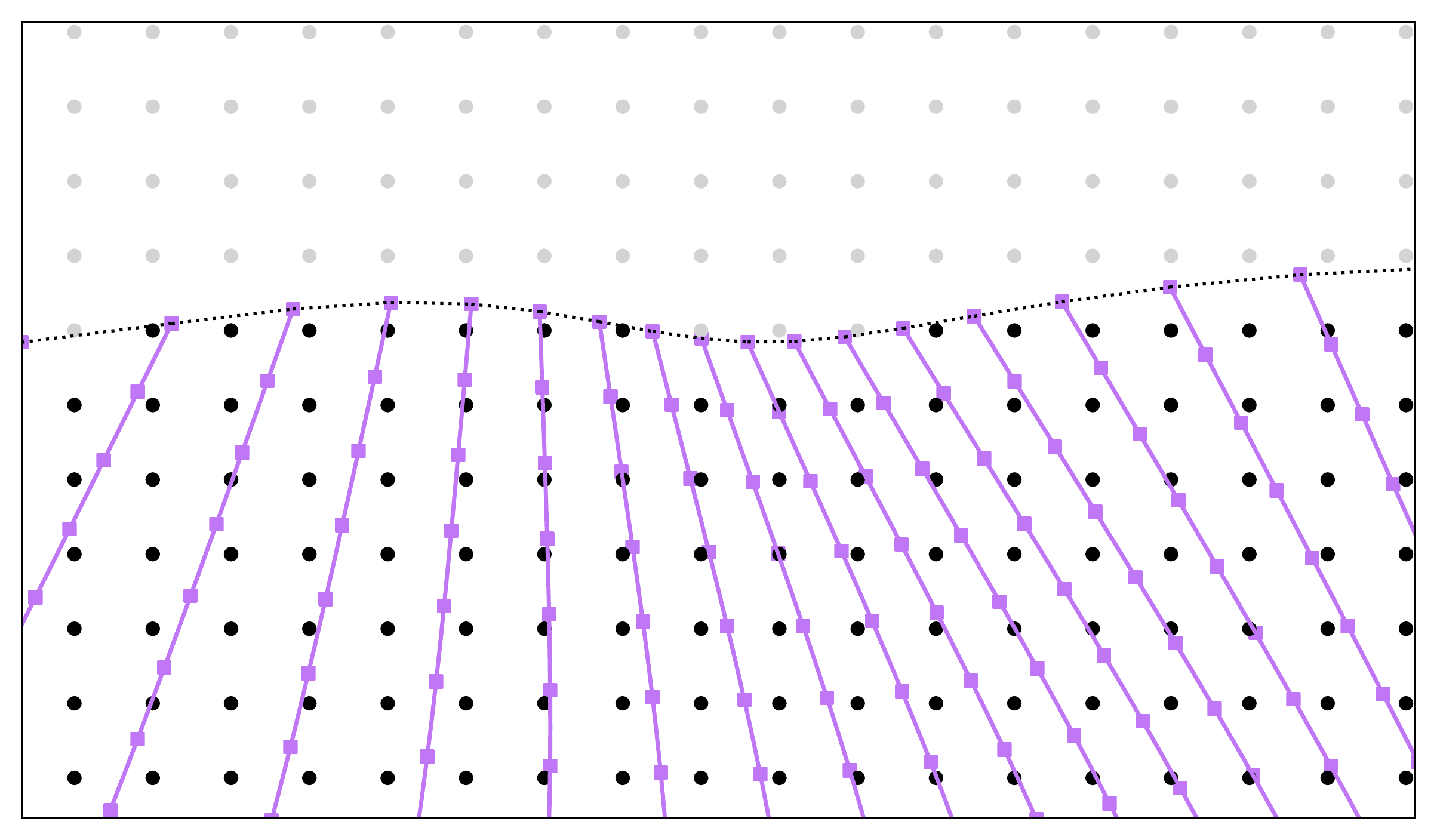}
    \caption{ }
    \label{figure:interpolation:zoom_in}
  \end{subfigure}
  \caption{Panel (a) shows a possible configuration of fibers. The smooth force density $\F$ is easily computable on the discretized fibers, shown in purple. In order to facilitate rapid solution of the Stokes equation, these forces are interpolated to the nodes of a regular grid. Panel (b) shows a zoomed in view of the gray rectangle in Panel (a). Forces are known at the discretization points of the fiber field, shown as purple squares. Nodes of the regular grid are shown as circles. A line separating the fiber region from the purely fluid region is formed by connecting the tips of the fibers. Points are determined to be inside of the fiber region if they fall within a polygon enclosing the fiber region (in this geometry, simply under the dashed line). Points within the fiber region are shown in black. Forces are interpolated from the known values at the purple squares to the black circles.}
  \label{figure:interpolation}
\end{figure}

The second direction is slightly trickier. Recall that due to the coarse-graining, the force density $\F$ is a \emph{smooth function}, except across the interface between the fluid-only region and the fluid-fiber region. Thus representing the forces $\F$ on an Eulerian grid requires two steps: (1) identifying those points that lie within the fiber region, and (2) interpolating the forces that are known at the irregular points $\X(\balpha)$ to the regular points within the fiber region. Points are determined to be in the fiber region if they lie within a polygon connecting the extremal points of the discrete fibers. Interpolation is done by forming a triangulation of the discrete nodes of the fibers, after which bi-linear or bi-cubic interpolants may be rapidly computed and evaluated at the grid nodes. This interpolation operator will be denoted by $\ILE$:
\begin{equation}
	(\ILE\F)(\x) = \F(\X^{-1}(\x)).
\end{equation}
See \Cref{figure:interpolation} for a graphical depiction of the interpolation process.
\begin{remark}
Using the interpolation schemes as described will lead to a discretization that is accurate to only first order in $\Delta x$. The dominant errors arise due to the sharp truncation of the fiber forces across the fluid-fiber interface. It is possible to improve this error by adopting finite-volume like corrections to the representation near the interface; these are somewhat complicated in general and will be presented in a forthcoming contribution. A detailed description of these corrections in a simplified one-dimensional setting is given in \Cref{section:1d_model:improved_interpolation}.
\end{remark}

With this notation, the simplified model in \Cref{equation:simplified_model} may be rewritten as:
\begin{subequations}
	\label{equation:simplified_model_interpolated}
	\begin{align}
    	-\mu\Delta\u + \grad p &= \rho\ILE\F,\qquad \grad\cdot\u = 0,             			\label{equation:simplified_model_interpolated:momentum} \\
        \eta(\partial_t\n-\partial_\s\IEL\u) &= \partial_\s(\A\F).
    \end{align}
\end{subequations}

\begin{remark}
In direct regularized methods, such as the immersed boundary method, communication between the Lagrangian and Eulerian frames takes the form of \emph{spreading}, in which the singular forces are communicated to the regular Eulerian grid by convolution with regularized delta functions. This results in a regularized but near-singular force density on the Eulerian grid. This is significantly different than what is done here: because of the coarse-graining process, the force density $\F$ is a smooth field, except across the fluid-fiber interface. This smooth field is known only at the coordinates $\x=\X(\balpha)$, which are irregular points in the Eulerian frame. Obtaining a representation of the smooth force $\F$ on a regular grid (or some subset of that regular grid) is thus an interpolation problem, where the function being interpolated is smooth. 
\end{remark}

\subsubsection{Discretization in time}
\label{section:numerics:timestepping}

Because the fluid is governed by low-Reynolds number dynamics, $\u$ is slaved to the fiber configuration. In order to make this explicit, let us first define the inverse of the Stokes operator $\mathcal{L}^{-1}$ by $\u=\mathcal{L}^{-1}\f$, where $\u$ is given by the solution to:
\begin{align}
-\Delta\u + \grad p = \f,	\qquad\grad\cdot\u = 0,
\end{align}
subject to appropriate boundary conditions for $\u$. We may now eliminate $\u$ from \Cref{equation:simplified_model_interpolated} to give:
\begin{subequations}
	\label{equation:simplified_model_reduced}
	\begin{align}
        \eta(\partial_t\n-\partial_\s\IEL\mathcal{L}^{-1}\ILE\rho\F) &= \partial_\s(\A\F).
    \end{align}
\end{subequations}
Recall that $\F$ contains a term of the form $\n_{\s\s\s}$, and thus this is a fourth order system, with stiffness arising from both bending rigidity and the inextensibility constraint that $\alpha$ is material to the flow $\X_t$, which is enforced by treating $\alpha$ as a material parameter and enforcing that it remains an arclength parameterization, that is, $\n\cdot\n=1$. In order to avoid a quartic time-step restriction, we discretize this equation implicitly in time using the second-order backward differentiation formula:
\begin{subequations}
	\begin{align}
      3\n(t) - 4\n(t-\Delta t) + \n(t-2\Delta t) &= 2\mathcal{G},		\\
      \n(t)\cdot\n(t) &= 1.
	\end{align}
    \label{equation:discrete_time}
\end{subequations}
where $\mathcal{G}$ is given by:
\begin{equation}
	\mathcal{G} = \partial_\s\IEL^{t-\Delta t}\mathcal{L}^{-1}\ILE^{t-\Delta t}\rho(t-\Delta t)\F(\n(t),T(t)) + \eta^{-1}\partial_\s(\A(\n(t))\F(\n(t),T(t))).
\end{equation}
Note that the interpolation operators $\ILE$ and $\IEL$, as well as the density $\rho$ are evaluated at time $t-\Delta t$. This nonlinear equation is solved for $\{\n(t), T(t)\}$ using Newton's method. The computation of the full Jacobian is numerically expensive, and even if it can be formed, requires $\mathcal{O}(N_f^3N^3)$ time to invert. Instead, let $\tilde{\mathcal{G}}$ be defined by:
\begin{equation}
	\tilde{\mathcal{G}} = \eta^{-1}\partial_\s(\A(\n(t))\F(\n(t),T(t))).
\end{equation}
The Jacobian to \Cref{equation:discrete_time} with $\mathcal{G}$ replaced by $\tilde{\mathcal{G}}$ may be easily and efficiently computed. Because the hydrodynamic interaction between fibers is neglected, the $N_f$ discrete fibers are independent, resulting in a block-diagonal Jacobian that may be inverted in $\mathcal{O}(N_f N^3)$ time. The full Jacobian to \Cref{equation:discrete_time} may be applied via numerical differentiation, and inverted using GMRES, preconditioned by the inverse to the simpler Jacobian. It is clear that when $\rho$ is small, the hydrodynamic feedback term will have little effect and this should yield a good approximation and rapid convergence. Surprisingly, rapid convergence is achieved across a wide range of densities. Numerical examples will be presented in \Cref{section:deflection:behavior} and \Cref{section:waves}. In the numerical experiments we have run, the Newton iteration typically requires 3-4 iterations to converge to a tolerance of $10^{-8}$, and GMRES requires 3-20 iterations to invert the Jacobian to a tolerance of $10^{-10}$, with more iterations required when the fiber density is higher. The scheme appears to be nearly unconditionally stable, although we note that each step of GMRES requires one Stokes solve, and thus 10-80 Stokes solves are required per time-step.

\subsection{A simplified 1-D model}
\label{section:1d_model}

For certain geometries and problems, the numerical tasks outlined in the method above can be dramatically simplified. In this section, we consider the simple case of a bi-infinite fiber bed with no dependence on the $x$- and $y$-directions. Although seemingly trivial, this model is capable of providing significant information and insight for many problems, including the deflection of dense fiber beds subject to shear (\Cref{section:deflection}), basic rheology of fiber beds (\Cref{section:rheology}), the gravity driven buckling of fiber beds (\Cref{section:buckling}), the design of soft micro-fluidic rectifiers (\Cref{section:rectification}), and the appearance of stable deformed states leading to streaming flows in beds of microtubules undergoing buckling due to the action of molecular motors~\cite{CSLGS2018}.

Reducing the problem in this way implies that the fiber field $\X$ may be fully described by a single fiber, with Lagrangian coordinate $\s$, and simplifies the Stokes equations, leading to the simpler equation:
\begin{subequations}
	\begin{align}
    	-\mu \partial_{zz}u &= \ILE\rho\mathcal{F}^x,	\\
        -\mu \partial_{zz}v &= \ILE\rho\mathcal{F}^y,	\\
        \eta(\X_t - \IEL\u)		 &= \mathcal{A}\F,				\\
        \d\cdot\d		 &= 1.
    \end{align}
\end{subequations}
The Poisson equations for $u$ and $v$ are uncoupled, and may be solved rapidly with a simple finite difference discretization. The interpolation operators are both simply implemented using standard bi-linear or bi-cubic interpolation in one dimension, with the adjustment described in \Cref{section:1d_model:improved_interpolation} to the operator $\ILE$ providing second-order accuracy in space. The equations are inverted as described before. We note that in this simple case, the density $\rho=J^{-1}\rho_0=\rho_0/n_z$. This reduced model allows for solutions to be found in an extremely rapid manner (typically seconds, as for the examples presented in \Cref{section:deflection:behavior}).

\section{Deflection of fiber beds in a shear flow}
\label{section:deflection}

To validate the model and obtain basic insights into how fiber density affects fiber dynamics, we examine how a uniform bed of fibers deforms when subjected to a shear flow. In \Cref{section:deflection:validation}, we compare simulations done using the simplified one-dimensional BE model described in \Cref{section:1d_model} to full three-dimensional simulations done using an Immersed Boundary method with periodic boundary conditions. In \Cref{section:deflection:behavior}, we look at the behavior of the Brinkman-Elasticae model across a wide range of densities. The dependence in the deformation of the fibers depends in a non-trivial way on the density of fibers. Below a critical density, deformation is effectively independent of density. For densities above the critical density, fiber deformation scales with inversely with the fiber bed density.

\subsection{Validation}
\label{section:deflection:validation}

To validate the BE model, we compute the response of a bi-infinite uniform bed of fibers subjected to a shear flow. Results computed using the simplified one-dimensional model given in \Cref{section:1d_model} are compared to results obtained using an Immersed Boundary method in three-dimensions, where a single fiber is simulated in a periodic box, with fiber density controlled by adjusting the dimensions of the periodic domain. We compare results for three measures: (1) fiber tip deflection, (2) time to $95\%$ deflection, and (3) percent flow occlusion. Percent flow occlusion is defined by the total flow at steady state divided by the total flow if no fibers were present. Parameters are chosen so that the fiber bed undergoes non-trivial deformation (at low fiber density, approximately 70\% of the length of the fiber).

\subsubsection{Immersed Boundary simulations}

For the immersed boundary simulations, a single initially straight fiber of length $L=0.5$ is clamped in the $\mathbf{\hat z}$ direction at the location $(0,0,0)$. Channel walls are positioned at $z=0$ and $z=0.8$. The velocity $\u(z=0)$ and $\u(z=0.8)$ are fixed to be $0$ and $0.8$, respectively, subjecting the fiber to a shear rate $\dot\gamma=1$. The initial fiber bed density $\rho_0$ is controlled by choosing the domain over which periodic boundary conditions are specified in the $x$ and $y$ directions. These coordinates are taken to be periodic over  the domain $[-\delta/2, \delta/2]$, and thus $\rho_0=1/\delta^2$. The bending rigidity of the fiber is fixed to be $E=0.01$, and the viscosity of the fluid is set to $\mu=1$. The setup for these simulations is depicted in \Cref{figure:validation:ib_setup}.

In traditional IB simulations, the constitutive model of elasticae is approximated using a network of springs. To facilitate direct comparison, we instead model the fiber as an inextensible Euler-Bernoulli beam. This results in the formulation:
\begin{subequations}
  \begin{gather}
      -\mu\Delta\u + \grad p + \mathcal{S}\mathbf{F} = 0; \qquad\grad\cdot\u = 0,    \\
        \mathcal{S}^*\u = \X_t,    \\
        \mathbf{F} \textbf{}= -E\X_{\s\s\s\s} + (T\X_\s)_\s; \qquad \X_\s\cdot\X_\s = 1,
  \end{gather}
\end{subequations}
where $\mathcal{S}$ and $\mathcal{S}^*$ are spread and interpolation operators, respectively \cite{Peskin2002}. The fluid equations are discretized using uniform grids with a grid-spacing $h$ in all directions; the Stokes equations on the channel domain with velocity boundary conditions are imposed using the 2nd-order Immersed Boundary Smooth Extension (IBSE) method described in \cite{Stein2017}. The fiber is discretized in space using the same discretization as used in the BE model, with $\delta\s\approx h$. Time discretization is again done using the second-order Backward Differentiation scheme, and the resultant non-linear system is solved using a  Newton-Krylov method. Unlike when solving the BE model, effective preconditioning of the Jacobian is challenging, especially for dense systems where the periodic boundary conditions become important. At time $t=0$, we directly form the Jacobian. This computation is expensive, and instead of repeating it at all later timesteps, its inverse is used as the preconditioner at all times $t>0$. This preconditioner becomes ineffective when the fiber has significantly deformed, but is sufficient for this validation study.

In order to facilitate comparison with the BE model, we estimate the radius of the fiber simulated using the IB method to be $r_\text{fiber} = r_\textnormal{hydro} h$, where $r_\textnormal{hydro}$ is the hydrodynamic radius\footnote{The \emph{hydrodynamic radius} of a regularized delta function is a measure of its effective physical size, similar to the small parameter $\epsilon$ used in defining the blobs in the Method of Regularized Stokeslets~\cite{Cortez2001}. It may be computed numerically by exerting a (regularized) point force $\mathcal{S}\f$ at a location $\x$ in space, inverting the Stokes equations, measuring the velocity $\u(\x)\approx \mathcal{S}^*\u$, and solving for $r_\textnormal{hydro}$ in the Stokes drag relation $\f=6\pi\mu r_\textnormal{hydro}\u$ \cite{bringley2008validation}.} of the regularized delta function and $h$ is the grid spacing for the Cartesian grid on which the fluid equations are solved. For these simulations, we use the 6-point delta function defined in \cite{bao2016gaussian}, whose hydrodynamic radius is $r_\textnormal{hydro}\approx 1.47h$. Note that because the radius of the fiber is set by the grid-spacing $h$, it is not possible to perform a refinement study at a fixed fiber radius.

\begin{remark}
A straightforward Immersed Boundary method, as described above, is not sufficient for studying this system across a wide range of densities. When the fiber bed is sparse and the fibers are thin, a large number of Fourier modes $N=\delta r_\textnormal{hydro}/r_\textnormal{fiber}$ must be used, resulting in computationally expensive simulations. This issue could be alleviated by using an adaptive method, e.g. IBAMR~\cite{griffith2014ibamr}. When the fiber bed is dense, only a few Fourier modes $N$ can be used, and the fluid flows are poorly resolved. Obtaining accurate solutions in these regimes requires a full discretization of the surface of the fiber.
\label{remark:IB_limitation}
\end{remark}

\subsubsection{Comparison to the Brinkman-Elasticae model}

\begin{figure}[h!]
  \centering
  \begin{tabular}[t]{cc}
      \begin{subfigure}[c]{0.38\textwidth}
      	\includegraphics[trim={1.2cm 0.7cm 0.0cm 1.6cm},clip,width=\textwidth]{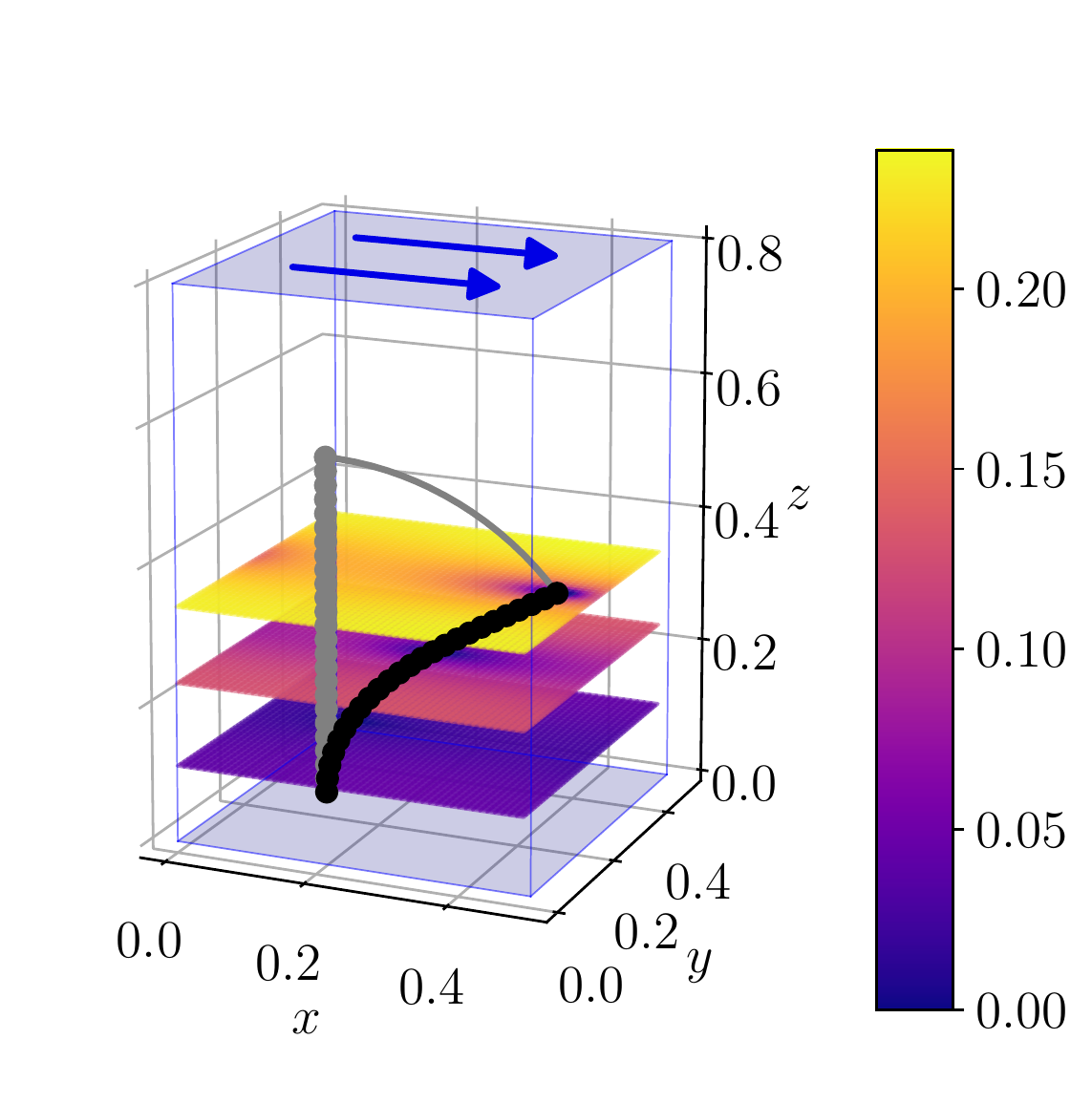}
        \caption{ }
        \label{figure:validation:ib_setup}
      \end{subfigure}
      &
      \begin{subfigure}[c]{0.48\textwidth}
      	\includegraphics[width=\textwidth]{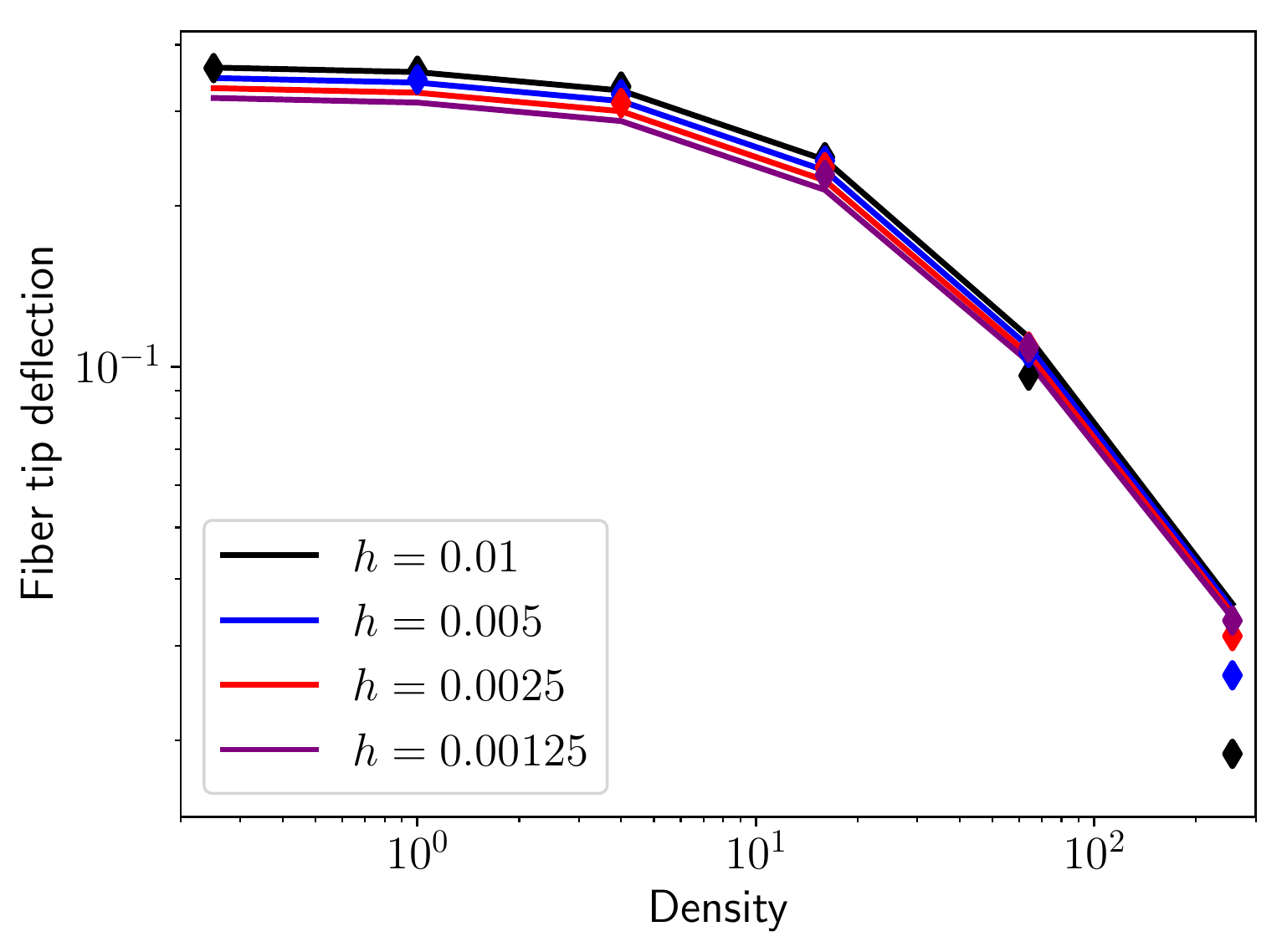}
        \caption{ }
        \label{figure:validation:deflection}
      \end{subfigure}
      \\
      \begin{subfigure}[c]{0.48\textwidth}
      	\includegraphics[width=\textwidth]{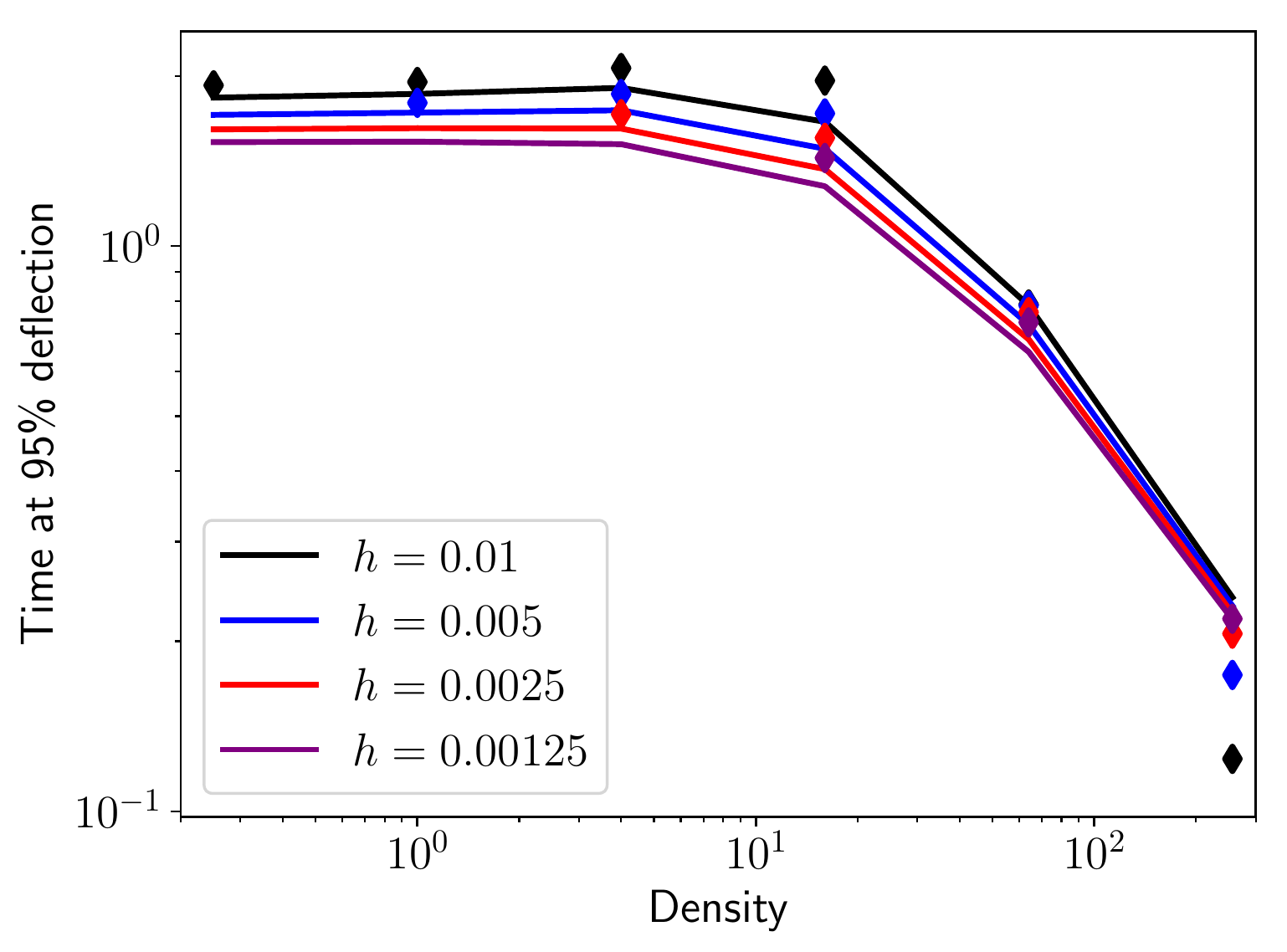}
        \caption{ }
        \label{figure:validation:t95}
      \end{subfigure}
      &
      \begin{subfigure}[c]{0.48\textwidth}
      	\includegraphics[width=\textwidth]{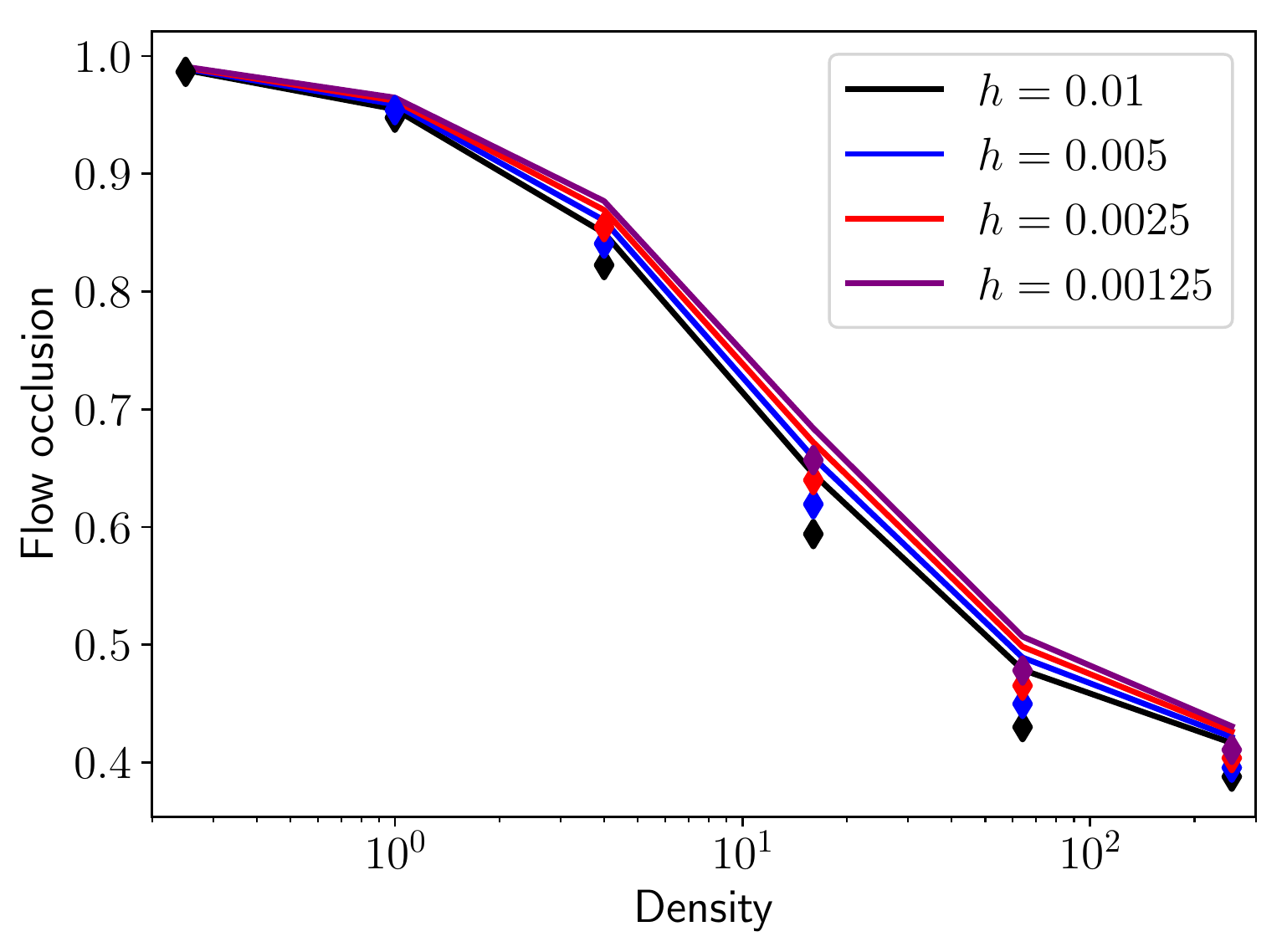}
        \caption{ }
        \label{figure:validation:flow_occlusion}
      \end{subfigure}
  \end{tabular}
  \caption{Panel (a) shows the shear-flow setup used for Immersed Boundary simulations when $\rho_0=0.25$. The initial and final positions of the fiber are shown in gray and black, respectively, and the gray curve shows the path of the fiber tip. The three contour plots show the $x$-component of the velocity at $z=Z(L)/3$, $z=2Z(L)/3$, and $z=Z(L)$. Panels (b)-(d) show a comparison of results for Immersed Boundary simulations (markers) and the Brinkman-Elasticae model (lines), for a bed of fibers subjected to a uniform shear, across a range of densities. Panel (b) shows the total fiber deflection ($X(L)-X(0)$) at steady state, Panel (c) shows the time at which the fiber was deflected by $95\%$ of its steady state deflection, and Panel (d) shows the flow occlusion at steady state (i.e. the ratio of flow through the channel with the deformed fibers present to the flow through the channel with no fibers present).}
  \label{figure:validation}
\end{figure}

Solutions to the BE model are computed using the simplified 1D model described in \Cref{section:1d_model} with an initial fiber density $\rho_0=\delta^{-2}$, and a fiber radius $r(L/2)=\frac{4}{\pi}r_\textnormal{hydro} h$, so that the \emph{average} radius of the slender fiber is equal to the radius of the fiber from the IB simulations. All other parameters are the same. Results are shown as curves in \Cref{figure:validation}. Results from the IB simulations are shown as diamonds, these are not available when the fiber density is sparse and the fibers are thin, see \Cref{remark:IB_limitation}. For the fiber tip deflection (\Cref{figure:validation:deflection}), the BE model and the IB simulations agree closely, except for dense suspensions of thick fibers. At the highest density and largest fiber radius shown, the fiber radius is nearly equal to $\delta$, the Immersed Boundary simulations are poorly resolved and some of the assumptions made in deriving the BE model are not met. For the sparsest of beds, the inter-fiber spacing is greater than $L$. This parameter range also breaks some of the assumptions from the BE model: in order to obtain convergent averages, a large value of $r$ with respect to $L$ would need to be used in the coarse-graining process; and certain neglected terms would be nontrivial. Nevertheless, the results from the model are consistent with the Immersed Boundary simulations.

Time to $95\%$ deflection (\Cref{figure:validation:t95}) is measured by computing the arclength of the curve traversed by the tip of the fiber (shown as the gray curve in \Cref{figure:validation:t95}), and measuring the time at which the tip of the fiber has traversed $95\%$ of the arclength along this path. The trends are in good agreement, and both models capture the fact that at intermediate densities, the time to $95\%$ deflection increases slightly from the low-density case. The IB simulations show a more significant trend, especially in the case when fibers are thick. This slight disagreement between the BE model and the IB simulations appears only when the fiber is soft enough to deflect substantially, as in this test. When the fiber elasticity is doubled, agreement in the time to $95\%$ deflection measure is comparable with the agreement in the tip-deflection measure.

For flow occlusion (\Cref{figure:validation:flow_occlusion}), the trends between the BE model and IB simulations are similar, although quantitative agreement is less than in the other two measures. The BE model builds on top of a theory developed from fibers that have a specific and non-constant radius, while the IB simulations are of an effectively constant radius cylinder. Although the radius of the BE model is set so that the average radii of the two are the same, the fastest flow is near the tip of the fiber, and the shape used in the BE model may allow additional flow. Additionally, errors could simply come from the first-order accuracy of the IB method.

\subsection{Behavior of the model for sparse and dense fiber beds}
\label{section:deflection:behavior}
In \Cref{section:deflection:validation}, we examined fiber deflection in a shear flow over a relatively narrow range of fiber densities, limited by practical constraints of the Immersed Boundary method; see \Cref{remark:IB_limitation}. The Brinkman-Elasticae model is free from these considerations, and may, in principle, be run at any density. That said, at very low or very high densities, some of the assumptions used in deriving the BE model may fail to hold. 

\begin{figure}[h!]
\centering
\begin{subfigure}[c]{0.45\textwidth}
	\includegraphics[width=1.0\textwidth]{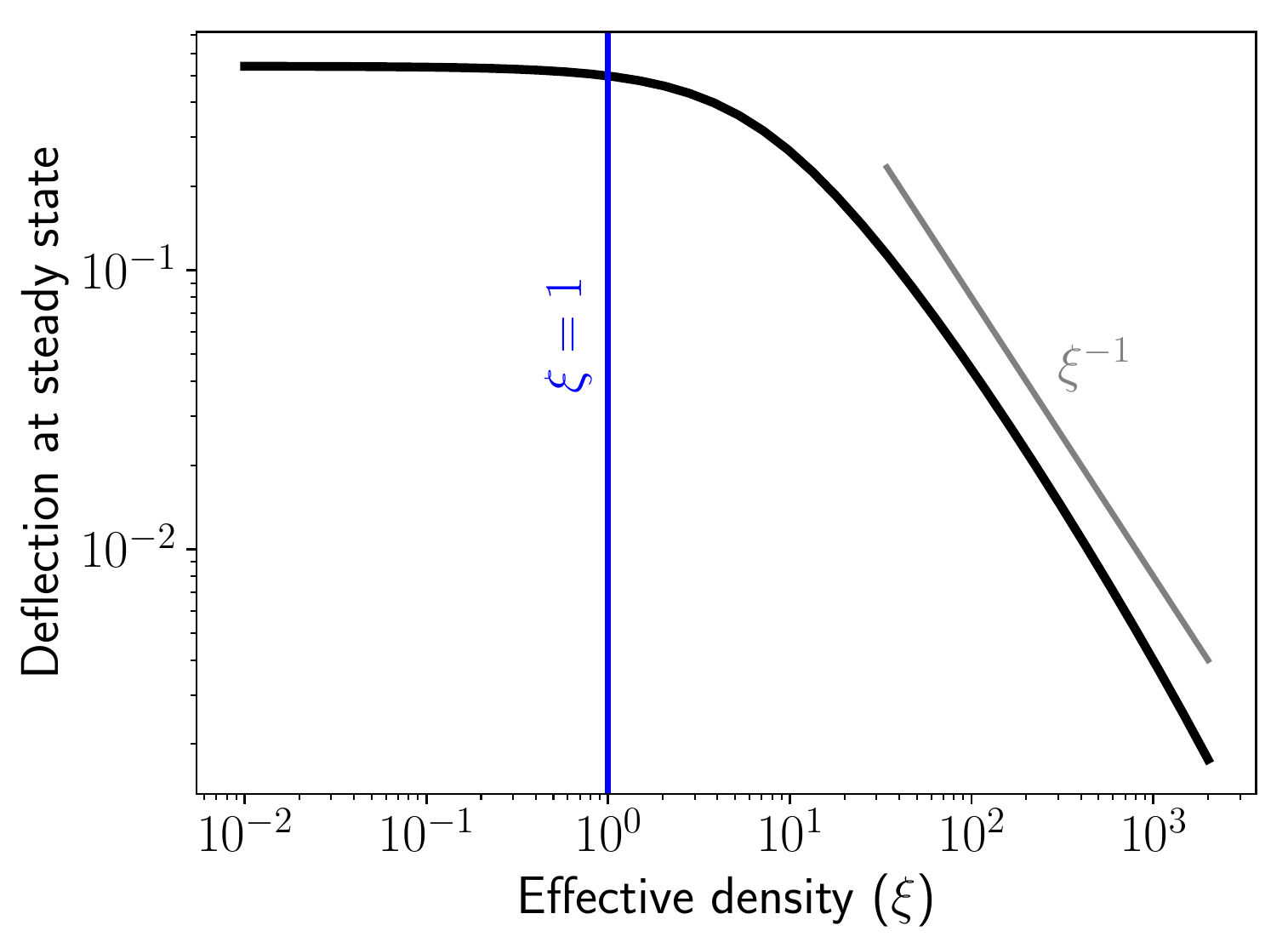}
    \caption{}
    \label{figure:deflection:homogenized}
\end{subfigure}
\begin{subfigure}[c]{0.45\textwidth}
	\includegraphics[width=1.0\textwidth]{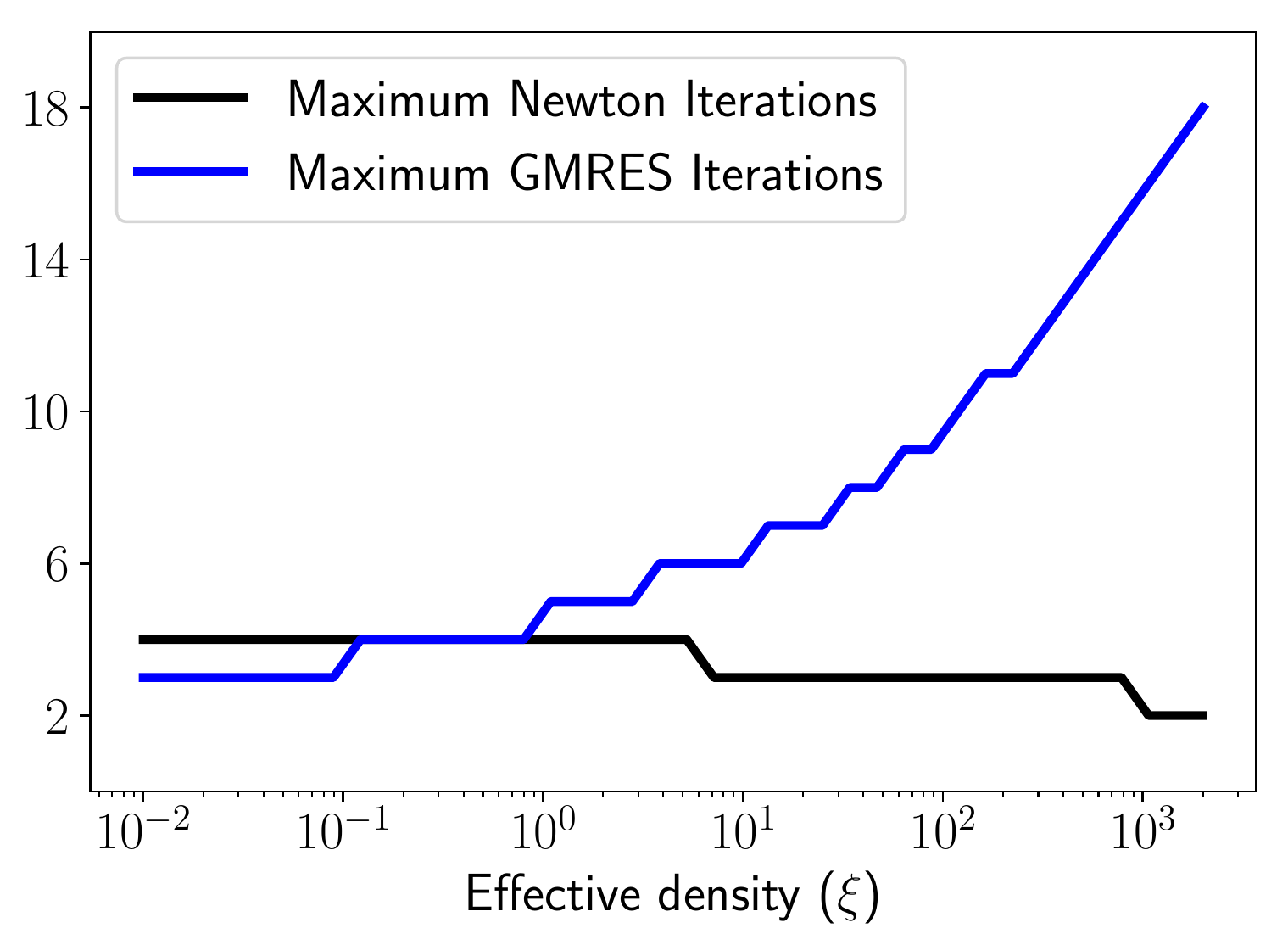}
    \caption{}
    \label{figure:deflection:gmres}
\end{subfigure}
\caption{Panel (a) shows the deflection of a fiber bed subject to a shear flow across a range of effective densities $\xi$. The blue line denotes $\xi=1$, where the bed transitions from a regime with behavior dominated by single-fiber dynamics to a regime where the behavior is dominated by many fiber dynamics. The value $\xi=1$ at which this transition occurs corresponds to when the average inter-fiber spacing is approximately half of the fiber length. Panel (b) shows the maximum number of Newton iterations required per timestep and the maximum number of GMRES iterations required to invert the Jacobian in each Newton iteration.}
\end{figure}

For these simulations, we let the channel height be given by $\lambda L$, where $L$ is the fiber length and $\lambda>1$. The fluid velocity at $z=0$ is $\u=0$ by the no-slip condition, and the velocity at $z=\lambda L$ is taken to be $\u=\NDU\mathbf{\hat x}$. Scaling space by $L$ and time by the inverse shear rate $\dot\gamma^{-1} = \lambda L/\NDU$ gives the non-dimensional system:
\begin{subequations}
\begin{alignat}{2}
	-\Delta\u + \grad p  &= \xi J^{-1}(-\tilde E\X_{\s\s\s\s} + (T\X_\s)_\s), \qquad&\grad\cdot\u &= 0,	\\
    \V - \u &= \mathcal{A}(-\tilde E\X_{\s\s\s\s} + (T\X_\s)_\s), &\V&=\X_t,
\end{alignat}
\end{subequations}
with $\u(0)=\mathbf{0}$, $\u(\lambda)=\lambda\mathbf{\hat x}$; the pressure $p$ and tension $T$ have been scaled by $(\mu\dot\gamma)^{-1}$ and $(L^2\eta\dot\gamma)^{-1}$, respectively. The effective density $\xi$ and effective rigidity $\tilde E$ are defined as
\begin{equation}
	\xi = \frac{8\pi\rho_0 L^2}{c},\qquad \tilde E = \frac{E}{\eta\dot\gamma L^4}.
	\label{eq:shear_parameters}
\end{equation}
Note that $\rho_0$ is simply the fiber number density per unit area at $z=0$ (the base of the fibers). We fix the effective rigidity to $\tilde E=0.1$, which is sufficiently soft to allow for large-scale deformation of the fibers when the bed is sparse. The effective density $\xi$ is varied over a wide range, and the system is evolved to steady state. The results from these simulations are shown in \Cref{figure:deflection:homogenized}. For $\xi\lesssim1$, deflection is nearly independent of $\xi$. For $\xi\gtrsim1$, the deflection scales as $\xi^{-1}$. Although it is physically clear that geometric constraints (due to crowding) require the deflection to go to $0$ as the density goes to infinity, the steric interactions that would enforce those constraints are not included in this model. The effective stiffening observed here is instead due purely to hydrodynamic interactions among the fibers.

To get a sense for what $\xi=1$ means, let us consider microtubules as a model fiber. Microtubules have a diameter of approximately $24$nm, and although their lengths may vary significantly they are typically on the order of a micron in length, giving $c\approx 8$. Let $\overline\delta$ denote the average inter-fiber spacing. The density $\rho_0$ is then given by $\rho_0=C/\overline\delta^2$, where the constant $C$ depends on the geometric packing of the fibers and is typically close to $1$. Thus for a bed of hexagonally packed microtubules, we have $\xi\approx\frac{2\pi}{\sqrt{3}}\left(\frac{L}{\overline\delta}\right)^2$, and when $\xi=1$, solving for the inter-fiber spacing gives $\overline\delta\approx L/2$. When fibers are separated by more than approximately half their length, the dynamics of the fiber bed are governed primarily by single-fiber dynamics. When fibers are closer together than this, the system transitions to a regime where the collective hydrodynamic effects dominate the dynamics. Although the value of $\overline\delta=L/2$ will change slightly depending on the packing and aspect ratio of the fiber, $c$ depends only logarithmically on the aspect ratio, and the conclusion is surprisingly robust. The importance of hydrodynamic interactions between fibers when inter-fiber separation is on the same order as the fiber length has been observed in other contexts, for example beat-synchronization between two nearby flagella \cite{brumley2014flagellar}.

In \Cref{figure:deflection:gmres}, we show the maximum number of Newton and GMRES iterations required during the simulation. For these simulations, the fluid is discretized with $100$ gridpoints, the fiber is discretized with $50$ gridpoints, and relatively large timesteps were taken, with approximately 100 timesteps required to reach steady state. The tolerance for the Newton solver is set at $10^{-8}$, and the tolerance for GMRES is set at $10^{-10}$. At low densities, the approximate Jacobian described in \Cref{section:numerics:timestepping} well approximates the actual Jacobian, and the inverse is computed in only a few iterations of GMRES. At higher densities, the approximate Jacobian is a less ideal preconditioner, but even at extremely high densities only 18 iterations are required to invert the Jacobian. In all simulations the number of Newton iterations is at most 4; for denser fiber beds this is slightly decreased since the bed is effectively stiffer and hence deforms less at each timestep. At the lowest density, the entire simulation takes only 2.7 seconds; at the highest density 3.2 seconds; simulations were run on a Macbook pro with an Intel\textsuperscript{\textregistered} Core\texttrademark \ i7-5557U CPU @ 3.10GHz and 16GB of RAM.  The code has not been carefully optimized to facilitate fast simulations, instead the rapid timings come from the reduction to the one-dimensional system that is possible in the BE model.

\section{Oscillatory Rheology of Fiber Beds}
\label{section:rheology}
\begin{figure}[h!]
\centering
\begin{subfigure}[c]{0.45\textwidth}
	\includegraphics[width=1.0\textwidth]{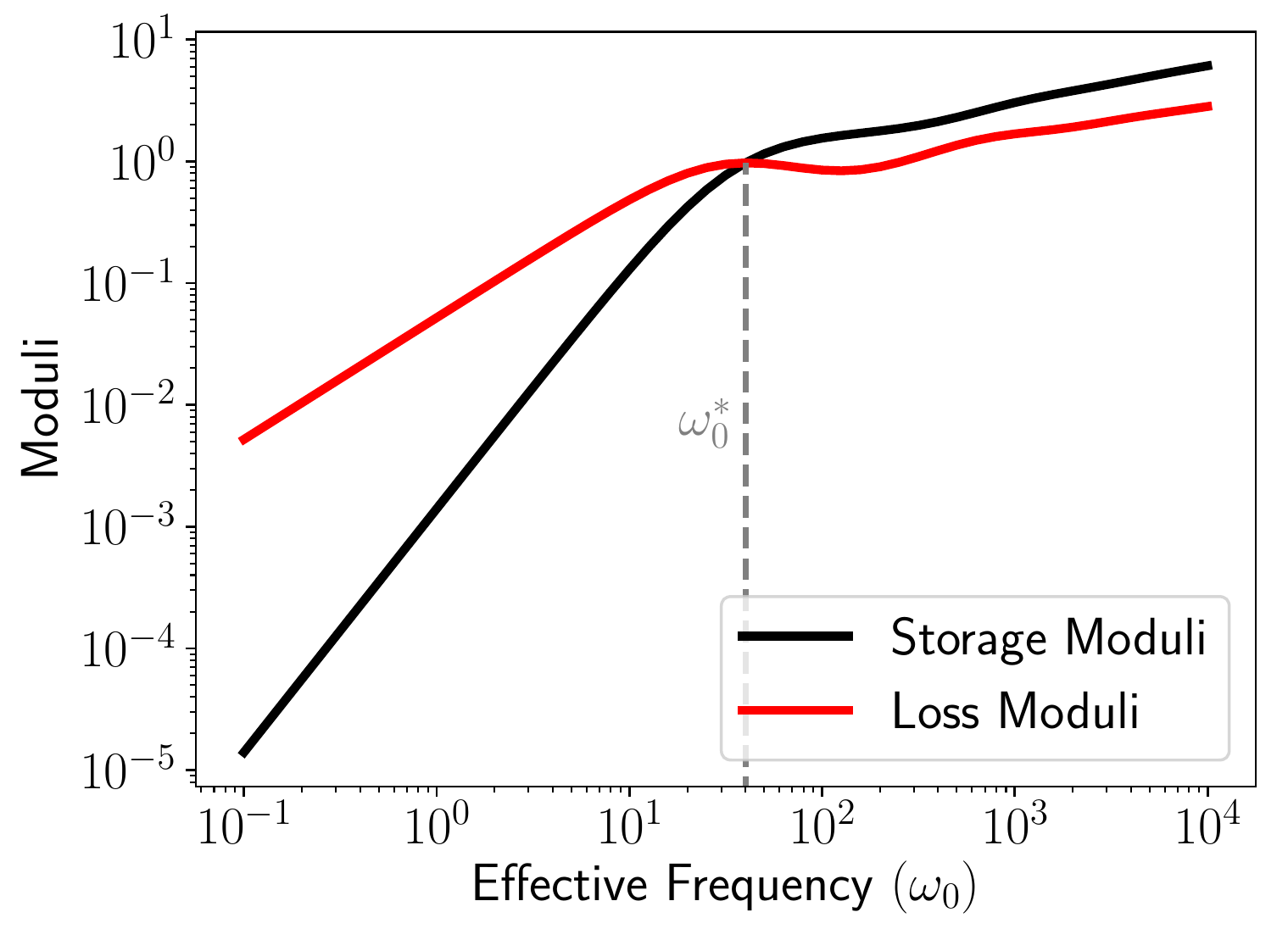}
    \caption{}
    \label{figure:rheology:xi10}
\end{subfigure}
\hfill
\begin{subfigure}[c]{0.45\textwidth}
	\includegraphics[width=1.0\textwidth]{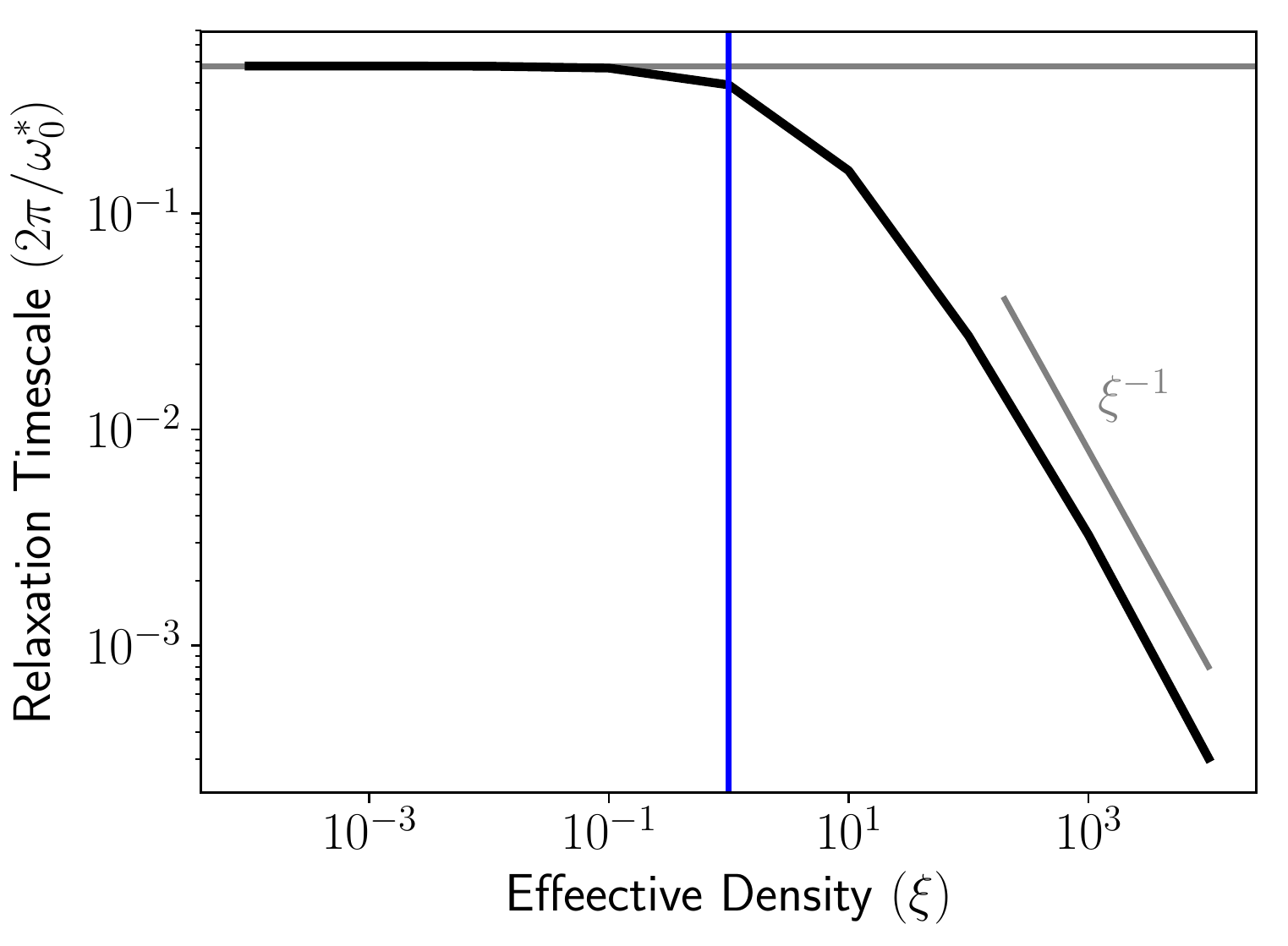}
    \caption{}
    \label{figure:rheology:timescales}
\end{subfigure}
\caption{Oscillatory rheology of a fiber bed. Panel (a) shows the storage ($G'$) and loss ($G''$) moduli for the fiber bed, computed at $\xi=10$. The frequency at which $G'$ and $G''$ intersect, $\omega_0^*$, is shown with a dashed gray line; this frequency gives a relaxation timescale $t^*=2\pi/\omega_0^*$ that depends on $\xi$. Panel (b) shows how the relaxation time $t^*$ depends on the effective density $\xi$. The $\xi=0$ limit, computed analytically, is shown as the horizontal gray line, and is in close agreement with the numerically computed values for low $\xi$. The blue line denotes $\xi=1$, and separates the single-fiber regime, which has a timescale independent of $\xi$, and the many-fiber regime, which has a timescale that scales with the inverse of $\xi$.}
\end{figure}

The deflection under shear simulations in the preceding section gave us some insight into the behavior of the BE system: in that numerical experiment, apparent stiffness of the fiber bed was independent of $\xi$ for $\xi\lesssim1$, and decreased as $\xi^{-1}$ for $\xi\gtrsim1$. To obtain more refined information regarding the fully coupled fluid-fiber behavior, we perform oscillatory rheology experiments. The setup is effectively the same as described in \Cref{section:deflection}: a uniform bed of fibers with length $L$ are clamped to a stationary lower wall at $z=0$ at a right angle, and an upper wall at $z=2L$ is subjected to time oscillatory motion. The velocity at the upper wall is set to $\u_\textnormal{upper}(t)=2 \epsilon L\omega\cos(\omega t)\mathbf{\hat x}$. This gives a maximum strain rate of $\epsilon\omega$, and $\epsilon=0.001$ is chosen to ensure a linear response. Scaling space by the fiber length $L$ and time by the elasto-viscous relaxation time $E/\eta L^4$ in \Cref{equation:simplified_model} gives the non-dimensional system:
\begin{subequations}
	\begin{alignat}{2}
		-\Delta\u + \grad p &= \xi J^{-1}\F(\X, T),\qquad &\grad\cdot\u	&= 0,	\\
        \V - \u	&= \mathcal{A}\F(\X, T), &\X_t&=\V,
	\end{alignat}
\end{subequations}
where $\F(\X, T) = -\X_{\s\s\s\s} + (T\X_\s)_\s$. Boundary conditions on the fiber are unchanged; boundary conditions on the fluid are given by $\u(z=0)=\mathbf{0}$ and $\u(z=2)=2\epsilon\omega_0\cos(\omega_0 t)$. The two non-dimensional parameters governing this system are the effective density $\xi$, defined in the same way as in \Cref{eq:shear_parameters}, and the effective frequency $\omega_0$:
\begin{equation}
    \omega_0 = \frac{8\pi\mu L^4}{cE}\omega.
\end{equation}
The shear stress measured by a rheometer is $\oldsigma=u_z(2)$. When the fiber bed is sparse and $\xi$ is small, fluid stresses dominate fiber stresses. To more easily extract information regarding relaxation of the fiber bed, we consider instead the fiber-induced shear stress:
\begin{equation}
    \oldsigma_\textnormal{fiber}(t) = u_z(2) - \epsilon\omega_0\cos(\omega_0 t) = -\xi\int_0^1\X_{zzzz}(t)\,dz + \mathcal{O}(\epsilon^2).
    \label{equation:shear_stress}
\end{equation}
The stress $\oldsigma_\textnormal{fiber}$ may be decomposed as $\oldsigma_\textnormal{fiber}=\epsilon[G'\sin(\omega_0 t) + G''\cos(\omega_0 t)]$, where $G'$ and $G''$ are the storage and loss moduli, respectively. For $\xi=10$, \Cref{figure:rheology:xi10} shows $G'(\omega_0)$ and $G''(\omega_0)$ for $\omega_0$ spanning 4 decades of frequency. The frequency at which $G'$ and $G''$ cross is denoted as $\omega_0^*$, and implies a relaxation timescale $t^*=2\pi/\omega_0^*$. The curves $G'(\omega_0)$ and $G''(\omega_0)$ depend on the effective density $\xi$. The relaxation timescale $t^*$ is shown as a function of $\xi$ in \Cref{figure:rheology:timescales}, along with the $\xi=0$ limit ($t^*\approx0.478$), which may be computed analytically for small $\epsilon$. In accord with the results shown in \Cref{section:deflection:behavior}, the relaxation timescale is effectively independent of $\xi$ for $\xi\lesssim1$, and scales with $\xi^{-1}$ for $\xi\gtrsim1$. Recall, that $\xi$ is $\mathcal{O}(1)$ when the average separation length between fibers is approximately half of the fiber length $L$.

In \cite{Nazockdast2016}, full simulations based on slender-body theory were used to study the behavior of a ball with flexible fibers clamped to its exterior undergoing oscillatory forcing. These simulations found a surprising result: the relaxation time of the system was approximately 25 times faster than the relaxation time estimated from the elasto-viscous response of a single fiber. This mechanism is explained here: when $\xi$ is large, the effects of other fibers become dominant over the single fiber dynamics. Because the fibers are attached to a sphere and oriented in a radial fashion, their density is not uniform: the effective density in their simulations varies from $\xi=512$ at the surface of the ball to $\xi=57$ at the free ends of the fibers. The average effective density in the spherical shell is $\overline\xi=118$. Our rheological experiments indicate that for $\overline\xi=118$, the relaxation time should be reduced from the single fiber relaxation time by a factor of $\approx 23$, in close agreement with the result from the full numerical simulations.

\section{Buckling of fiber beds}
\label{section:buckling}

Consider a bed of fibers in a fluid subjected to a compressive force density due to their own weight. An individual fiber in a fluid will buckle when the gravitational load exceeds a critical size that depends on the bending rigidity of the fiber. In this section we ask how this critical load changes in a bed of fibers, and how the growth rate of the instability is changed as the density is varied. In \Cref{section:deflection,section:rheology}, shear experiments demonstrated that a dense bed of fibers has an effective rigidity that is increased with respect to the rigidity of a single fiber. From this result, it is natural to predict that for fibers of a given rigidity, one would need a greater gravitational load to induce buckling in a dense system. Inclusion of a gravitational force in \Cref{equation:simplified_model} gives:
\begin{subequations}
	\label{equation:massive}
	\begin{alignat}{2}
    	-\mu\Delta\u + \grad p &= \rho(-E\X_{\s\s\s\s}+(T\X_\s)_\s - \oldsigma g\mathbf{\hat z}),\qquad&
        \grad\cdot\u&= 0,	\\
        \eta(\V-\u) &= \A(-E\X_{\s\s\s\s}+(T\X_\s)_\s - \oldsigma g\mathbf{\hat z}),  &\X_t&=\V,     
    \end{alignat}
\end{subequations}
where $\oldsigma$ is the mass per unit length of an individual fiber, and $g>0$ is the gravitational constant. Nondimensionalization and rewriting the system in terms of the unit-tangent $\d$ yields:
\begin{subequations}
	\label{equation:massive:nondim}
	\begin{align}
    	-\Delta\u + \grad p &= \xi J^{-1}(-\d_{\s\s\s}+(T\d)_\s - \tilde g\mathbf{\hat z}),\qquad
        \grad\cdot\u= 0,	\\
        \d_t-\u_\s &= \partial_\s\A(-\d_{\s\s\s}+(T\d)_\s - \tilde g\mathbf{\hat z}),
    \end{align}
\end{subequations}
where the effective density $\xi=\frac{8\pi\rho_0 L^2}{c}$, the effective buckling force $\tilde g=\frac{\oldsigma gL^3}{E}$, and the Lagrange multipliers $T$ and $p$ have been rescaled by $L^2/E$ and $\eta L^4/E$, respectively. Note that the effective density is defined as in \Cref{section:deflection} and \Cref{section:rheology}.

A primary benefit of the BE model is that it leads to a set of simple equations that are easy to linearize. The state $\n=\mathbf{\hat z}$ clearly satisfies \Cref{equation:massive:nondim}, with $\u=\mathbf{0}$, $p=0$, and $T=-\tilde g(1-z)$. Letting $\d(\s)=\epsilon f^x(\s)\mathbf{\hat x} + \epsilon f^y(\s)\mathbf{\hat y} + \mathbf{\hat z}$, we find that, to first order in $\epsilon$, $f^x$ and $f^y$ are decoupled and each evolves according to:
\begin{equation}
	f_t = -f_{zzzz} + [\xi - \tilde g(1-z)]f_{zz} + 2\tilde g f_z + \xi \tilde g (1-z) f = 0,
\end{equation}
subject to the boundary conditions that $f(0)=f_{zzz}(0)=f_z(1)=f_{zz}(1)=0$. Gathering terms together that depend on $\xi$ and reorganizing, we find that:
\begin{equation}
	f_t = (\xi-\partial_{zz})[f_{zz} + \tilde g(1-z)f].
    \label{eq:massive_perturbation}
\end{equation}
Due to the boundary conditions on $f$, the operator $\xi-\partial_{zz}$ is symmetric positive-definite for all positive $\xi$, and thus the stability of the equilibrium is determined by the classical buckling competition between gravitational force and bending rigidity, with no dependence on the fiber density $\xi$. This allows us to semi-analytically compute the bifurcation point, as a function of $\tilde g$.  The general solution to $f_{zz} + \tilde g(1-z)f=0$ is given by:
\begin{equation}
	f(z) = c_1\textnormal{Ai}[\tilde g^{1/3}(z-1)] + c_2\textnormal{Bi}[\tilde g^{1/3}(z-1)],
\end{equation}
where Ai and Bi are the Airy functions of the first and second kinds, respectively, and then using the boundary condition that $f_y(1)=0$ gives that $c_1=\sqrt{3}c_2$. Using the boundary condition that $f(0)=0$ gives that:
\begin{equation}
	0 = \sqrt{3}\textnormal{Ai}(-\tilde g^{1/3}) + \textnormal{Bi}(-\tilde g^{1/3}).
\end{equation}
The equation $\sqrt{3}\textnormal{Ai}(x) + \textnormal{Bi}(x)=0$ can then be solved numerically to high accuracy. This has an infinite family of solutions, the smallest of which is $x=-1.98635270743047$, implying that the first bifurcation occurs at $\tilde g=7.837347438943452$. This is consistent with the numerics for both the full nonlinear system, and an eigen-decomposition of the right-hand side to \Cref{eq:massive_perturbation}. 

This is a significantly different result than our intuition from the numerical experiments in \Cref{section:deflection} and \Cref{section:rheology} might have led us to expect. In those simulations, the fiber beds were externally forced: the bending response of the fiber beds scaled with the density $\xi$, but the forces didn't, and thus the fiber beds appeared effectively more rigid. In this experiment, the forcing is intrinsic to the fiber beds: as density is increased, both the effective bending rigidity and effective buckling forces increase in the same way with $\xi$, and so there is no change in the critical gravitational load at which the bed buckles. Another way to think of this is that there are two competitions of bending rigidity vs. buckling forces at work here: one in the individual fibers, and one due to the many-fiber hydrodynamics, and these competitions flip sign at precisely the same point.

\begin{figure}[h!]
    \centering
    \begin{subfigure}[c]{0.49\textwidth}
        \includegraphics[width=\textwidth]{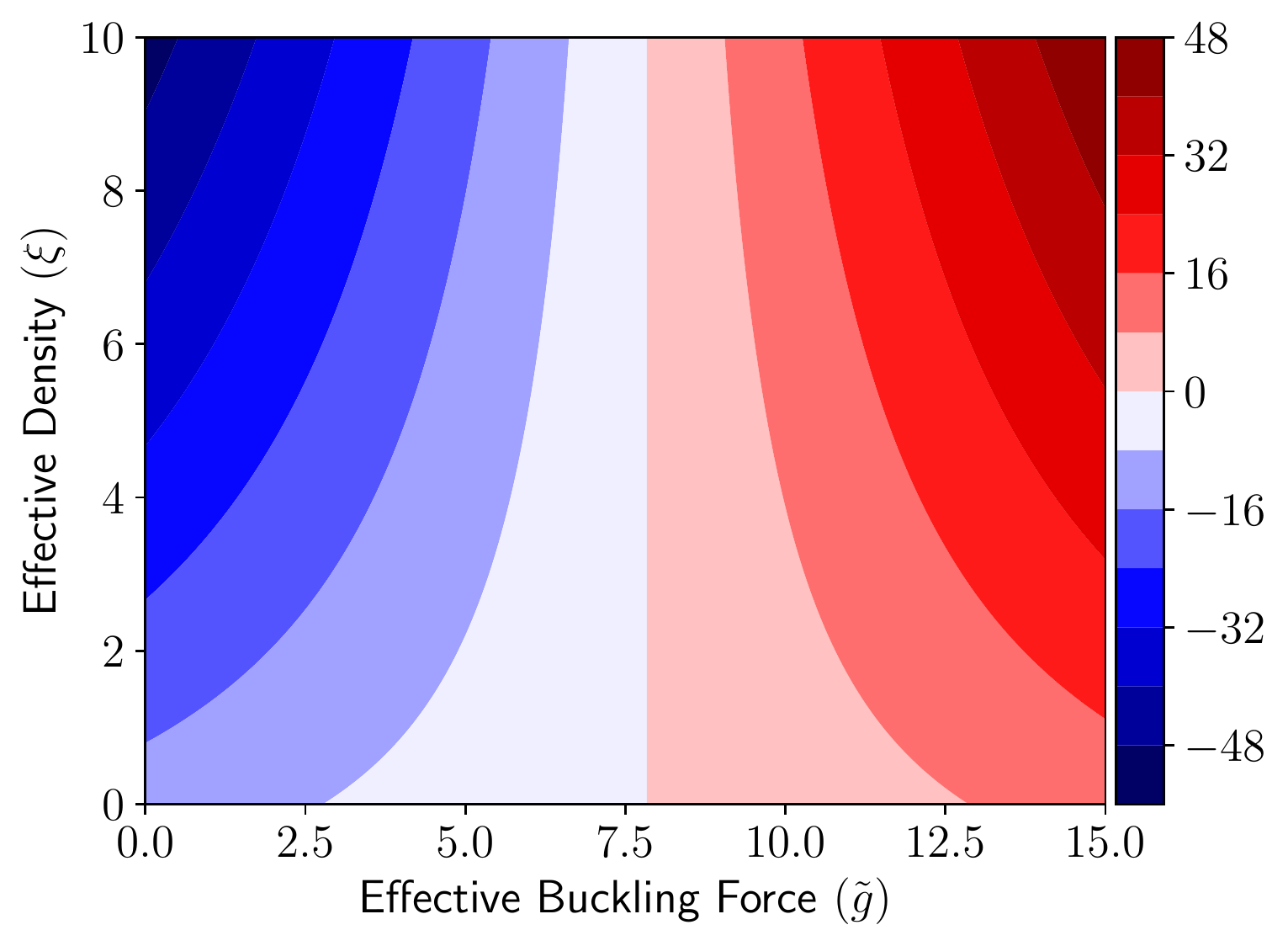}
        \caption{}
        \label{figure:growth:rates}
    \end{subfigure}
    \begin{subfigure}[c]{0.49\textwidth}
        \includegraphics[width=\textwidth]{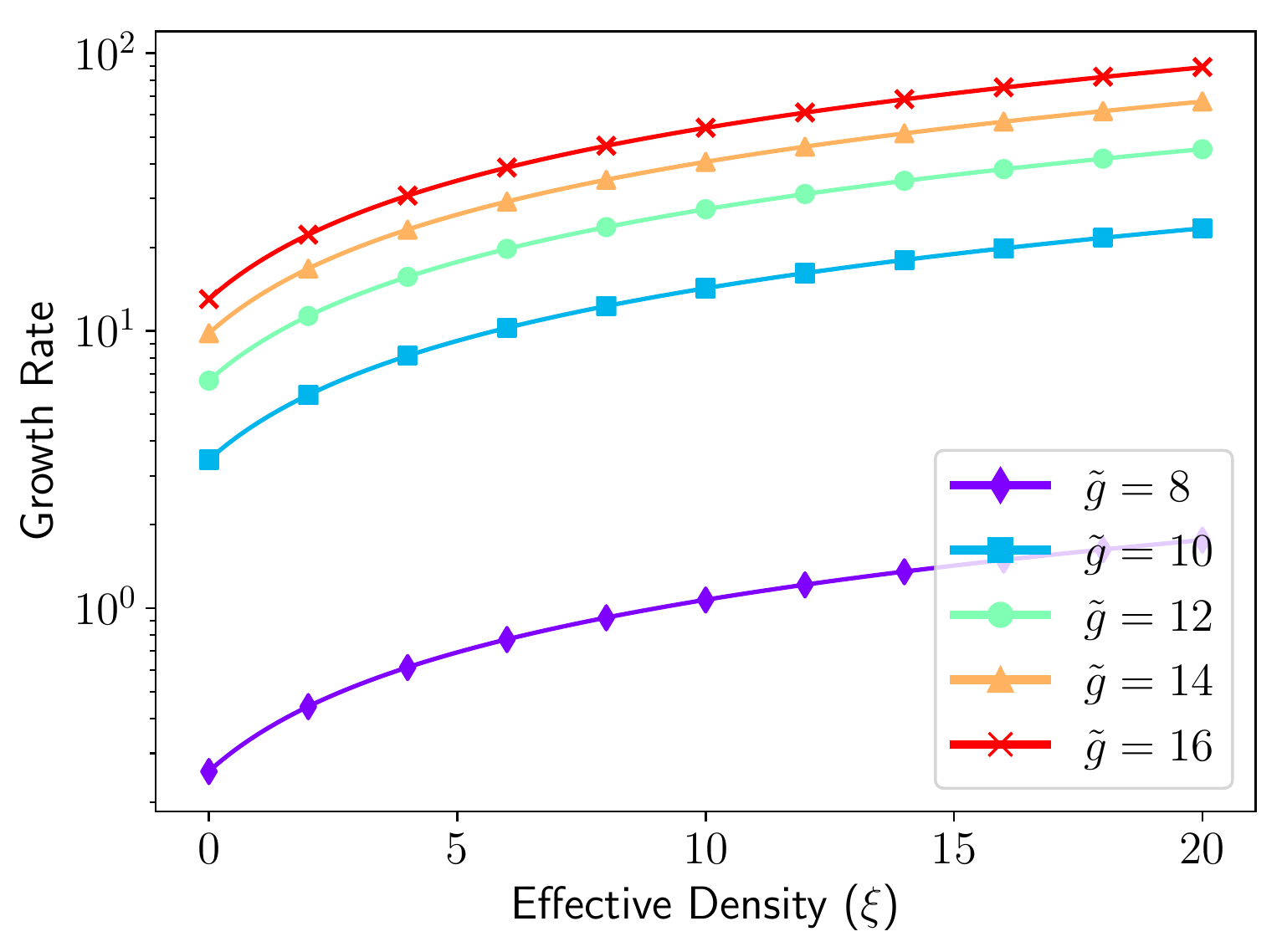}
        \caption{}
        \label{figure:growth:comparison}
    \end{subfigure}
    \caption{Stability of fiber beds under gravitational load. Panel (a) shows the growth rate of a small perturbation $f$, as a function of the effective buckling force $\tilde g$ and the effective density $\xi$. Panel (b) shows the growth rate computed from both the eigen-decomposition (lines), and from numerical simulation of the full nonlinear system (markers), as a function of the effective density $\xi$, for several values of the effective gravitational force $\tilde g$).}
    \label{figure:growth}
\end{figure}

Although the bifurcation between stable and buckling behavior may occur at the same value of $\tilde g$ regardless of the density $\xi$, this does not mean that there is no difference in the behavior of the beds. Once one side wins in the competition  between rigidity and buckling forces, the amount by which it wins \emph{does depend on $\xi$}, and thus the rate at which the instability grows in a bed of fibers depends on its density. Due to the non-constant coefficient nature of \Cref{eq:massive_perturbation}, these rates are non-trivial to solve for analytically. Instead, we discretize the right hand side of \Cref{eq:massive_perturbation}, subject to the given boundary conditions, and compute its eigen-decomposition. The growth rate, as a function of $\xi$ and $\tilde g$ is shown in \Cref{figure:growth:rates}. \Cref{figure:growth:comparison} shows growth rates computed from both the eigen-decomposition of \Cref{eq:massive_perturbation}, as well as from simulation of the full non-linear system \Cref{equation:massive:nondim}, as a function of the effective density $\xi$ for several values of $\tilde g$.

\section{Modeling a Soft Rectifier}
\label{section:rectification}
\begin{figure}[h!]
	\centering
	\includegraphics[width=0.6\textwidth]{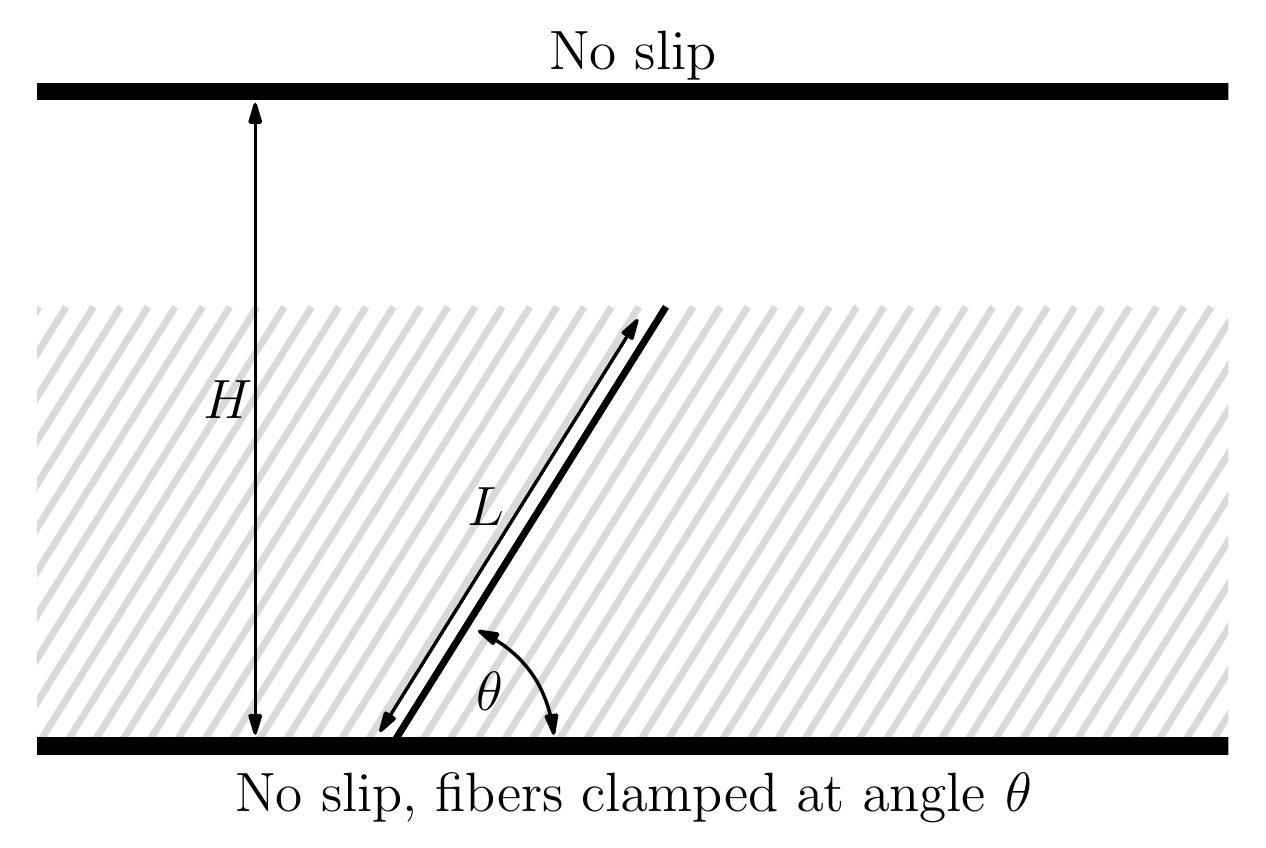}
    \caption{Physical setup for fiber-bed flow rectifier. Fibers of length $L$ are placed at an angle $\theta$ relative to the base of a channel of height $H$. Flow in different directions induces an asymmetric response in the fiber bed, allowing more fluid to pass in the forward direction than the reverse direction (see \Cref{figure:valve}).}
    \label{figure:valve:schematic}
\end{figure}

In \Cref{section:deflection}, we see that beds of fibers deform and inhibit flow in response to an imposed shear in a manner that depends on the fiber bed density. It has been recently observed \cite{Alvarado2017} that this may be exploited for flow rectification by angling the beds with respect to the channel. In this section we examine this problem in detail, with a focus on optimizing flow rectification. The physical setup is shown in \Cref{figure:valve:schematic}.
A bed of fibers of length $L$ are mounted in a channel of height $H$, at an angle $\theta$ with respect to the channel axis. The pressure gradient that would drive a flow of $u_0$ at the centerline in the absence of fibers is applied. The resultant flow is unidirectional and the resultant fiber deformation is asymmetric in the sign of the applied gradient. We assume that the fibers are made of a material with Young's modulus $Y$, have a radius $r_\textnormal{fiber}$, and that the bed has a number density per unit area $\rho_0$, measured at the base of the fibers. Scaling space by $H$ and time by $H/u_0$, we find that the fiber bed deformations satisfy:
\begin{subequations}
	\begin{align}
    	-\Delta\u + \grad p &= \xi J^{-1}(-\tilde E\d_{\s\s\s} + (T\d)_\s),\qquad\grad\cdot\u-0, \\
        \d_t - \u_\s &= \mathcal{A}(-\tilde E\d_{\s\s\s} + (T\d)_\s),
    \end{align}
\end{subequations}
where $\u(z=0)=\u(z=1)=0$, forced by a pressure gradient of $8$, so that when $\xi=0$, $\u(z)=4(z^2-z)\mathbf{\hat x}$ and the average flow across the channel is $2/3$. The effective density is given by $\xi=8\pi\rho_0H^2/c$, and the effective rigidity $\tilde E=\frac{E}{\eta H^3 u_0}$, with $E$ the bending modulus of a fiber, given by $E=\frac{\pi}{4}Y r_\textnormal{fiber}^4$. The height of the channel may be effectively scaled out of the problem by defining the radius, length, and density of the fibers, as well as the flow speed $u_0$, relative to $H$. Let $\tilde r=r_\textnormal{fiber}/H$, $\tilde L = L/H$,  $\tilde\rho=\rho_0 H^2$, and $\tilde u=u_0/H$. Then we have that:
\begin{equation}
	\xi = \frac{8\pi\tilde\rho}{c}, \qquad \tilde E = \frac{\pi Y\tilde r^4}{4\eta\tilde u},
\end{equation}
where as before $\eta = 8\pi\mu/c$ and $c=-\log(e(\tilde r/\tilde L)^2)$.

\begin{figure}[h!]
\centering
\begin{subfigure}[c]{0.32\textwidth}
	\includegraphics[width=1.0\textwidth]{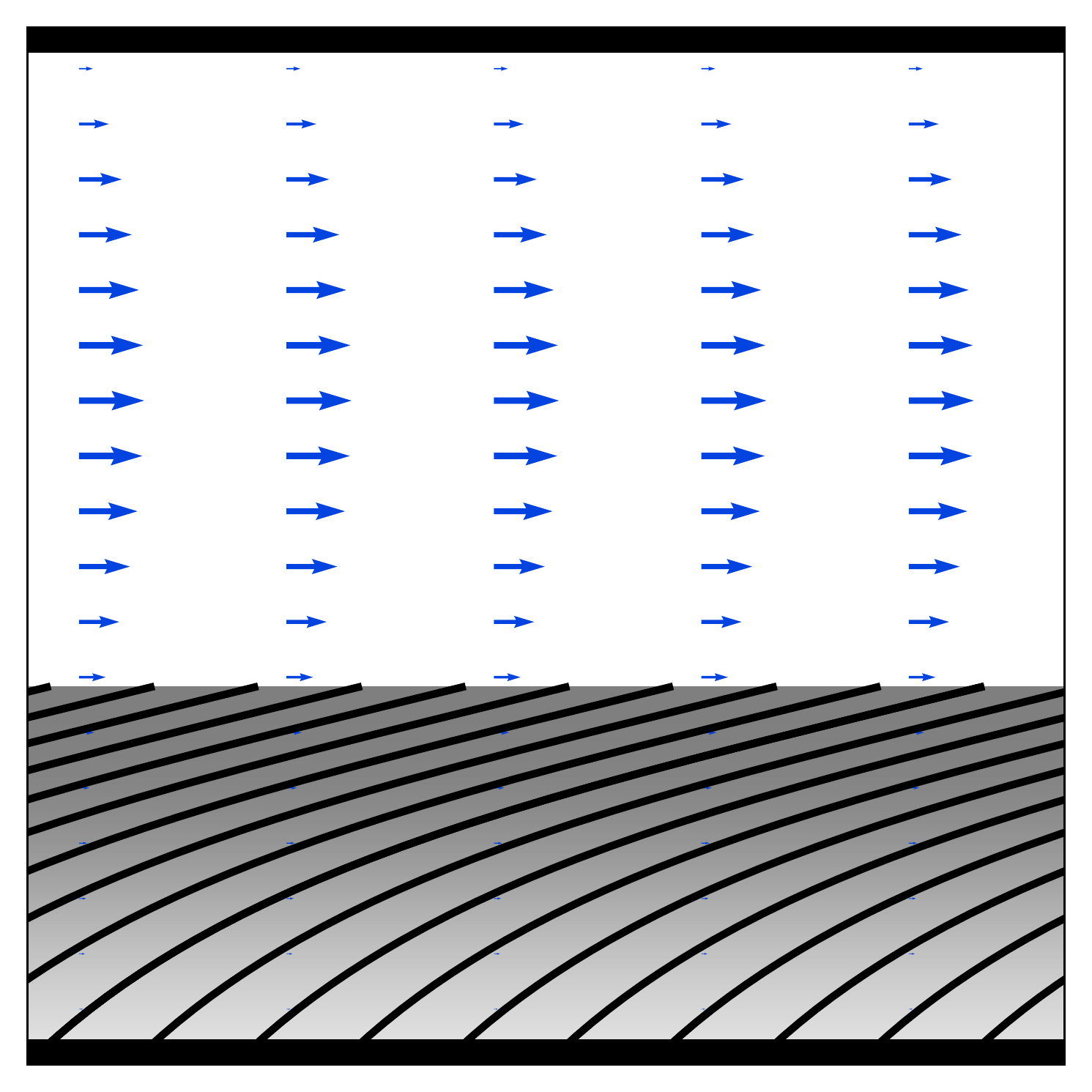}
    \caption{}
    \label{figure:valve:forward}
\end{subfigure}
\begin{subfigure}[c]{0.32\textwidth}
	\includegraphics[width=1.0\textwidth]{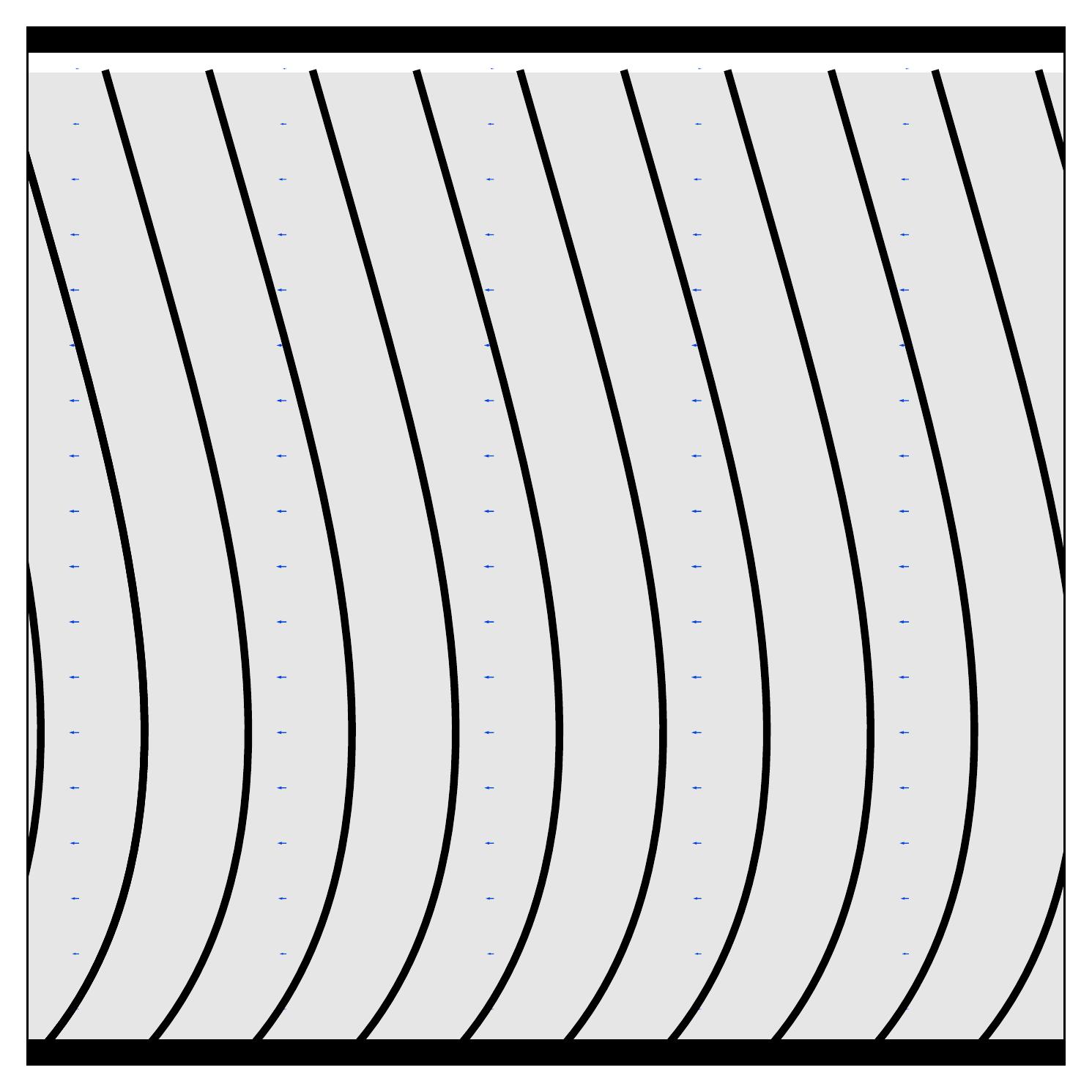}
    \caption{}
    \label{figure:valve:backward}
\end{subfigure}
\begin{subfigure}[c]{0.32\textwidth}
	\includegraphics[width=1.0\textwidth]{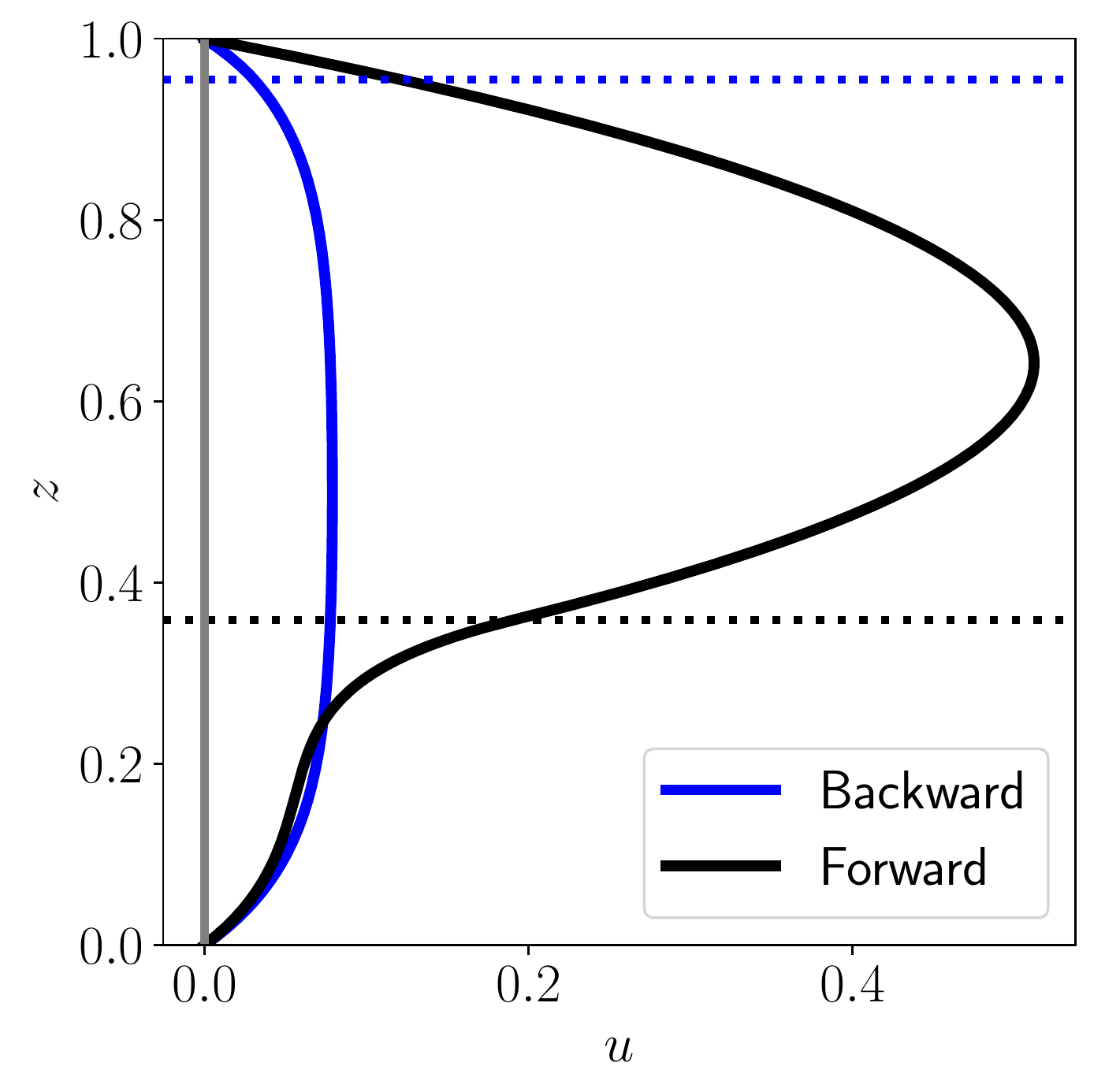}
    \caption{}
    \label{figure:valve:u}
\end{subfigure}
\caption{Panel (a) and Panel (b) show the fiber deformation and flow fields for a flow rectifier with $\theta=45^\circ$, $\tilde L=1$, $\xi=100$, and $\tilde E=0.01$. Blue arrows show the flow field and the grayscale shows the fiber density, scaled consistently. Panel (c) shows the velocity field as a function of $z$ corresponding to Panel (a), in black, and Panel (b), in blue; $u$ for the `backward' direction has been multiplied by $-1$ to make comparison easier.}
\label{figure:valve}
\end{figure}

We now fix $\theta=45^\circ$ and $\tilde L=1$ in order to gain an intuition for how the system provides flow rectification, and how this depends on $\xi$ and $\tilde E$. \Cref{figure:valve} shows the behavior of the fiber bed and velocity profile of the fluid at steady state, for $\xi=100$ and $\tilde E=0.01$. Here we can see the primary mechanism behind flow rectification: asymmetric deformation gives a much larger channel for fluid flow unimpeded by the fibers in one direction as compared with the other direction. Let us define two quantities: the Impedance Ratio, given by $\mathcal{I}_r=-\int_0^1\u_{\textnormal{forward}}(z)\,dz/\int_0^1\u_{\textnormal{backward}}(z)\,dz$, and the Forward Impedance, given by $\mathcal{I}_f=\frac{3}{2}\int_0^1\u_{\textnormal{forward}}(z)\,dz$. The velocity $\u_{\textnormal{forward}}(z)$ denotes the steady-state solution when the flow direction coincides with the direction of forward inclination of the fibers; $\u_{\textnormal{backward}}(z)$ denotes the same quantity when the flow direction is opposed to the inclination direction of the fibers. Note that the Forward Impedance is the ratio of the flow in the forward direction to the flow in an unimpeded channel.  An optimal rectifier will have $\mathcal{I}_r=\infty$ and $\mathcal{I}_f=1$. In the example shown in \Cref{figure:valve}, $\mathcal{I}_r=3.96$ and $\mathcal{I}_f=0.38$. We note that a non-trivial percentage of the flow is \emph{in the fiber region}: $\approx 9\%$ for the forward direction and $\approx 99\%$ for the backward direction. Although this depends on $\xi$, in general the flow in the fiber-region cannot be neglected. 

\begin{figure}[h!]
\centering
\begin{subfigure}[c]{0.325\textwidth}
	\includegraphics[height=0.7\textwidth]{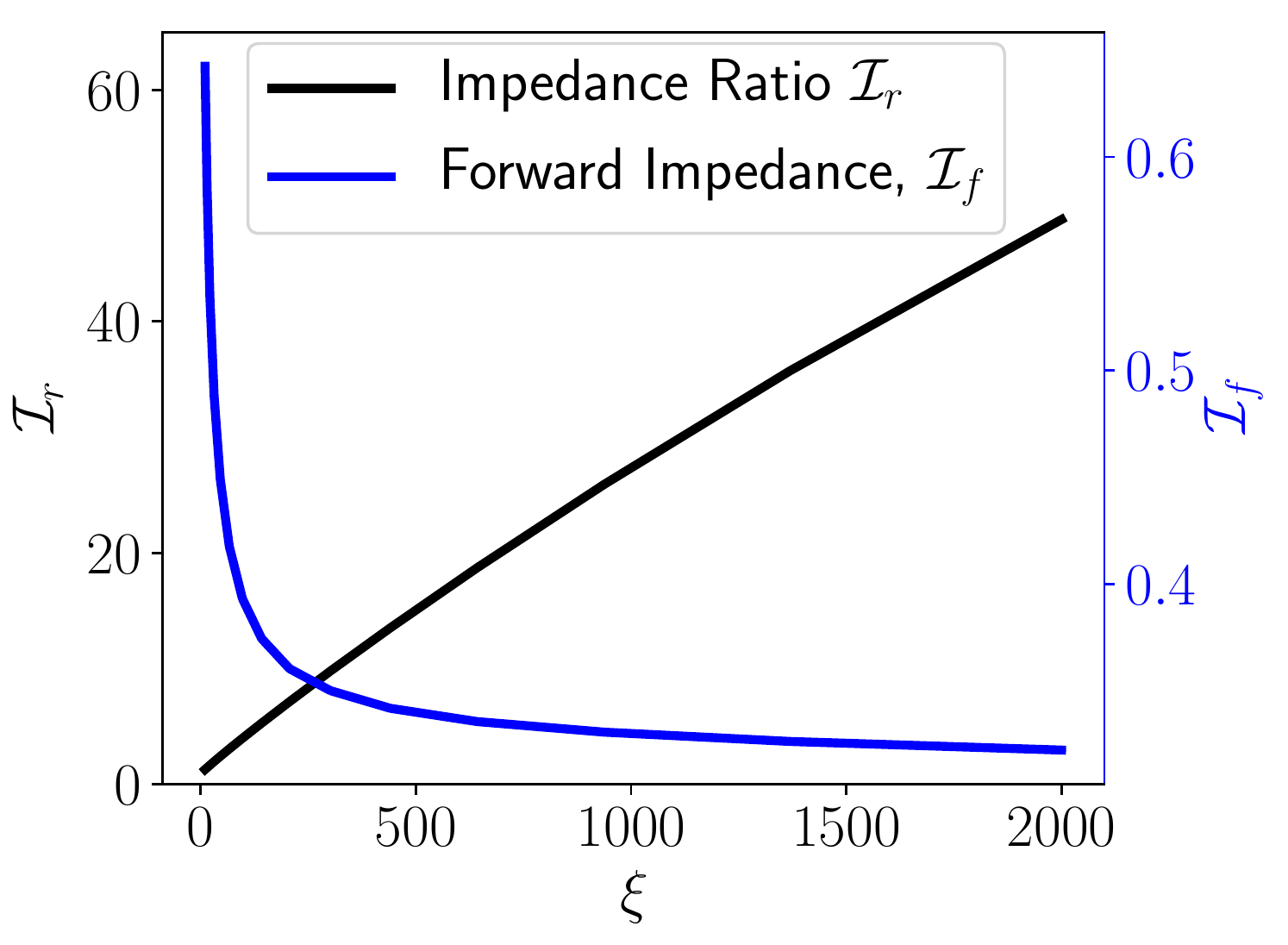}
    \caption{}
    \label{figure:valve_many:Is}
\end{subfigure}
\begin{subfigure}[c]{0.325\textwidth}
	\includegraphics[height=0.7\textwidth]{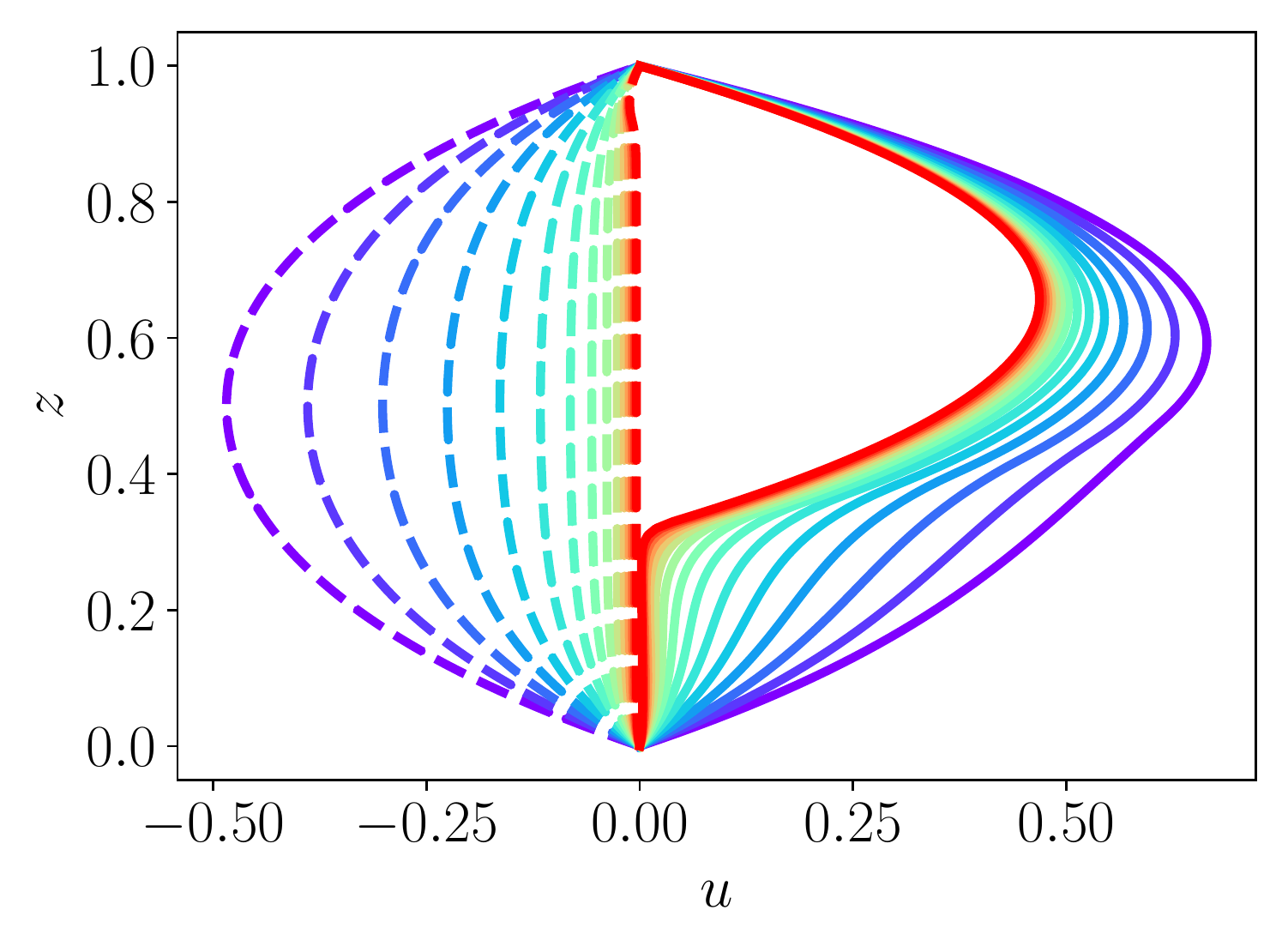}
    \caption{}
    \label{figure:valve_many:us}
\end{subfigure}
\begin{subfigure}[c]{0.325\textwidth}
	\includegraphics[height=0.7\textwidth]{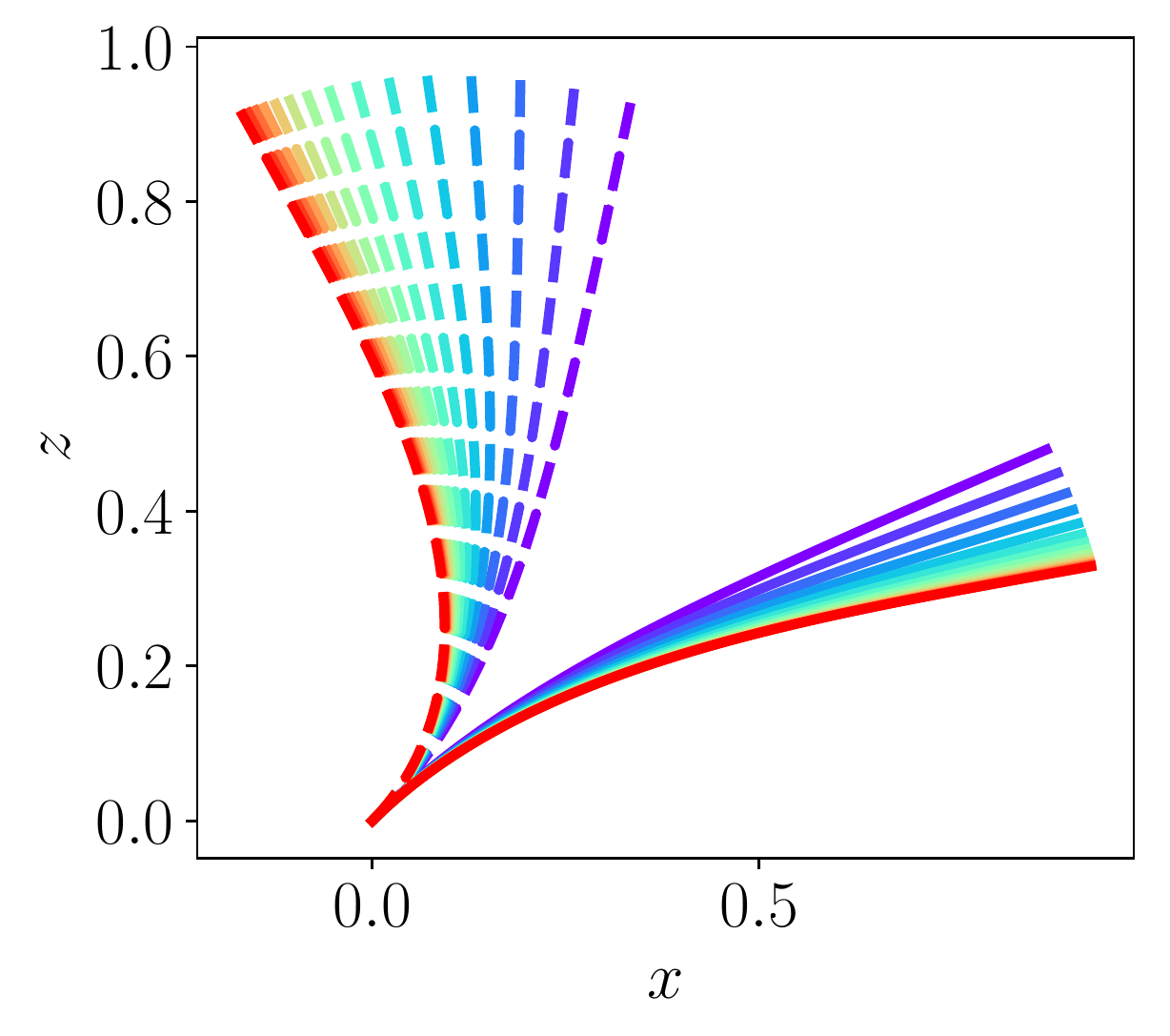}
    \caption{}
    \label{figure:valve_many:Xs}
\end{subfigure}
\caption{Panel (a) shows the Impedance Ratio ($\mathcal{I}_r$), and the Forward Impedance ($\mathcal{I}_f$), for $\theta=45^\circ$ and $\tilde L=1$, across a wide range of $\xi$ with $\xi\tilde E=1$. Panel (b) and Panel (c) show the velocity field and  fiber deformation, respectively, with solid lines denoting results in the Forward direction and dashed lines denoting results in the Backward direction. Colors denote density; purple is the least dense ($\xi=10$); red is the most dense ($\xi=2000$). The colors are equispaced on a log scale in $\xi$; the two most purple lines are much closer together in $\xi$ than the two most red lines.}
\label{figure:valve_many}
\end{figure}

It is clear that in order to obtain a large Impedance Ratio, the deformation of the fibers should be tuned so that the fibers occupy most of the channel, as in \Cref{figure:valve:backward}. From \Cref{section:deflection:behavior}, we know that fiber deflection should scale as $\xi^{-1}$ for $\xi\gtrsim 1$, and thus fixing $\tilde E\xi$ should yield similar fiber deformations across a wide range of $\xi$. With $\theta$ and $\tilde L$ still set to $45^\circ$ and $1$, respectively, we fix $\tilde E\xi=1$ and vary $\xi$ from $\xi=10$ to $\xi=2000$. The results are shown in \Cref{figure:valve_many}. As $\xi$ is increased, $\mathcal{I}_f$ rapidly decreases from $\approx 0.65$, asymptoting to a value of $\approx 0.3$. The Impedance Ratio, however, scales very nearly linearly with $\xi$ across the entire range. \Cref{figure:valve_many:us} explains this: as $\xi$ is increased, permittivity in the fiber region is decreased. Because $\tilde E$ has been scaled to keep $\tilde E\xi$ fixed as $\xi$ is varied, the deformation is nearly unchanged (see \Cref{figure:valve_many:Xs}), and the fiber region is effectively constant. Due to the asymmetry in the setup, the final configuration in the forward direction has a far larger region of the channel that remains fiber free, and hence the Impedance Ratio increases.

\subsection{Rectifier optimization}
\begin{figure}[h!]
    \centering
	\includegraphics[width=0.6\textwidth]{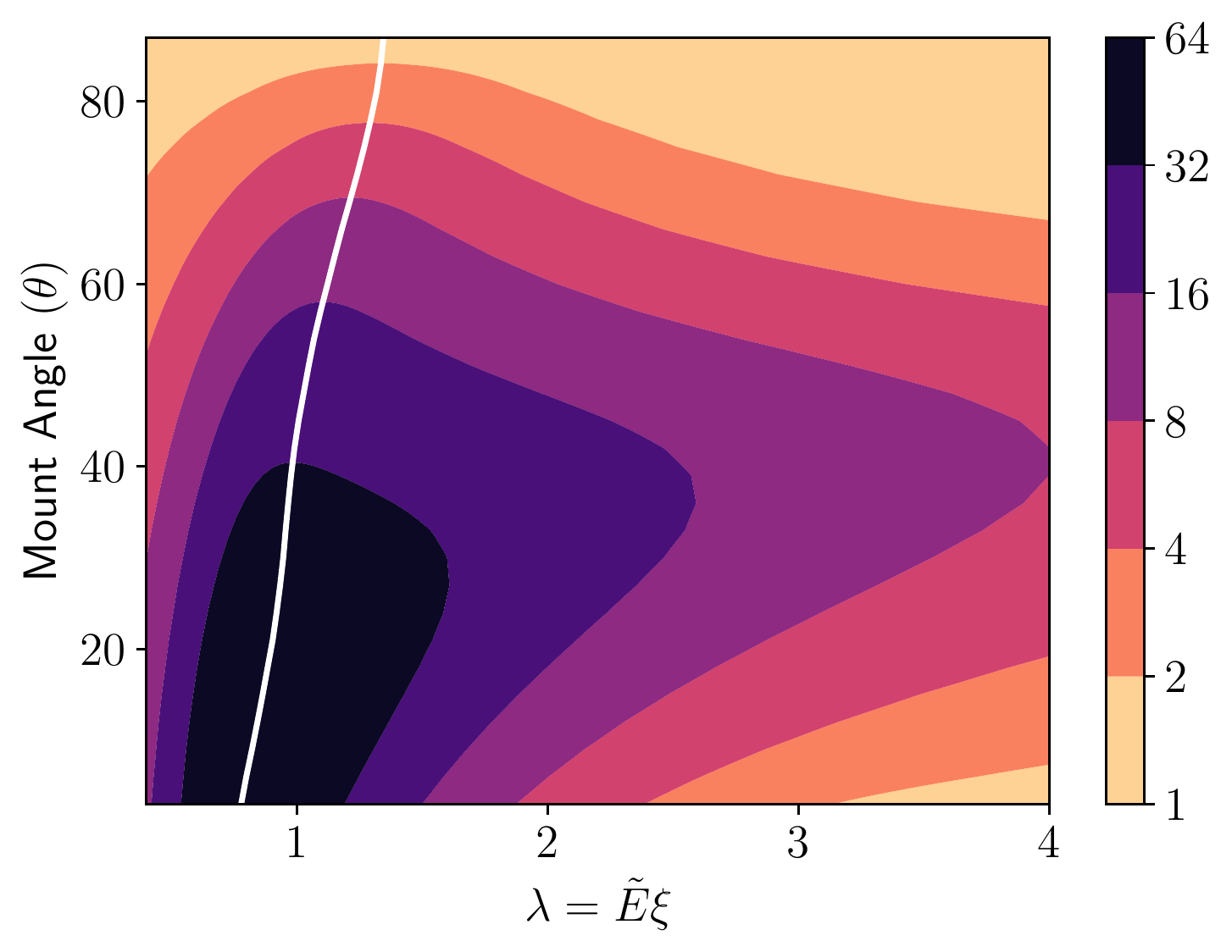}
    \caption{The rectification ratio $\mathcal{I}_r$, as a function of the mount angle $\theta$ and $\lambda=\xi\tilde E$ with $\xi=1000$. The white line shows the value of $\lambda$ that produces the largest rectification ratio, as a function of $\theta$.}
    \label{figure:optim}
\end{figure}

We now consider designing an optimal flow rectifier (that maximizes $\mathcal{I}_r$) for a given flow rate $u_0$. In general, this problem involves an optimization over the variables $\tilde L$, $\theta$, $\xi$, and $\tilde E$. The optimal value for $\tilde L$ should clearly be slightly greater than $1$ so that the tips of the fibers are coincident with the top of the channel in the backflow state (see \Cref{figure:valve_many:Xs}). However, as the flow rate varies, this is likely to cause sticking and damage to the fibers (as well as issues with our numerical scheme). We will thus fix $\tilde L=1$. The results from \Cref{figure:valve_many} provide a simple strategy for optimizing $\xi$ and $\tilde E$: fix $\xi$ to be relatively large, and compute the rectification ratio with $\lambda=\xi\tilde E$ over a range of $\lambda$. We fix $\xi=1000$, and compute the rectification ratio $\mathcal{I}_r$ for $0.1\leq\lambda\leq 4$ and for $3^\circ\leq\theta\leq87^\circ$. The rectification ratios are shown in \Cref{figure:optim}, and the value of $\lambda$ for which $\mathcal{I}_r$ is optimized, as a function of $\theta$, is shown by a white line. The optimal value of $\lambda$ varies only slightly as $\theta$ varies. When $\theta$ is close to $90^\circ$, there is little geometric asymmetry, leading to weak rectification. Small values of $\theta$ are most efficient, but provide high rectification only over a small range of $\lambda$. Note that because $\tilde E$ depends on $u_0$, changes in the flow rate manifest themselves as changes in $\lambda$. If the flow rate is inconsistent in the intended application, angles in the range $40-60^\circ$ provide effective rectification over a wide range of flow rates.

Once the target value of $\lambda$, denoted by $\lambda^*$ is determined, based on the desired rectification and robustness to flow rate, $\xi$ should be taken as large as possible by decreasing $\tilde r$ (see \Cref{figure:valve_many:Is}). For a fiber of Young's modulus $Y$, $\tilde r$ cannot be scaled to be arbitrarily small: as $\tilde r$ is decreased with $\lambda$ held constant, the area fraction $\phi$ of the fibers increases. If the maximum area fraction that we will accept is $\phi_\textnormal{max}$, then $\tilde r_\textnormal{min}$ is given by:
\begin{equation}
	\tilde r_\textnormal{min}^2 = \frac{4\mu\tilde u\lambda^*}{\phi_\textnormal{max}Y\sin\theta}, \qquad r_\textnormal{min}^2 = \frac{4\mu u_0 H\lambda^*}{\phi_\textnormal{max}Y\sin\theta}.
\end{equation}
In practice, it is unlikely that this minimal limit will ever be met. For example, consider a worst case scenario: a channel with $H=1mm$, and a high viscosity fluid ($\mu=10 Pa\,s$) at a free flow rate $u_0=10mm/s$, and fibers made of a relatively weak elastomer ($Y=0.01 GPa$), with a maximum allowed area fraction $\phi=0.1$. Note that this flow has Reynolds number $1$, and is hence at the far edge of applicability of the theory. Nevertheless, $r_\textnormal{min}=20\sqrt{\lambda^*}\mu m$. Since $\lambda(\theta)$ is typically $\mathcal{O}(1)$, $r_\textnormal{min}$ is likely to be impractically small for manufacture. Thus $r_\textnormal{min}$ should be chosen to be as small as practically manufacturable, and $\rho_0$ should be chosen so that $\xi\tilde E=\lambda^*$.

\section{Waves in an Actuated Fiber Bed}
\label{section:waves}
Finally, we demonstrate the applicability of the model in a fully two-dimensional setting. Consider a simple model for fluid pumping by a fiber bed with defined actuation given by a metachronal wave in the base angle of the fibers. The physical setup is shown in \Cref{figure:diagram}. Fibers are clamped at an angle $\theta$ with respect to a lower channel wall, where the angle $\theta$ varies as a function of time and space according to:
\begin{equation}
	\theta(x, t) = \frac{\pi}{2}\left[1 - \gamma\cos(kx - \omega t)\right].
\end{equation}
Generating initial data consistent with all boundary conditions is nontrivial. Instead, we begin simulations with a bed uniformly oriented in the $\mathbf{\hat z}$ direction, and smoothly increase the size of the oscillation magnitude from $0$ to $\gamma$ over one time unit. The actuated fibers have a bending rigidity $E$, radius $r$, length $L$, and drive a fluid with viscosity $\mu$ in a channel with half-width $H$. We assume the fluid velocity $\u=u\mathbf{\hat x}+w\mathbf{\hat z}$ obeys no-slip boundary conditions at $z=0$, symmetry conditions ($u_z=0$, $w=0$) at $z=H$, and is periodic in the $\mathbf{\hat x}$ direction. Scaling space by $L$ and time by $\omega^{-1}$ gives the non-dimensional system:
\begin{subequations}
  \begin{align}
  	-\Delta\u + \grad p &= \xi J^{-1}(-\tilde E\X_{\s\s\s\s} + (T\X_\s)_\s),\qquad \grad\cdot\u = 0,	\\
    \V - \u &= \mathcal{A}(-\tilde E\X_{\s\s\s\s} + (T\X_\s)_\s),
  \end{align}
\end{subequations}
with $u(0)=w(0)=u_z(\beta)=w(\beta)=0$ and $\theta=\tfrac{\pi}{2}[1-\gamma\cos(\tilde k x - t)]$, where:
\begin{equation}
	\xi = \frac{8\pi\rho_0 L^2}{c},\qquad \tilde E = \frac{E}{\omega\eta L^4}, \qquad \beta=\frac{H}{L}, \qquad\tilde k=\frac{k}{L}.
\end{equation}
The behavior of this system depends on five parameters: the density ($\xi$), effective fiber rigidity ($\tilde E$), channel to fiber-length ratio ($\beta$), effective wavenumber ($\tilde k$), and actuation range ($\gamma$). A full analysis of this system is beyond the scope of this paper, and will be presented in a forthcoming contribution that also considers a physically motivated ciliary actuation model. Instead, in this section we (1) demonstrate that the continuum model allows for the efficient simulation of the fully-coupled fluid-structure problem for dense fiber beds undergoing complex motions in multiple dimensions, (2) demonstrate that the timescale over which the transient dynamics relax to steady state behavior primarily follows the timescale $(\xi\tilde E)^{-1}$, in accord with the results of \Cref{section:rheology}, (3) show that this basic model qualitatively reproduces the pumping action of a ciliary bed undergoing metachronal beating, and  (4) that complex nonlinear phenomena arise at large amplitude in this simple model, including a transition from a steady pumping regime to a pulsatile pumping regime.

\subsection{Numerical performance in two dimensions}

\begin{figure}
    \centering
    \begin{subfigure}[c]{0.48\textwidth}
        \centering
        \vspace{3.88em}
        \begin{tabular}{c|ccc}
            $\xi\downarrow\ n_z\rightarrow$   &   $2^6$     &   $2^7$     &   $2^8$     \\
            \hline
            10                                &   0.62      &   1.60      &    5.57   \\
            100                               &   0.76      &   1.61      &    7.35   \\
            1000                              &   1.00      &   1.83      &    7.02   \\
            10000                             &   1.47      &   3.26      &   13.86
        \end{tabular}
        \vspace{3.88em}
        \caption{}
        \label{fig:2d_timing_stuff:timings}
    \end{subfigure}
    \begin{subfigure}[c]{0.48\textwidth}
        \includegraphics[width=\textwidth]{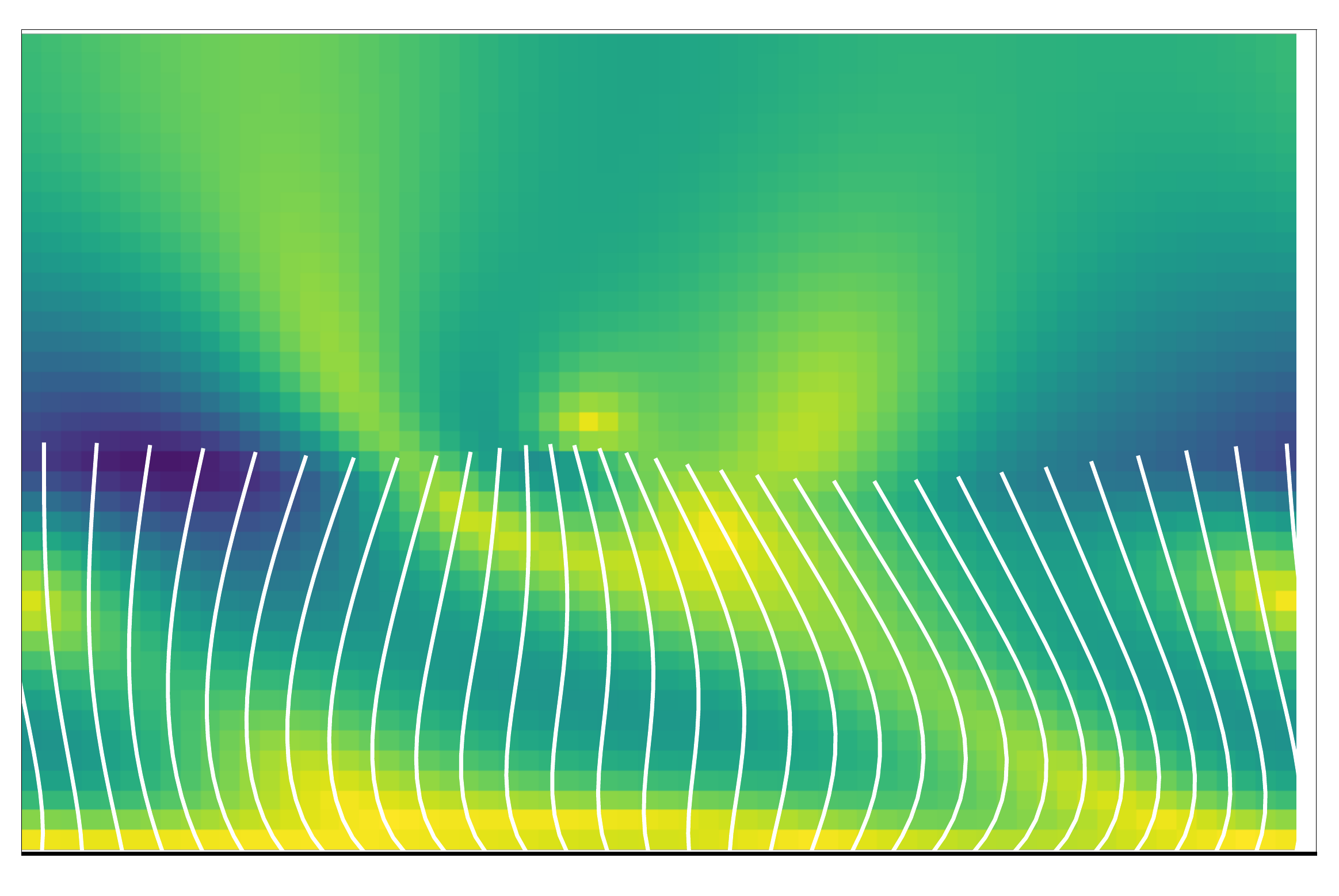}
        \caption{$n=2^6$}
        \label{fig:2d_timing_stuff:6}
    \end{subfigure}
    \begin{subfigure}[c]{0.48\textwidth}
        \includegraphics[width=\textwidth]{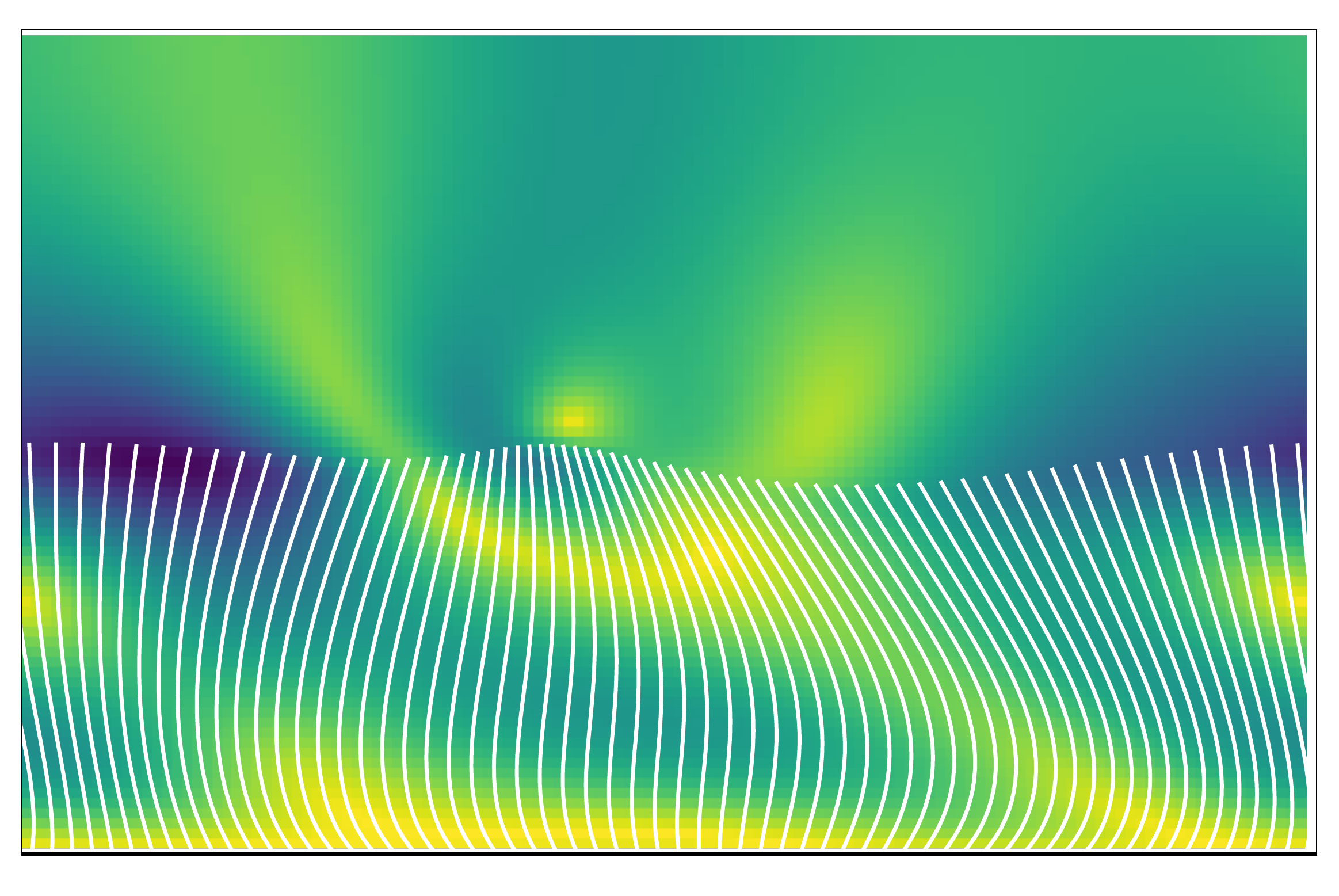}
        \caption{$n=2^7$}
        \label{fig:2d_timing_stuff:7}
    \end{subfigure}
    \begin{subfigure}[c]{0.48\textwidth}
        \includegraphics[width=\textwidth]{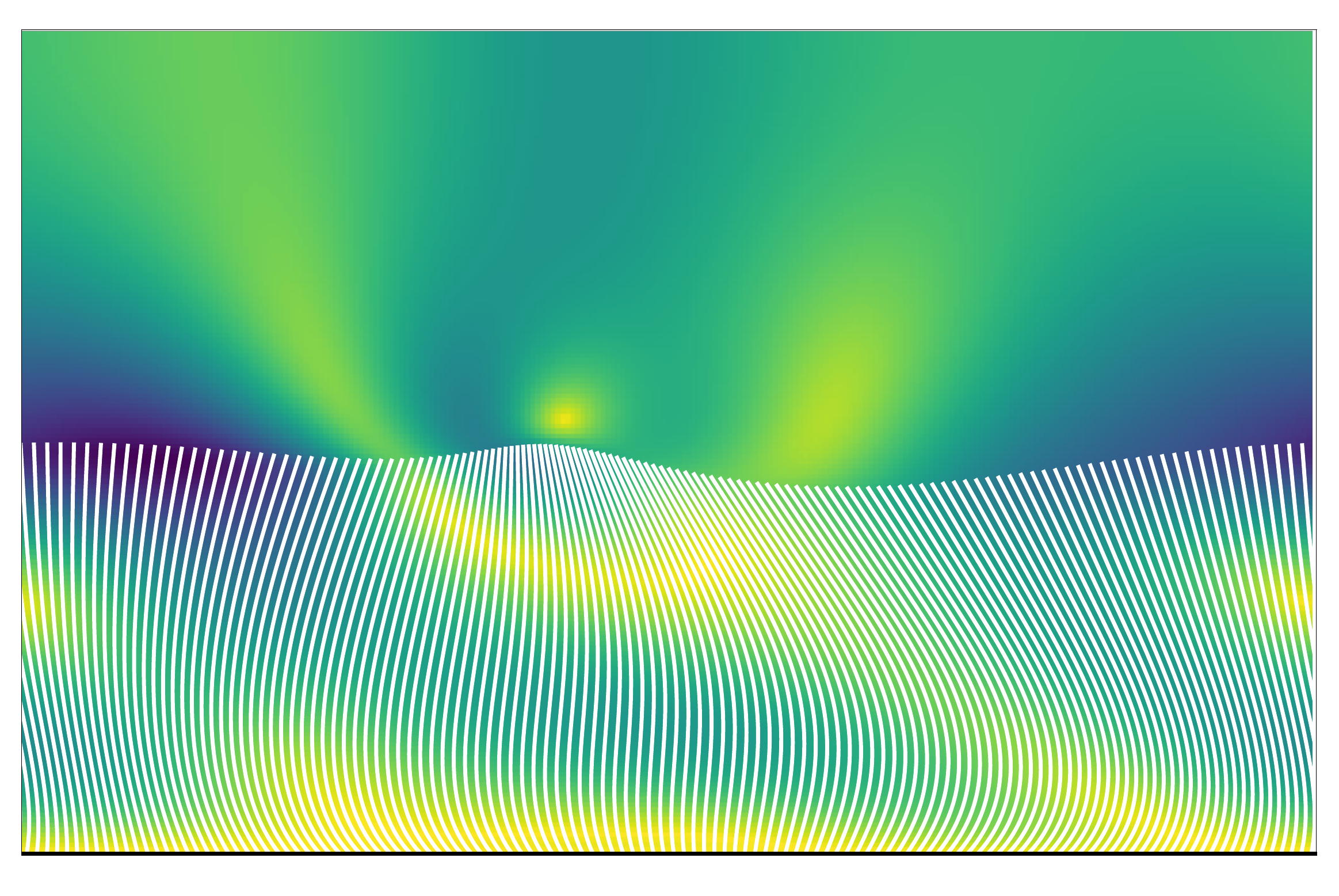}
        \caption{$n=2^8$}
        \label{fig:2d_timing_stuff:8}
    \end{subfigure}
    \caption{Numerical performance and accuracy in two dimensions. Panel (a) shows the wall clock time-per-timestep, in seconds, measured as the average over the first ten timesteps after $t=1$, for a range of densities and discretizations. Panel (b) through (d) show fiber deformation (in white) and velocity magnitude (color scale, larger magnitudes are darker) at time $t=5$, for coarser (b) to finer (d) discretizations. All discrete fibers used in the computation are shown.}
    \label{fig:2d_timing_stuff}
\end{figure}

For simplicity, let $\beta=k=2$ and $\gamma=\frac{\pi}{4}$, and fix $\xi\tilde E=1$. As discussed in \Cref{section:buckling}, the fiber bed will undergo similar size deformations for all sufficiently large $\xi$. We run simulations across a wide range of effective densities: $\xi=10$, $100$, $1000$, and $10000$, with $n=2^6$, $2^7$, and $2^8$ points discretizing the domain in each spatial dimension. The timestep is set to $\Delta t=0.01$, so that the oscillatory timescale ($2\pi)$ and the relaxation timescale $(\tilde E\xi)^{-1}=1$ associated with the dense fiber system are well resolved. The wall clock time-per-timestep, measured as an average over the first ten timesteps after $t=1$ when the oscillation amplitude $\gamma$ is stable, is reported in \Cref{fig:2d_timing_stuff:timings}. Computations were done on a 12-core workstation with two Intel\textsuperscript{\textregistered} Xeon\textsuperscript{\textregistered} CPU E5-2643 v3 processors clocked at 3.40GHz and 128GB of RAM. Run times are linear in the number of unknowns, and increase slowly with $\xi$, as both the GMRES iteration for inverting the Jacobian and the Newton iteration typically require more steps to converge. Nevertheless, even at extremely high densities, runtime is only increased by a factor of 2-3 over the runtime at low densities. In \Cref{fig:2d_timing_stuff:6,fig:2d_timing_stuff:7,fig:2d_timing_stuff:8}, for $\xi=1000$, we show the fiber field and velocity magnitude at time $t=5$ for the three different discretization sizes. Although minor differences are visible between the coarsest and finest discretizations, even the coarsest discretization adequately captures all qualitative features of the fluid velocity and fiber field deformation.

\subsection{Steady and pulsatile pumping}

\begin{figure}
    \centering
    \begin{subfigure}[c]{0.45\textwidth}
        \includegraphics[width=\textwidth]{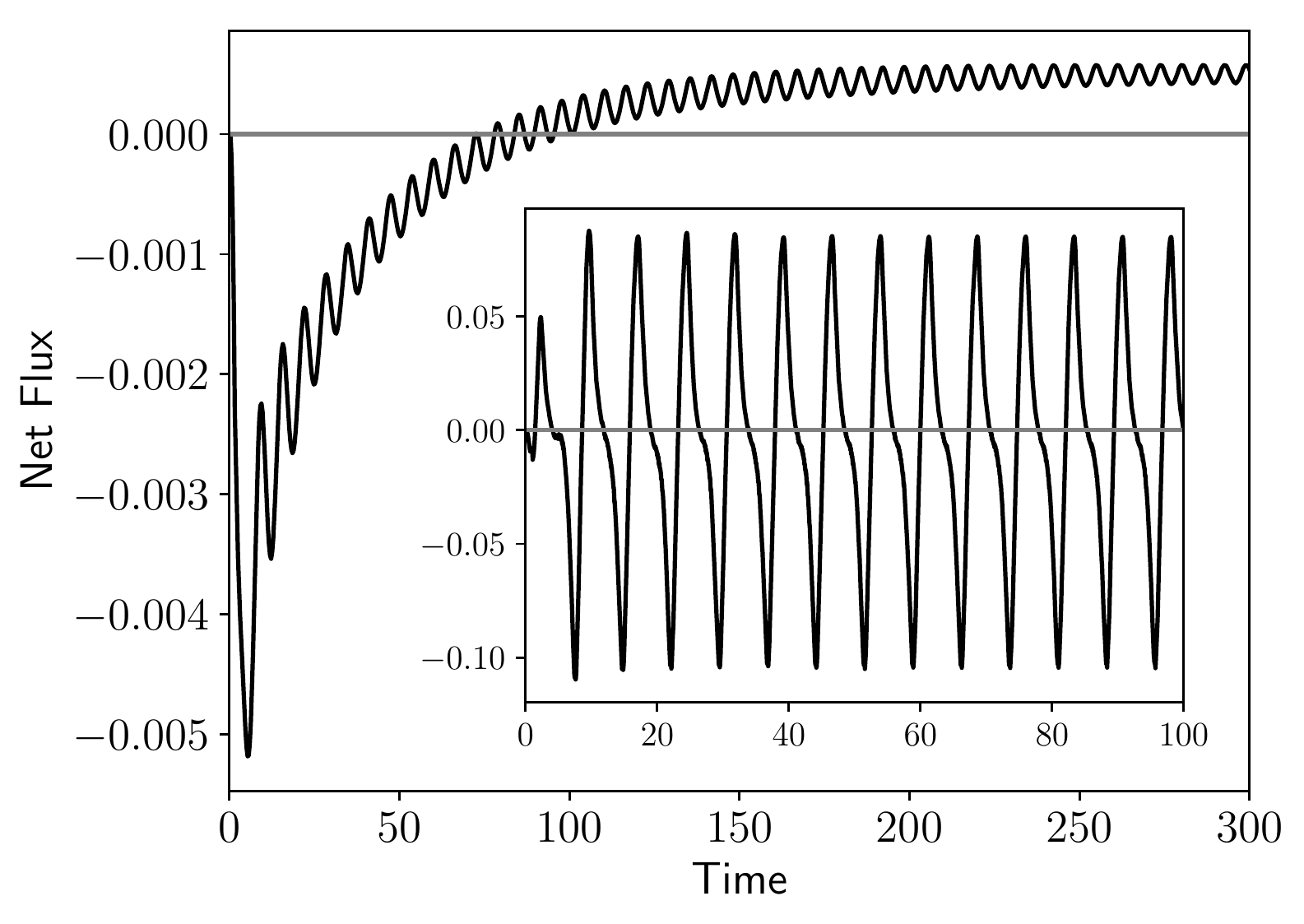}
        \caption{}
        \label{fig:2d_timescales:a}
    \end{subfigure}
    \begin{subfigure}[c]{0.45\textwidth}
        \includegraphics[width=\textwidth]{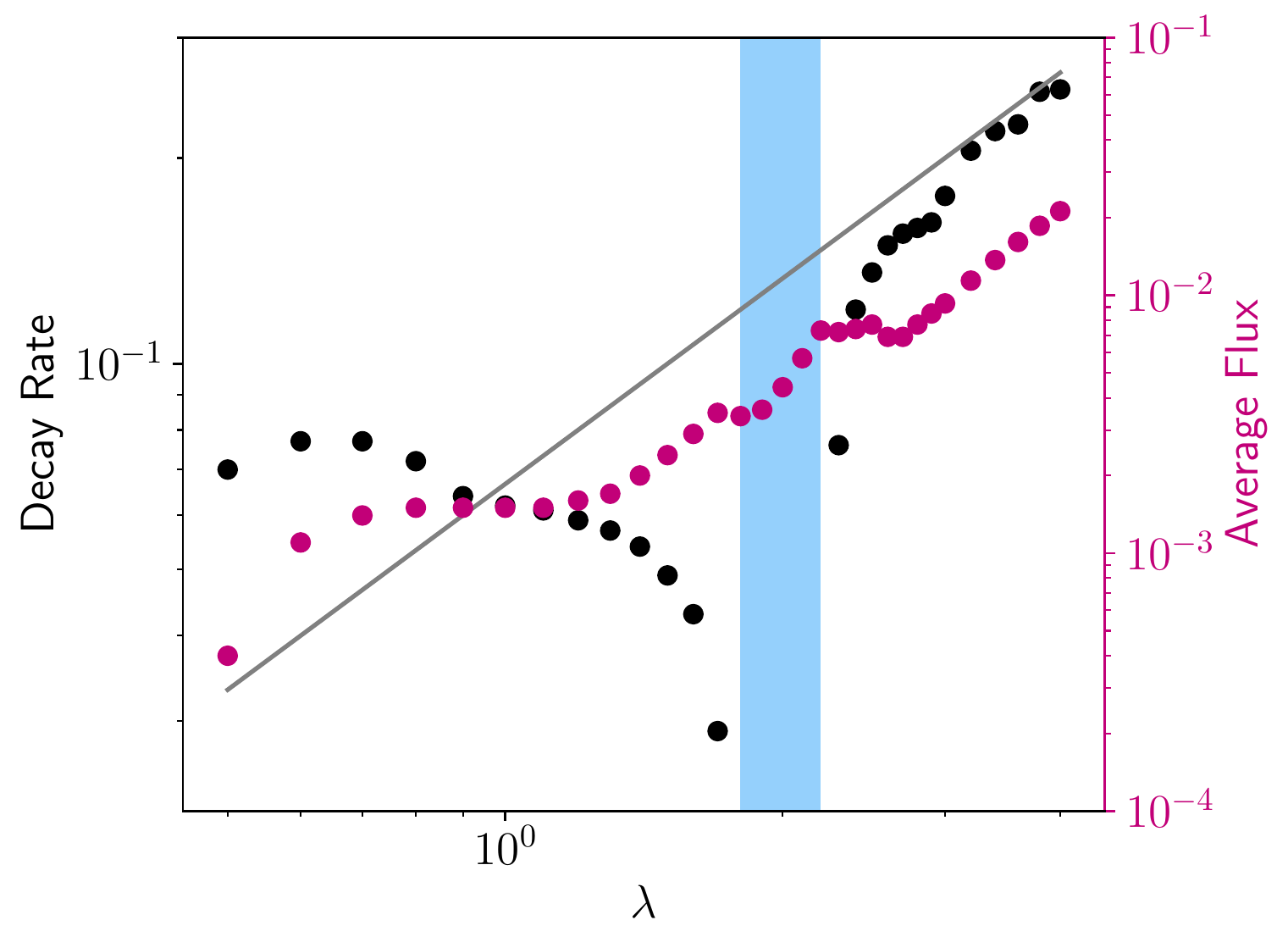}
        \caption{}
        \label{fig:2d_timescales:b}
    \end{subfigure}
    \caption{Panel (a) shows the transient behavior and decay of oscillations in the net flux due to small amplitude ($\gamma=0.01$) actuation. The density and effective rigidity for the main figure are $\xi=100$ and $\tilde E=10^{-4}$, respectively, and thus $\lambda=\xi\tilde E=0.01$. Transient behavior relaxes with a timescale $\lambda^{-1}=100$, while a persistent oscillation at the driving frequency damps with the timescale $2\pi\lambda^{-1}$. The inset shows the net flux at large amplitude ($\gamma=\pi/4$) and in the pulsatile regime ($\lambda=2$). In this case, the persistent oscillations are large and do not decay. Panel (b) shows the rate of decay of the persistent oscillations for large amplitude ($\gamma=0.5$) actuation with $\xi=1000$ over a range of $\tilde E$. Although both the decay rate and flux scale with $\lambda$ for large $\lambda$, a pulsatile regime exists in a range about $\lambda\approx 2$ in which these oscillations do not decay.}
    \label{fig:2d_timescales}
\end{figure}

In \Cref{section:rheology}, we performed rheological experiments on beds of fibers and found that for sufficiently dense fibers, the associated relaxation of the fluid-fiber system scaled with the density. With these results guiding our intuition, we expect a non-dimensional relaxation timescale $(\xi\tilde E)^{-1}$. To examine this, we again fix $\beta=k=2$ and $\gamma=\pi/4$. The density $\xi$ is set to $\xi=100$, and $\tilde E$ is chosen to be $10^{-4}$ so that the timescales associated with the oscillation ($2\pi$), the single fiber relaxation time $\tau_f=\tilde E^{-1}=10,000$ and the timescale associated with the dense fluid-fiber system $\tau_\lambda=(\tilde E\xi)^{-1}=\lambda^{-1}=100$ are all well-separated. Simulations are performed with $\Delta t=2\pi/20$ and $n=2^7$. The average flux in the purely fluid region above the fibers is computed and shown, as a function of time, in \Cref{fig:2d_timescales:a}. The $2\pi$ oscillation is clearly visible, as is the relaxation with timescale $\tau_\lambda=100$ to an equilibrium that drives a net symplectic\footnote{A \emph{symplectic} flow is one that moves in the direction of the metachronal wave; an \emph{antipleptic} flow moves in the opposite direction of the metachronal wave. Both flow directions can be realized in the system considered here, depending on parameters.} flux. For $t\gtrsim100$, the driven bed drives a very nearly steady flow, with persistent small oscillations decaying with a timescale $2\pi\lambda^{-1}$. This constant net flux is a reflection of the behavior of the fiber bed: the metachronal wave in the base angle of the fibers is translated into a traveling wave of deformation in the fibers that appears fixed in a frame moving at the speed of the driving metachronal wave.

At low amplitude, this is the generic behavior: transient decay to a pseudo-steady state traveling wave solution occurs with a timescale $\tau_\lambda=\lambda^{-1}$. For large amplitude, the picture becomes more complicated. To examine the behavior in this regime, we fix the effective density $\xi=1000$, and examine the effectiveness of a fiber carpet in propelling flow as the effective stiffness of the fibers is varied. We vary $\lambda=\tilde E\xi$ in the range of $[0.5, 4]$, and thus the highest value of $\tilde E$ that is probed is $4\times10^{-3}$. At these low values of effective rigidity, a single driven fiber would fail to cause significant flow, with transient behavior damping out over extremely long timescales. Nevertheless, in the many-fiber regime probed here, we expect the bed to act in a dramatically stiffer manner, driving strong flows, with large scale transients relaxing on the timescale $\lambda^{-1}$. \Cref{fig:2d_timescales:b} shows both the decay rate of the persistent oscillations, and the average net flux, as a function of $\lambda$. For sufficiently large $\lambda$, the same type of traveling wave solutions as in the small amplitude regime are observed, and both the decay rate and average flux scale linearly with $\lambda$. At lower system rigidity, complicated nonlinear phenomena arise. In particular, for a small range of $\lambda$ centered about $\lambda\approx2$, the persistent oscillations grow, rather than decay, and have minimum and maximum values an order of magnitude larger than the average flux. A plot of the net flux, as a function of time, for a bed with $\lambda=2.0$, is shown in the inset to \Cref{fig:2d_timescales:a}. In the range of $\lambda$ highlighted in blue in \Cref{fig:2d_timescales:b}, the system drives a pulsating, rather than steady flow. Despite the large fluctuations in the net flux driven by the pulsating beds as a function of time, the \emph{average} net flux driven is nearly unchanged from the trend seen in the non-pulsatile beds.

\section{Concluding Remarks}

We have presented a coarse-grained model for the fluid-structure interaction of well-aligned assemblies of inextensible, elastic fibers immersed in a Stokesian fluid, along with an efficient numerical scheme for evolving these systems in time. The coarse-grained system, given in \Cref{equation:simplified_model}, which we call the Brinkman-Elasticae (BE) model, is simple in form and allows for the development of analytical insights, while the numerical scheme provides rapid simulation of dense fluid-fiber systems, facilitating numerical experimentation and parameter sweeps in regimes inaccessible to direct simulation. The model (and numerical scheme) have been validated against fully three-dimensional Immersed Boundary simulations, and provide quantitative agreement with results found in other studies \cite{Nazockdast2016}. In this paper, we have modeled the fibers as inextensible Euler-Bernoulli elasticae. The choice of constitutive model is not critical to the development of the theory or numerics, and other choices could be treated with only minor modifications.

In \Cref{section:deflection,section:rheology}, we performed basic numerical studies of the properties of dense fiber beds. The primary insight from studying these simple systems is that \emph{their fundamental properties depend on the fiber density}. This behavior has also been observed in direct simulations \cite{Nazockdast2016}, where the relaxation timescale of a spherical fiber assembly was found to be approximately 25 times faster than the single-fiber elasto-viscous relaxation time. The Brinkman-Elasticae model provides a simple explanation for this behavior that agrees quantitatively with the direct simulations. Many important biophysical phenomena may be modeled via fluid-fiber systems, and these have been widely studied using direct simulation. Due to their high-computational cost, it is typical to reduce the number of fibers in direct simulations from the biologically relevant number to an (often significantly) smaller number of fibers that is feasible to simulate. The results of \Cref{section:deflection,section:rheology} should be a cause of some concern for researchers doing this: in the dense regime, using 100 fibers to represent 1000 fibers will change the effective relaxation timescale by a factor of 10! It is likely that judicious adjustment of parameters (e.g. by using fibers with an effective stiffness greater than the real fiber stiffness to compensate for the reduced density) will provide improved results for direct simulations, although this will have to be rigorously tested. The results of our analysis provide a simple framework to determine how parameters should be altered.

In \Cref{section:buckling,section:rectification}, we demonstrate how the model can be used to reduce the dimensionality of problems in simple geometries while still capturing the effects of fiber-bed density. In the first case, this allowed us to compute semi-analytical solutions to a buckling problem, while in the second case we were able to make use of fast numerics to compute optimal parameters for the design of a soft rectification device. We expect that this technique will be widely applicable. In a forthcoming contribution \cite{CSLGS2018}, we study the formation of large-scale vortical flows in the Drosophila melanogaster oocyte, and use a variation of the BE  model to determine a critical density at which motor-protein driven buckling of fiber beds is able to drive the coherent motion observed \emph{in vivo}.

Finally, in \Cref{section:waves}, we demonstrate the feasibility of using the BE model, in conjunction with the numerical scheme presented in \Cref{section:numerics}, to study higher-dimensional phenomenon. In the example studied, the beams were passive, and actuated only at the base. Despite its simplicity, this system was able to give rise to complex behaviors, including a transition from steady to pulsatile wave motion as the fiber bed density was varied. This transition occurs only when the fiber bed is sufficiently dense. We suspect that such surprising phenomenon will play a role in the biological functioning of similar systems, including ciliary pumping and the locomotion of squirmer bacteria. The BE model developed in this paper assumed passive beams; cilia and flagella exert active moments along their length, and some modification to the theory and numerical scheme will be required to appropriately model these systems. Additionally, for many biological systems, three-dimensional effects are important, and the dimensionality of the systems will not be able to be reduced. Although the numerical scheme presented here is readily generalizable, it uses an interpolation scheme based on triangulation of the Lagrangian mesh in order to facilitate communication of functions between Eulerian and Lagrangian frames. In three dimensions, this will be the dominant cost, and careful choices will have to be made in order to develop efficient numerics.

\section{Acknowledgements}
The authors would like to thank Eva Kanso and Raymond Goldstein for illuminating discussions. MJS thanks the support from NSF Grants DMR-1420073 (NYUMRSEC), DMS-1463962, and DMS-1620331.

\bibliographystyle{abbrv}
\bibliography{ARFM}

\begin{appendices}
\crefalias{section}{appsec}

\section{An improved Lagrangian to Eulerian interpolation operator}
\label{section:1d_model:improved_interpolation}

Interpolating the fiber forces from the Lagrangian grid to the underlying Cartesian grid is complicated by the jump in fiber forces across the fluid-fiber interface. The simplest scheme is to simply interpolate the forces to all Cartesian gridpoints inside of the fiber regime. Even if this interpolation is done to high-accuracy, this scheme only achieves first-order accuracy in space due to errors in grid cells cut by the fluid-fiber interface. As a clear example of this, consider what happens as the top of the fiber bed moves from just above an Eulerian gridpoint to just below that gridpoint. In the simplest scheme, as the tip moves past the gridpoint, the Eulerian representation of the force jumps. To fix this, we instead compute the value at the cell-centers of the grid to be the average of the force over the cell:
\begin{equation}
	\f_{i+1/2} = \frac{1}{\Delta z}\int_{z_i}^{z_{i+1}}\f(z)\,dz = \frac{1}{2}[\f(z_{i+1}) + \f(z_{i})] + \mathcal{O}(\Delta z^2),
\end{equation}
for all cells completely within the fiber region. For the cell split by the fluid-fiber interface, the force is instead computed to be $\f_{i+1/2}=\frac{Z-z_i}{2\Delta z}[\f(Z) + \f(z_i)]$, where $Z$ is the height of the fiber. Nodal values for $\f$ are then computed, to second order in $\Delta z$, by centered averages.

\end{appendices}

\end{document}